\documentclass[11pt,aps,axodraw,showpacs,nofootinbib,prd,aps,epsf,floats]{revtex4}
\usepackage{amsmath, amssymb, graphics,axodraw}

\newcommand{\mathsym}[1]{{}}

\usepackage{graphicx}
\usepackage{amsmath}
\usepackage{amssymb}
\usepackage{bm}
\newcommand{\be}{\begin{equation}}
\newcommand{\ee}{\end{equation}}
\newcommand{\bea}{\begin{eqnarray}}
\newcommand{\eea}{\end{eqnarray}}

\newcommand{\rem}[1]{}
\newsavebox{\PSLASH}
 \sbox{\PSLASH}{$p$\hspace{-1.8mm}/}
 
\renewcommand{\theequation}{\thesection.\arabic{equation}}
\newcounter{saveeqn}
\newcommand{\add}{\addtocounter{equation}{1}}
\newcommand{\alpheqn}{\setcounter{saveeqn}{\value{equation}}%
\setcounter{equation}{0}%
\renewcommand{\theequation}{\mbox{\thesection.\arabic{saveeqn}{\alph{equation}}}}}
\newcommand{\reseteqn}{\setcounter{equation}{\value{saveeqn}}%
\renewcommand{\theequation}{\thesection.\arabic{equation}}}

 \newsavebox{\notrightarrow}
 \sbox{\notrightarrow}{$\to$\hspace{-4mm}/}
 
 \newsavebox{\PARTIALSLASH}
 \sbox{\PARTIALSLASH}{$\partial$\hspace{-1.6mm}/}
 
 \newsavebox{\ASLASH}
 \sbox{\ASLASH}{$A$\hspace{-2.1mm}/}
 
 \newsavebox{\KSLASH}
 \sbox{\KSLASH}{$k$\hspace{-1.8mm}/}
 
 \newsavebox{\LSLASH}
 \sbox{\LSLASH}{$\ell$\hspace{-1.8mm}/}
 
 \newsavebox{\QSLASH}
 \sbox{\QSLASH}{$q$\hspace{-1.8mm}/}
 
 \newsavebox{\DSLASH}
 \sbox{\DSLASH}{$D$\hspace{-2.2mm}/}
 
 \newsavebox{\DbfSLASH}
 \sbox{\DbfSLASH}{${\mathbf D}$\hspace{-2.8mm}/}
 
 \newsavebox{\DELVECRIGHT}
 \sbox{\DELVECRIGHT}{$\stackrel{\rightarrow}{\partial}$}
 
 \newcommand{\blue}{\IfColor{\textCadetBlue}{}}
\newcommand{\black}{\IfColor{\textBlack}{}}
\newcommand{\red}{\IfColor{\textRed}{}}
\newcommand{\green}{\IfColor{\textOliveGreen}{}}
\newcommand{\lila}{\IfColor{\textRedViolet}{}}

\begin{document}

\begin{flushright}
SUT/P-2008/3a\\
0805.0078 [hep-ph]
\end{flushright}
\title{Improved ring potential of QED at finite temperature and\\
 in the presence of weak and strong magnetic field
}

\author{N. Sadooghi}\email{sadooghi@physics.sharif.edu}
\author{K. Sohrabi Anaraki}
\affiliation{Department of Physics, Sharif University of Technology,
P.O. Box 11155-9161, Tehran-Iran}

\begin{abstract}
Using the general structure of the vacuum polarization tensor
$\Pi_{\mu\nu}(k_{0},\mathbf{k})$ in the infrared (IR) limit,
$k_{0}\to 0$, the ring contribution to QED effective potential at
finite temperature and non-zero magnetic field is determined beyond
the static limit, $(k_{0}\to 0,\mathbf{k}\to \mathbf{0})$. The
resulting ring potential is then studied in weak and strong magnetic
field limit. In the limit of weak magnetic field, at high
temperature and for $\alpha\to 0$, the improved ring potential
consists of a term proportional to $T^{4}\alpha^{5/2}$, in addition
to the expected $T^{4}\alpha^{3/2}$ term arising from the static
limit. Here, $\alpha$ is the fine structure constant. In the limit
of strong magnetic field, where QED dynamics is dominated by the
lowest Landau level (LLL), the ring potential includes a novel term
consisting of dilogarithmic function
$(eB)\mbox{Li}_{2}\left(-\frac{2\alpha}{\pi}\frac{eB}{m^{2}}\right)$.
Using the ring improved (one-loop) effective potential including the
one-loop effective potential and ring potential in the IR limit, the
dynamical chiral symmetry breaking of QED is studied at finite
temperature and in the presence of strong magnetic field. The gap
equation, the dynamical mass and the critical temperature of QED in
the regime of LLL dominance are determined in the improved IR as
well as in the static limit. For a given value of magnetic field,
the improved ring potential is shown to be more efficient in
decreasing the critical temperature arising from one-loop effective
potential.
\end{abstract}
 \maketitle
\section{Introduction}
\subsection{Motivation}
The existence of phase transitions in the early universe has been a
question that has preoccupied a generation of cosmologists. Early
on, Kirzhnits \cite{kirzhnits} found that the symmetry between the
weak and electromagnetic interactions would be restored at high
temperatures. This result was soon complemented by similar works by
Weinberg \cite{weinberg}, Dolan and Jackiw \cite{dolan}, Kirzhnits
and Linde \cite{linde}. In particular, there has been much interest
in the nature of the electroweak phase transition (EWPT), which is
closely related to the still unsolved problem of baryogenesis. It
has been known since Sakharov's work that there are three necessary
(but not sufficient) conditions for the baryon asymmetry of the
Universe to develop \cite{sakharov}. \textit{First}, we need
interactions that do not conserve baryon number B, otherwise no
asymmetry could be produced in the first place. \textit{Second}, C
and CP symmetry must be violated, in order to differentiate between
matter and antimatter, otherwise the same rate of baryons and
antibaryons would be produced leading to zero net baryon number.
\textit{Third}, the universe in his history, must have experienced a
departure from thermal equilibrium. In other words, the above $C$
and $CP$ violating processes should have been occurred in a state
out of equilibrium, otherwise the net baryon number cannot change in
time. The Standard Model (SM) of electroweak interaction meets all
the above requirements to generate a baryon asymmetry during the
EWPT, provided that this last be of first order.
\par
The type of symmetry restoring phase transition is determined by the
behavior of the effective or thermodynamic potential. The fact that
the symmetry is restored at high temperatures is a result of the
$T^{2}m^{2}(v)$ term as the leading order contribution from the
thermal fluctuations of the field. This term appears in the
perturbatively calculable one-loop effective potential. Here, $T$ is
the temperature and $m^{2}$ is the mass squared proportional to the
expectation value of some classical scalar (Higgs) field $v$. As the
temperature is increased, the contribution from thermal fluctuation
dominates the negative-mass-squared term in the tree level potential
and symmetry will be restored. According to this one-loop
approximation, it can be shown that the phase transition is of
second order \cite{kirzhnits}-\cite{linde} and that the effective
potential includes terms proportional to $m^{3}(v)T$ and therefore
is imaginary when the mass squared is negative. As it was shown in
\cite{takahashi}, however, the appearance of imaginary terms in the
one-loop effective potential indicates the breakdown of the
semiclassical loop expansion through IR singularities. As it is then
argued in \cite{carrington}, these IR singularities are included in
the ring (plasmon or daisy) diagrams of the theory. In
\cite{carrington}, the nonperturbative ring contribution to the
effective potential is calculated. It is shown to have the general
structure
\begin{eqnarray}\label{XX1}
V_{ring}(v)=\frac{T}{12\pi}\mbox{Tr}\left(\bigg[m^{2}(v)+\Pi_{00}(0)\bigg]^{3/2}-m^{3}(v)\right),
\end{eqnarray}
where $\Pi_{00}(0)\equiv\Pi_{00}(n=0,{\mathbf{k}}\to {\mathbf{0}})$
is the vacuum polarization in the \textit{static (zero momentum)
limit}.\footnote{In \cite{bordag, demchik}, this limit is called the
`'Debye mass'' limit.} Adding this contribution to the one-loop
effective potential, it is then shown that the SM has indeed a first
order phase transition and the critical temperature is much lower
than the temperature arising from one-loop effective potential
\cite{carrington}. As for the question of baryon asymmetry, however,
it is known that neither the amount of CP violation within the
minimal SM nor the strength of the EWPT are enough to generate
sizable baryon number \cite{gavela-kajantie}.\footnote{Other
possibilities to explain the generation of baryon number during the
EWPT include minimal and non-minimal supersymmetric model.}
\par
In the recent years, due to the observation that magnetic fields are
able to generate a stronger first order EWPT
\cite{giovannini}-\cite{ayala-2}, the electroweak baryogenesis is
revisited within the minimal SM and in the presence of external
hypermagnetic fields (for a review see \cite{demchik}). In
\cite{bordag}, the ring improved effective potential of SM,
including one-loop effective potential and ring contributions, is
calculated explicitly. Here, as in \cite{carrington}, the ring
potential is determined in the \textit{static (zero momentum)
limit}, where, in the presence of external magnetic field $B$,
$\Pi_{00}(0)$  in (\ref{XX1}) is defnined by $\Pi_{00}(0)\equiv
\Pi_{00}(n=0,{\mathbf{k}}\to {\mathbf{0}};eB)$. It is found that for
the field strengths $10^{23}-10^{24}$ Gau\ss, the phase transition
is of first order but the baryogenesis condition $\frac{\langle
v\rangle}{T_{c}}>1-1.5$ is still not satisfied.\footnote{In the
electroweak SM at finite $T$, the existence of baryon number
violation is realized by means of its vacuum structure through
sphaleron mediated processes. The sphaleron transition between
different topological distinct vacua is associated to baryon number
$n_{B}-n_{\bar{B}}$ violation and can either induce or wash out a
baryon asymmetry. In order to satisfy the baryon asymmetry condition
during the baryogenesis process the rate of baryon violating
transitions between different topological vacua must be suppressed
in the broken phase, when the universe returns to thermal
equilibrium. In other words, the sphaleron transitions must be
slower than the expansion of the universe and this in turn
translates into the condition $\frac{\langle v\rangle
}{T_{c}}>1-1.5$, where $\langle v\rangle$ is the Higgs mass
\cite{gavela-kajantie, ayala}.} To improve this condition one is
looking for possibilities to decrease the critical temperature
$T_{c}$ of EWPT.
\par
Motivated by the previous facts and as the first step to improve the
results in \cite{bordag}-\cite{ayala-2} to solve the problem of
baryogenesis within the minimal SM, we will go in this paper
\textit{beyond the static (zero momentum) limit} and will calculate
the ring improved effective potential of QED at finite temperature
and in the presence of constant magnetic field in a certain
\textit{IR limit}. As we have seen above, the ring part of the ring
improved effective potential is given by QED vacuum polarization
tensor, $\Pi_{\mu\nu}(n,{\mathbf{k}})$, at finite temperature. In
this paper, in contrast to previous works, \textit{e.g.}
\cite{kapusta,carrington,bordag}, we will determine the ring
potential using the vacuum polarization tensor in the \textit{IR
limit}, which is particularly characterized by
$(n=0,{\mathbf{k}}\neq \mathbf{0})$.\footnote{See Sect. I.B for
technical details.}
\par
This paper consists of two parts: In the first part of the paper,
using the diagonalized form of the vacuum polarization tensor in the
IR limit $\Pi_{\mu\nu}\left(n=0,{\mathbf{k}}\neq
{\mathbf{0}}\right)$, the general structure of the ring potential
will be determined. Then, the ring potential in the improved IR
limit will be calculated explicitly in the presence of weak and
strong magnetic field and the resulting expressions will be compared
with the conventionally used static ring potential in these limits.
In the second part of the paper, the dynamical chiral symmetry
breaking of QED at finite temperature and in the presence of strong
magnetic field will be studied. Our main goal to study this example
is to answer the question how efficient the improved IR
approximation is in decreasing the critical temperature of the above
dynamical chiral symmetry breaking. Comparing the effect of the ring
potential in the IR limit numerically with the effect of the ring
potential in various static limits, we arrive at the conclusion that
the improved IR limit is more efficient in decreasing the critical
temperature arising from one-loop effective potential $T_{c}^{(1)}$.
Here, $T_{c}^{(1)}$ is the critical temperature that arises from
one-loop effective potential in the lowest order of $\alpha$
correction (ladder approximation). Defining a ring improved critical
temperature ${\cal{T}}_{c}$ containing the contribution of one-loop
effective potential in ladder approximation and the ring
contributions, it turns out that the difference of the efficiency
factors defined as $\eta\equiv 1-\frac{T_{c}^{(1)}}{{\cal{T}}_{c}}$
is more than $60\%$ for the magnetic field $B\approx 10^{16}$
Gau\ss. The above conclusion is promising in view of the problem of
EWPT in the electroweak SM in the presence of weak/strong
hypermagnetic field \cite{bordag}-\cite{ayala-2}. Here, one is
looking for a possibility to decrease the critical temperature of
EWPT in order to improve the baryogenesis condition $\frac{\langle
v\rangle}{T_{c}}>1-1.5$. Using the improved ring potential in the IR
limit in determining the critical temperature of EWPT in SM may
improve the results from \cite{bordag}-\cite{ayala-2}.
\par
The organization of this paper is as follows: In Sect. I.B, we will
review some technical details on ring diagrams in thermal field
theory without magnetic field. In particular, we will review the
well-known results of QED ring contribution to the effective
potential in the static limit \cite{kapusta}. In Sect. II, we will
determine the vacuum polarization tensor of QED in the IR limit,
$\Pi_{\mu\nu}(n=0,{\mathbf{k}}\neq {\mathbf{0}};eB)$. Here, we will
use some results from \cite{shabad} and \cite{alexandre}. In
particular, we will the method in \cite{shabad} to diagonalize the
vacuum polarization tensor in certain basis. In Sect. III, using the
diagonalized $\Pi_{\mu\nu}(n=0,{\mathbf{k}}\neq {\mathbf{0}};eB)$,
we will be able to determine the general structure of the ring
contribution to QED effective potential in the presence of external
magnetic field at finite temperature. In Sect. III.A and III.B, the
resulting ring improved effective potential in the IR limit will be
considered first in the weak and then in the strong magnetic field
limit.
\par
In the weak magnetic field limit, at high temperature and for
$\alpha\to 0$, the ring potential in the IR limit consists of a term
proportional to $T^{4}\alpha^{5/2}$, in addition to the expected
$T^{4}\alpha^{3/2}$ term arising from the static limit. Here,
$\alpha\equiv \frac{e^{2}}{4\pi}$ is the fine structure constant.
This term can be viewed as a nonperturbative correction to QED
effective potential in addition to the perturbative loop corrections
to this potential in the corresponding $\alpha^{5/2}$ order. Note
that, using Hard Thermal Loop (HTL) expansion \cite{braaten},
similar contributions of order $\alpha_{s}^{3/2}$ and
$\alpha_{s}^{5/2}$ are previously found in QCD effective potential
at finite temperature and zero magnetic field (see \cite{alpha5-2}
and the references therein).
\par
In the strong magnetic field limit, the ring potential of QED at
finite temperature includes a novel term consisting of a
dilogarithmic function
$(eB)\mbox{Li}_{2}\left(-\frac{2\alpha}{\pi}\frac{eB}{m^{2}}\right)$.
As in the weak magnetic field limit, similar contribution to QCD
effective potential at finite temperature and zero magnetic field is
calculated in \cite{toimela}. Here, going beyond the static limit,
it is shown that QCD effective potential consists of an unexpected
$g_{s}^{4}\ln g_{s}$ term. The appearance of a similar term in the
QED ring potential at finite temperature and in the presence of
strong magnetic field is however expected due to the well-known
phenomenon of  magnetic catalysis \cite{miransky-2,
miransky1-5}\footnote{The magnetic catalysis has wide applications
in condensed matter physics \cite{cond-mat} and cosmology
\cite{cosmology}.}; In the limit of strong magnetic field, QED
dynamics is believed to be dominated by the Lowest Landau Level
(LLL), where the chiral symmetry of the theory is broken by a
dynamically generated fermion mass. As a consequence of a
dimensional reduction from $D$ to $D-2$, four dimensional QED
exhibits confining properties like ordinary confining Abelian or
non-Abelian gauge theories without magnetic field \cite{miransky-2,
miransky1-5}.\footnote{A two dimensional Schwinger model is an
example of a confining Abelian gauge theory. It is known that four
dimensional QED in the presence of strong magnetic field is reduced
to a two dimensional theory, very similar to the ordinary Schwinger
model without external magnetic field.}
\par
To compare the ring potential in the IR limit from III.B with the
static ring potential in the LLL, we will calculate in  Sect. III.C
and III.D, the static ring potential in the strong magnetic field
limit in two different methods. In the first approach, we will
calculate the ring potential after taking the limit $eB\to \infty$.
In the second approach, we will take the limit $eB\to \infty$ after
calculating the ring potential mathematically. We will arrive at two
different results. This difference can be interpreted as a direct
consequence of the dynamics of QED in the LLL and the
above-mentioned dimensional reduction in the regime of LLL
dominance.
\par
In the second part of the paper, we will use the results from
III.B-III.D to study the dynamical chiral symmetry breaking of QED
at finite temperature and in the presence of strong magnetic field
in [see Sect. IV.]\footnote{Recently chiral transition in strong
magnetic field is studied in \cite{recent-chiral}.} As we have
mentioned before, we are indeed interested in the effect of our
improved ring potential in decreasing the critical temperature of
chiral symmetry breaking of QED at finite temperature and in the
presence of strong magnetic field. Using the ring improved effective
potentials in the IR and the static limit, the gap equation, the
dynamical mass and the critical temperature $T_{c}$ of QED in the
LLL are determined. To have a first estimate on the efficiency of
the improved IR limit in decreasing the critical temperature arising
from one-loop effective potential in the ladder approximation,
$T_{c}^{(1)}$, we will compare the ratio $u\equiv
T_{c}^{(1)}/{\cal{T}}_{c}$ for magnetic field $eB$ in the interval
$[10^{-8}, 1]$ GeV$^{2}$. Here, ${\cal{T}}_{c}$ is the improved
critical temperature in the ladder approximation. This range
corresponding to $B\in[1.7\times 10^{12}, 1.7\times 10^{20}]$
Gau\ss, is phenomenologically relevant in the astrophysics of
neutron stars, where the strength of the magnetic field is of order
$10^{13}-10^{15}$ Gau\ss\ (see \cite{shabad-neutron} and the
references therein). It is also relevant in the heavy ion
experiments, where it is believed that the magnetic field in the
center of gold-gold collision is $eB\sim 10^{2}-10^{3}$ MeV$^{2}$ or
$B\sim 10^{16}-10^{17}$ Gau\ss\ \cite{mclerran}. Defining further an
efficiency factor $\eta\equiv1-u^{-1}$ for the IR and static
approximation, we will be able to compare the IR limit with various
static limits. According to our numerical results presented in IV.E,
for a given value of the magnetic field, the IR limit seems to be
more efficient in decreasing the critical temperature arising from
one-loop effective potential. The maximum value of $\eta$ in the IR
limit $\sim 63\%$ for $B\approx 1.6\times 10^{16}$ Gau\ss.  Our
results are summarized in Sect. V.  In Appendix A, we will define
the one-loop effective potential and the ring potential in the LLL
at zero temperature. In Appendix B, the gap equation arising from
one-loop effective potential and ring improved effective potential
at zero temperature is derived. Appendix C generalizes the
definition from Appendix B to finite temperature.
\subsection{Technical details}
\subsubsection*{QED ring potential at $T\neq 0$ and
$B=0$ in the static limit}  \noindent In this section, we will
review the results in \cite{kapusta} on QED ring potential at finite
temperature using the static (zero momentum) limit. Eventually we
will argue why an approximation beyond the zero momentum limit is
necessary when we turn on a strong magnetic field.
\par
Let us just start with the partition function of QED at finite
temperature
\begin{eqnarray}\label{X1}
Z=\int{\cal{D}}c\ {\cal{D}}\bar{c}\ {\cal{D}}A_{\mu}\ {\cal{D}}\psi\
{\cal{D}}\bar{\psi}\ \exp\left(\int_{0}^{\beta} d\tau \int d^{3}x\
{\cal{L}}\right),
\end{eqnarray}
where ${\cal{L}}={\cal{L}}_{0}+{\cal{L}}_{I}$. Here, ${\cal{L}}_{0}$
is the free part of the Lagrangian
\begin{eqnarray*}
{\cal{L}}_{0}=\bar{\psi}\left(i\gamma_{\mu}\partial^{\mu}-m\right)\psi-\frac{1}{4}F_{\mu\nu}F^{\mu\nu}
-\frac{1}{2\xi}\left(\partial^{\mu}A_{\mu}\right)^{2}+\left(\partial^{\mu}\bar{c}\right)\left(\partial_{\mu}c\right),
\end{eqnarray*}
and ${\cal{L}}_{I}$ the interaction Lagrangian
\begin{eqnarray*}
{\cal{L}}_{I}=-e\bar{\psi}A_{\mu}\gamma^{\mu}\psi.
\end{eqnarray*}
Using the above Lagrangian the free photon propagator of the theory
is given by
\begin{eqnarray*}
D_{0}^{\mu\nu}&=&\frac{1}{k^{2}}\bigg\{g^{\mu\nu}-(1-\xi)\frac{k^{\mu}k^{\nu}}{k^{2}}\bigg\},\qquad
k_{0}\equiv 2\pi i nT;
\end{eqnarray*}
and $n\in ]-\infty,+\infty[$ labels the Matsubara frequencies for
the bosons. The photon-self energy at one-loop level is \
\begin{eqnarray}\label{X2}
\Pi_{\mu\nu}={\cal{D}}_{\mu\nu}^{-1}-D_{0\mu\nu}^{-1}.
\end{eqnarray}
Using the corresponding Ward identities arising from the gauge
invariance of the theory
\begin{eqnarray*}
k_{\mu}\Pi^{\mu\nu}=0,\qquad k^{\mu}k^{\nu}{\cal{D}}_{\mu\nu}=\xi,
\end{eqnarray*}
it is possible to determine the general structure of the photon
propagator ${\cal{D}}_{\mu\nu}$ and the corresponding photon
self-energy $\Pi_{\mu\nu}$ as a symmetric second-rank tensors. Here,
$\xi$ denotes the covariant gauge. At finite temperature, the most
general tensor of this type is a linear combination of $g_{\mu\nu}$,
$k_{\mu}k_{\nu}$, $u_{\mu}u_{\nu}$, and
$k_{\mu}u_{\nu}+k_{\nu}u_{\mu}$, where $u_{\mu}=(1,0,0,0)$ specifies
the rest frame of the many body system. Using the above properties
${\cal{D}}_{\mu\nu}$ and $\Pi_{\mu\nu}$ have the general form
\begin{eqnarray}\label{X3}
\Pi^{\mu\nu}&=&GP_{T}^{\mu\nu}+FP_{L}^{\mu\nu},\nonumber\\
{\cal{D}}^{\mu\nu}&=&\frac{1}{G-k^{2}}P_{T}^{\mu\nu}+\frac{1}{F-k^{2}}P_{L}^{\mu\nu}
+\frac{\xi}{k^{2}}\frac{k^{\mu}k^{\nu}}{k^{2}},
\end{eqnarray}
where $F$ and $G$ are scalar function of $k^{0}$ and $\omega\equiv
|{\mathbf{k}}|$. They are of order $e^{2}$ in the QED coupling
constant $e$. The projector operators $P_{T}$ and $P_{L}$ in
(\ref{X3}) are given by
\begin{eqnarray}\label{X4}
P_{T}^{00}&=&P_{T}^{0i}=P_{T}^{i0}=0,\nonumber\\
P_{T}^{ij}&=&\delta^{ij}-\frac{k^{i}k^{j}}{{\mathbf{k}}^{2}},\nonumber\\
P_{L}^{\mu\nu}&=&\frac{k^{\mu}k^{\nu}}{k^{2}}-g^{\mu\nu}-P_{T}^{\mu\nu}.
\end{eqnarray}
They have the properties
\begin{eqnarray}\label{X5}
P_{L}^{\mu\rho}P_{L\rho\nu}=-P^{\mu}_{L\nu},&\qquad&P_{T}^{\mu\rho}P_{T\rho\nu}=-P^{\mu}_{T\nu},\nonumber\\
k_{\mu}P_{T}^{\mu\nu}=k_{\mu}P_{L}^{\mu\nu}=0,&\qquad&P_{L}^{\mu\rho}P_{T\rho\nu}=0,\nonumber\\
P_{L\mu}^{\mu}=-1,&\qquad&P_{T\mu}^{\mu}=-2.
\end{eqnarray}
As it is shown in \cite{kapusta}, the lowest correction to the QED
thermodynamic potential $V_{2}$ is of order $e^2$
\begin{eqnarray}\label{X6}
\SetScale{0.35}
  \begin{picture}(20,20)(0,0)
    \SetWidth{2}
    \Text(-40,0)[l]{$\ln Z_{2}=-\frac{1}{2}$}
    \CArc(100,0)(50,270,630)
    \Photon(50,0)(150,0){7.5}{5}
    \Text(60,0)[l]{$=-\frac{1}{2}$}
    \CArc(310,40)(25,270,630)
    \PhotonArc(310,0)(46.25,122,414){-7.5}{9.5}
    \Text(105,11)[lb]{$\Pi$}
  \end{picture}
\end{eqnarray}
\par\vspace{0.3cm}
Note that these two diagrams are equivalent \cite{kapusta}. As for
the next correction to the thermodynamic effective potential, it is
not of order $e^{4}$ as expected, but of order $e^{3}$ when $T>0$.
It arises from the set of ring diagrams shown in Fig. 1.\footnote{In
some recent references, as in \cite{carrington} and \cite{ayala,
bordag, demchik} the ring potential $V_{ring}$ is defined by adding
(\ref{X6}) to the expression in (\ref{X7}), which is given by the
series of diagrams in Fig. 1. This means that the series over
$D_{0}^{\mu\rho}\left(n,{\mathbf{k}}\right)
\Pi_{\rho\mu}\left(n,{\mathbf{k}}\right)$ on the first line of
(\ref{X7}) starts in \cite{carrington} and \cite{ayala, bordag,
demchik} from $N=1$. In this paper, we will use in the latter
definition, where the sum over $N$ starts from $N=1$ on the first
line of (\ref{X7}) [see (\ref{A9}) for the definition of the ring
potential in this paper].}
\begin{eqnarray}\label{X7}
V_{ring}&=&-\frac{T}{2}\sum\limits_{n=-\infty}^{+\infty}\int\frac{d^{3}k}{(2\pi)^{3}}\sum\limits_{N=2}^{\infty}\frac{(-1)^{N-1}}{N}\bigg[
D_{0}^{\mu\rho}\left(n,{\mathbf{k}}\right)
\Pi_{\rho\mu}\left(n,{\mathbf{k}}\right)
\bigg]^{N}\nonumber\\
&=&-\frac{T}{2}\sum\limits_{n=-\infty}^{+\infty}\int\frac{d^{3}k}{(2\pi)^{3}}\bigg\{\ln[1+D_{0}^{\mu\rho}(n,{\mathbf{k}})\Pi_{\rho\mu}(n,{\mathbf{k}})]
-D_{0}^{\mu\rho}(n,{\mathbf{k}})\Pi_{\rho\mu}(n,{\mathbf{k}})\bigg\}.
\end{eqnarray}
\vspace{0.3cm}\begin{figure}[h] \SetScale{0.5}
  \begin{picture}(301,120) (100,-100)
    \SetWidth{1.2}
    \SetColor{Black}
   \CArc(389,-60)(15,270,630)
   \CArc(389,-150)(15,270,630)
   \Text(191.5,-33)[lb]{$\Pi$}
   \Text(191.5,-79)[lb]{$\Pi$}
    \Text(-159,-125)[lb]{$D_{0}(k)$}
    \PhotonArc(392.75,-105)(48.75,112.62,247.38){-7.5}{4.5}
    \PhotonArc(385.25,-105)(48.75,-67.38,67.38){-7.5}{4.5}
 \CArc(520,-90)(15,270,630)
 \CArc(600,-90)(15,270,630)
 \CArc(560,-150)(15,270,630)
 \Text(297.5,-48)[lb]{$\Pi$}
 \Text(257.2,-48)[lb]{$\Pi$}
 \Text(277,-79)[lb]{$\Pi$}
\PhotonArc(560.25,-80)(34.5,9,173){-4.5}{4.5}
\PhotonArc(554,-110)(36,172,258){-4.5}{2.5}
\PhotonArc(567,-110)(36,282,369){-4.5}{2.5}
\Text(350,-45)[lb]{\Large{\Black{$\cdots$}}}
  \end{picture}
  \vspace{-0.5cm}\caption{Diagrams contributing to the ring potential (\ref{X7}). In the LLL approximation, the wavy lines are free photon propagators
  $D_{0}^{\mu\nu}$ and the circles indicate
  the insertion of vacuum polarization tensor $\Pi_{\mu\nu}$. In this approximation, $\Pi_{\mu\nu}$ are calculated using the free fermion propagator in the LLL.}
\end{figure}
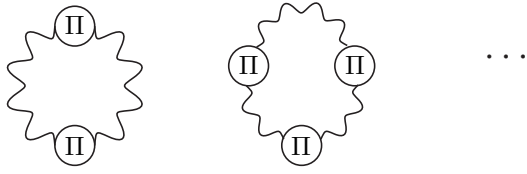
\par\noindent
Plugging $D_{0}^{\mu\nu}$ and $\Pi_{\mu\nu}$  in (\ref{X7}) and
using the properties of the projection operators $P_{L}$ and $P_{T}$
from (\ref{X4}) and (\ref{X5}), (\ref{X7}) is given by
\begin{eqnarray}\label{X8}
V_{ring}=-\frac{T}{2}\sum\limits_{n=-\infty}^{+\infty}\int\frac{d^{3}k}{(2\pi)^{3}}\bigg\{
2\bigg[ \ln\left(1-\frac{G(n,\omega)}{k^{2}}\right)
+\frac{G(n,\omega)}{k^{2}}
\bigg]+\ln\left(1-\frac{F(n,\omega)}{k^{2}}\right)+\frac{F(n,\omega)}{k^{2}}
\bigg\}.
\end{eqnarray}
Here, $-k^{2}=\omega^{2}+4\pi^{2}n^{2}T^{2}$. To determine the next
to leading order term in the effective potential, $V_{ring}$ in the
\textit{static (zero momentum) limit}, an IR divergent term,
$V_{e^{3}}$, is added to and subtracted from the ring potential
(\ref{X8}) \cite{kapusta}. It is therefore given by
$$V_{ring}={{V}}_{e^3}+{{V}}_{e^{4}},$$
where
\begin{eqnarray}\label{X9}
{{V}}_{e^{3}}=-\frac{T}{2}\int\frac{d^{3}k}{(2\pi)^{3}}\bigg\{
2\bigg[ \ln\left(1+\frac{G(0,{\mathbf{0}})}{\omega^{2}}\right)
-\frac{G(0,{\mathbf{0}})}{\omega^{2}}
\bigg]+\ln\left(1+\frac{F(0,{\mathbf{0}})}{\omega^{2}}\right)-\frac{F(0,{\mathbf{0}})}{\omega^{2}}
\bigg\},
\end{eqnarray}
and
\begin{eqnarray}\label{X10}
{{V}}_{e^4}=-\frac{T}{2}\sum\limits_{n=-\infty}^{+\infty}\int\frac{d^{3}k}{(2\pi)^{3}}\bigg\{
2\bigg[ \ln\left(1-\frac{G(n,\omega)}{k^{2}}\right)
+\frac{G(n,\omega)}{k^{2}}
\bigg]+\ln\left(1-\frac{F(n,\omega)}{k^{2}}\right)+\frac{F(n,\omega)}{k^{2}}
\bigg\}\nonumber\\
+\frac{T}{2}\int\frac{d^{3}k}{(2\pi)^{3}}\bigg\{ 2\bigg[
\ln\left(1+\frac{G(0,{\mathbf{0}})}{\omega^{2}}\right)
-\frac{G(0,{\mathbf{0}})}{\omega^{2}}
\bigg]+\ln\left(1+\frac{F(0,{\mathbf{0}})}{\omega^{2}}\right)-\frac{F(0,{\mathbf{0}})}{\omega^{2}}
\bigg\}.
\end{eqnarray}
Carrying out the three-dimensional integration over $\mathbf{k}$,
$V_{e^{3}}$ is given by
\begin{eqnarray}\label{X11}
V_{e^{3}}=\frac{T}{12\pi}\bigg[2G^{3/2}\left(0,{\mathbf{0}}\right)+
F^{3/2}\left(0,{\mathbf{0}}\right)\bigg]=\frac{T}{12\pi}F^{3/2}\left(0,{\mathbf{0}}\right),
\end{eqnarray}
where $G\left(0,{\mathbf{0}}\right)=0$ is used. This is the
well-known nonperturbative $e^{3}$ (or equivalently $\alpha^{3/2}$)
contribution to the thermodynamic potential arising from the ring
(plasmon) part of this potential. As for $V_{e^{4}}$ from
(\ref{X10}), it can now be expanded in the orders of $e$ and is thus
given by
\begin{eqnarray}\label{X12}
V_{e^{4}}&=&\frac{T}{4}\int\frac{d^{3}k}{(2\pi)^{3}}\left\{\sum\limits_{n\neq
0}\bigg[2\left(\frac{G(n,\omega)}{k^{2}}\right)^{2}+
\left(\frac{F(n,\omega)}{k^{2}}\right)^{2}\bigg]+\bigg[2\left(\frac{G(0,\omega)}{\omega^{2}}\right)^{2}+
\left(\frac{F(0,\omega)}{\omega^{2}}\right)^{2}\bigg]\right.\nonumber\\
&&\left.-\bigg[2\left(\frac{G(0,{\mathbf{0}})}{\omega^{2}}\right)^{2}+
\left(\frac{F(0,{\mathbf{0}})}{\omega^{2}}\right)^{2}\bigg]\right\},
\end{eqnarray}
which is of order $e^{4}$ in the QED coupling constant. Note that
expanding the logarithms in $V_{e^{4}}$ in the orders of $e$, is
only possible because it remains IR-finite. This is the case in QED
but not in the confining gauge theories like QCD. The ring
contribution to the effective potential of QCD is calculated in
\cite{toimela}, where it is shown that, in particular $F(0,\omega)$
in (\ref{X12}) is a so called ``dangerous'' term containing a
logarithmically divergent part. To remove this term, it is necessary
to go beyond the static limit. This is done in \cite{toimela}, where
it is shown that the plasmon potential in QCD effective potential
contains besides the $g_{s}^{3}$-term a contribution of order
$g_{s}^{4}\ln g_{s}$ at any non-zero temperature. This contribution
arises from the term $\Pi_{\mu\nu}(n=0, {\mathbf{k}}\neq
\mathbf{0})$ in the IR limit.
\par
In the present paper, we are in particular interested in the ring
improved effective thermodynamic potential of QED in a
\textit{strong} magnetic field. It is believed that at weak
coupling, the QED dynamics is dominated by the LLL. It is known that
in the regime of LLL dominance the ordinary four dimensional QED is
reduced to a two dimensional confining theory, like QCD, where the
original chiral symmetry is broken by a dynamically generated
fermion mass. Comparing to \cite{toimela}, we expect therefore to
have some ``dangerous'' logarithmically divergent terms in the
plasmon potential, if we use the static (zero momentum) limit
$(n=0,\mathbf{k}=0)$. To avoid these types of difficulties, we will
determine, in the next section, the general structure of QED vacuum
polarization tensor in the IR limit $n=0$ (or equivalently $k_{0}\to
0$) as a function of finite three-momentum ${\mathbf{k}}$. To
determine the ring potential in the limit of \textit{weak} and
\textit{strong} magnetic field, we will use (\ref{X7}), where only
$n=0$  is considered.
\section{QED
in a constant magnetic field at zero and finite temperature}
\setcounter{equation}{0}\noindent In the first part of this section,
we will briefly review the characteristics of QED in a constant
magnetic field at zero temperature. In particular, we will consider
the fermion and photon propagators in a \textit{strong} magnetic
field, where the QED dynamics is dominated by LLL. Then, in the
second part, we will determine QED vacuum polarization tensor in a
constant magnetic field at finite temperature and in a certain IR
limit.
\subsection{Fermions and photons in a strong magnetic field at zero and non-zero temperature}
Let us start with QED Lagrangian density at zero temperature
\begin{eqnarray}\label{Z1}
{\cal{L}}=\bar{\psi}\left(i\gamma^{\mu}D_{\mu}^{ext.}-m\right)\psi-\frac{1}{4}F_{\mu\nu}F^{\mu\nu},
\end{eqnarray}
where $D_{\mu}^{ext.}\equiv \partial_{\mu}+ieA_{\mu}^{ext.}(x)$,
with the gauge field $A_{\mu}\equiv A_{\mu}^{ext.}$ describing an
external magnetic field. In this paper, we will choose the symmetric
gauge
\begin{eqnarray*}
A_{\mu}^{ext.}=\frac{B}{2}\left(0,x_{2},-x_{1},0\right),
\end{eqnarray*}
that leads to a magnetic field in the $x_{3}$ direction. From now
on, the longitudinal $\left(x_0,x_3\right)$ components will be
indicated as ${\mathbf{x}}_{\|}$ and the transverse directions
$\left(x_1,x_2\right)$ components by ${\mathbf{x}}_{\perp}$. The
free (bare) fermion propagator of a four dimensional QED in a
constant magnetic field at zero temperature can be found using the
Schwinger's proper-time formalism \cite{schwinger} from
$(i\gamma^{\mu}D_{\mu}^{ext.}-m)^{-1}$. In the above symmetric
gauge, the free fermion propagator in a constant magnetic field is
given by
\begin{eqnarray}\label{Z2}
{\cal{S}}_{F}(x,y)&=&\exp\left(\frac{ie}{2}\left(x-y\right)^{\mu}A_{\mu}^{ext.}(x+y)\right)S(x-y)\nonumber\\
&=& e^{\frac{ieB}{2}\epsilon^{ab}x_{a}y_{b}} S(x-y),\qquad a,b=1,2.
\end{eqnarray}
Here, the first factor containing the external $A_{\mu}^{ext.}$ is
the Schwinger line integral \cite{schwinger} and $S(x-y)$ is a
translationally invariant part, whose Fourier transform is given by
\begin{eqnarray}\label{Z3}
\tilde{S}(k)&=&i\int\limits_{0}^{\infty}ds\ e^{-ism^{2}}\
\exp\left(is\big[k_{\|}^{2}-\frac{k_{\perp}^{2}}{eBs\cot(eBs)}\big]\right)
\nonumber\\
&&\times \bigg\{\left(m+\mathbf{\gamma}^{\|}\cdot
\mathbf{k}_{\|}\right)\left(1+\gamma^{1}\gamma^{2}\tan(eBs)\right)-\mathbf{\gamma}^{\perp}\cdot
\mathbf{k}_{\perp}\left(1+\tan^{2}(eBs)\right)\bigg\}.
\end{eqnarray}
Here, $\mathbf{k}_{\|}=(k_{0},k_{3})$ and
$\mathbf{\gamma}_{\|}=(\gamma_{0},\gamma_{3})$ and
$\mathbf{k}_{\perp}=(k_{1},k_{2})$ and
$\mathbf{\gamma}_{\perp}=(\gamma_{1},\gamma_{2})$.  After performing
the integral over $s$,  $\tilde{S}(k)$ can be decomposed in Landau
levels, that are labeled by $n$
\begin{eqnarray}\label{Z4}
\tilde{S}(k)=ie^{-\frac{\mathbf{k}_{\perp}^{2}}{|eB|}}\sum\limits_{n=0}^{\infty}(-1)^{n}\
\frac{D_{n}(eB,k)}{k_{\|}^{2}-m^{2}-2|eB|n}.
\end{eqnarray}
Here, $D_{n}(eB,k)$ are expressed through the generalized Laguerre
polynomials $L_{m}^{\alpha}$ as
\begin{eqnarray}\label{Z5}
D_{n}(eB,k)=(\gamma^{\|}\cdot \mathbf{k}_{\|}+m)\ \bigg\{ 2\
{\cal{O}}\bigg[L_{n}\left(2\rho\right)-
L_{n-1}\left(2\rho\right)\bigg]+4\gamma^{\perp}\cdot
k_{\perp}L_{n-1}^{(1)}\left(2\rho\right) \bigg\},
\end{eqnarray}
where we have introduced $\rho\equiv \frac{k_{\perp}^{2}}{|eB|}$ and
\begin{eqnarray*}
{\cal{O}}\equiv\frac{1}{2}\left(1-i\gamma^{1}\gamma^{2}\mbox{sign}(eB)\right).
\end{eqnarray*}
Relation (\ref{Z5}) suggests that in the IR region, with
$|\mathbf{k}_{\|}|, |\mathbf{k}_{\perp}|\ll \sqrt{|eB|}$, all the
higher Landau levels with $n\geq 1$ decouple and only the LLL with
$n=0$ is relevant. In strong magnetic field limit, the free fermion
propagator (\ref{Z3}) can therefore be decomposed into two
independent transverse and longitudinal parts \cite{miransky-2,
miransky1-5}
\begin{eqnarray}\label{Z6}
{\bar{S}}_{\mbox{\tiny{LLL}}}(x,y)=\bar{S}_{\|}(\mathbf{x}_{\|}-\mathbf{y}_{\|})P(\mathbf{x}_{\perp},\mathbf{y}_{\perp}),
\end{eqnarray}
with the longitudinal part
\begin{eqnarray}\label{Z7}
\bar{S}_{\|}(\mathbf{x}_{\|}-\mathbf{y}_{\|})=\int
\frac{d^{2}k_{\|}}{(2\pi)^{2}}\
e^{i\mathbf{k}_{\|}\cdot(\mathbf{x}-\mathbf{y})^{\|}}\
\frac{i{\cal{O}}}{\gamma^{\|}\cdot \mathbf{k}_{\|}-m},
\end{eqnarray}
and the transverse part
\begin{eqnarray}\label{Z8}
P(\mathbf{x}_{\perp},\mathbf{y}_{\perp})=\frac{|eB|}{2\pi}\exp\left(\frac{ieB}{2}\epsilon^{ab}x^{a}y^{b}-
\frac{|eB|}{4}\left(\mathbf{x}_{\perp}-\mathbf{y}_{\perp}\right)^{2}\right),\qquad
a,b=1,2.
\end{eqnarray}
As for the photon propagator ${\cal{D}}_{\mu\nu}$ of QED in an
external constant magnetic field, it is calculated explicitly in
\cite{miransky-2, miransky1-5, loskutov} in the LLL at one-loop
level. It is given by
\begin{eqnarray}\label{Z9}
i{\cal{D}}_{\mu\nu}(q)=\frac{g_{\mu\nu}^{\perp}}{q^{2}}+\frac{q_{\mu}^{\|}q_{\nu}^{\|}}{q^{2}q_{\|}^{2}}+
\frac{\left(g_{\mu\nu}^{\|}-q_{\mu}^{\|}q_{\nu}^{\|}/q_{\|}^{2}\right)}{q^{2}+q_{\|}^{2}
\Pi(q_{\perp}^{2},q_{\|}^{2})}-{\xi}\
\frac{q_{\mu}q_{\nu}}{(q^{2})^{2}},
\end{eqnarray}
where $\xi$ is an arbitrary gauge parameter. Since the LLL fermions
couple only to the longitudinal $(0,3)$ components of the photon
fields, no polarization effects are present in the transverse
$(1,2)$ components of ${\cal{D}}_{\mu\nu}(q)$. Therefore, the photon
propagator in the LLL approximation including the one-loop
correction is given by the Feynman-like covariant propagator
\cite{miransky-2, miransky1-5}
\begin{eqnarray}\label{Z10}
i\widetilde{{\cal{D}}}_{\mu\nu}(q)=\frac{g_{\mu\nu}^{\|}}{q^{2}+\mathbf{q}_{\|}^{2}
\Pi\left(\mathbf{q}_{\|}^{2},\mathbf{q}_{\perp}^{2}\right)},
\end{eqnarray}
with $\Pi(q_{\perp}^{2},q_{\|}^{2})$ having the form
\cite{kuznetsov}
\begin{eqnarray}\label{Z11}
\Pi(q_{\perp}^{2},q_{\|}^{2})=\frac{2\alpha|eB|N_{f}}{\mathbf{q}_{\|}^{2}}e^{-\frac{q_{\perp}^{2}}{2|eB|}}
H\left(\frac{\mathbf{q}^{2}_{\|}}{4m^{2}_{dyn.}}\right),
\end{eqnarray}
where $N_{f}$ is the number of fermion flavors. Here, $H(z)$ is
defined by
\begin{eqnarray}\label{Z12}
H(z)\equiv
\frac{1}{2\sqrt{z(z-1)}}\ln\left(\frac{\sqrt{1-z}+\sqrt{-z}}{\sqrt{1-z}-\sqrt{-z}}\right)-1.
\end{eqnarray}
Expanding this expression for $|{\mathbf{q}}_{\|}^{2}|\ll
m_{dyn.}^{2}\ll |eB|$ and $m_{dyn.}^{2}\ll
|\mathbf{q}_{\|}^{2}|\ll|eB|$, we arrive at
\begin{eqnarray}
\Pi({\mathbf{q}}_{\perp}^{2},{\mathbf{q}}_{\|}^{2})\simeq\
+\frac{\alpha|eB|N_{f}}{3\pi
m_{dyn.}^{2}}e^{-\frac{q_{\perp}^{2}}{2|eB|}}&\qquad\mbox{for}\qquad&
|\mathbf{q}_{\|}^{2}|\ll m_{dyn.}^{2}\ll|eB|,\label{Z13}\\
\Pi({\mathbf{q}}_{\perp}^{2},{\mathbf{q}}_{\|}^{2})\simeq
-\frac{2\alpha|eB|N_{f}}{\pi\
\mathbf{q}_{\|}^{2}}e^{-\frac{q_{\perp}^{2}}{2|eB|}}&\qquad\mbox{for}\qquad&m_{dyn.}^{2}\ll
|\mathbf{q}_{\|}^{2}|\ll|eB|.\label{Z14}
\end{eqnarray}
In \cite{miransky-2, miransky1-5}, it is shown that the kinematic
region mostly responsible for generating the fermion mass is with
the dynamical mass, $m_{dyn.}$, satisfying  $m_{dyn.}^{2}\ll
|\mathbf{q}_{\|}^{2}|\ll|eB|$. Plugging (\ref{Z14}) in the photon
propagator (\ref{Z10}) and assuming that
$|\mathbf{q}_{\perp}^{2}|\ll |eB|$, we get
\begin{eqnarray}\label{Z15}
\widetilde{\cal{D}}_{\mu\nu}(q)\approx
-\frac{ig_{\mu\nu}^{\|}}{q^{2}-M_{\gamma}^{2}},\qquad
\mbox{with}\qquad M_{\gamma}^{2}=\frac{2\alpha|eB|N_{f}}{\pi},
\end{eqnarray}
where $M_{\gamma}$ is the finite photon mass, whose appearance is
the result of the dimensional reduction $3+1\to 1+1$ in the presence
of a constant magnetic field.
\par
The dynamically generated fermion mass in the LLL approximation is
determined in \cite{miransky-2} by solving the ladder Bethe-Salpeter
(BS) equation in the LLL approximation. It is given by
\begin{eqnarray}\label{Z16}
m_{dyn.}^{(1)}=C\sqrt{eB}\exp\left(-\sqrt{\frac{\pi}{\alpha}}\right),
\end{eqnarray}
where the constant $C$ is of order one. In \cite{shovkovy}, the same
result is determined by solving the corresponding SD equation in the
ladder LLL approximation. In both methods, to determine (\ref{Z16})
in the ladder approximation, the free fermion propagator in the LLL
approximation (\ref{Z6})-(\ref{Z8}) with $m=m_{dyn.}$ is used. In
this approximation, the full photon propagator are also replaced by
free photon propagator in the covariant Feynman gauge
\begin{eqnarray}\label{AB7}
D_{\mu\nu}^{(0)}(q)=-\frac{ig_{\mu\nu}}{q^{2}}
\end{eqnarray}
In \cite{miransky-cornwall} two-loop contribution to the
corresponding SD equation arising from composite effective action
$\grave{\mbox{a}}$ la Cornwall-Jackiw-Tomboulis \cite{cornwall} in
LLL approximation is considered. Here, the full fermion propagator
in the LLL and the full photon propagator in a covariant and a
non-covariant gauge is taken into account in the two-loop level. In
the improved rainbow approximation, defined by the photon propagator
in a non-covariant gauge, the expression for $m_{dyn.}$ takes the
following form \cite{miransky-cornwall}
\begin{eqnarray}\label{Z17}
m_{dyn.}=\tilde{C}\sqrt{|eB|}F(\alpha)\exp\left(-\frac{\pi}{\alpha\ln\left(C_{1}/\alpha
N_{f}\right)}\right),
\end{eqnarray}
where $F(\alpha)\simeq (N_{f}\alpha)^{1/3}$, $C_{1}\simeq 1.82\pm
0.06$ and $\tilde{C}\sim O(1)$.
\par
In this paper, we work with fermion and photon propagators at finite
temperature. To find the free fermion propagator at finite
temperature, we will turn into the Euclidean space by replacing
$s\to -is$ and $k_{0}\to i\hat{\omega}_{\ell}$ in (\ref{Z3}) and
find \cite{alexandre}
\begin{eqnarray}\label{Z18}
S_{\ell}(\vec{k})&=&-i \int_{0}^{\infty} ds\
e^{-s(\hat{\omega}_{\ell}^{2}+k^{2}_{3}+\mathbf{k}^{2}_{\perp}\frac{\tanh(|eB|s)}{|eB|s}+m^{2})}\nonumber\\
&&\times[(-\hat{\omega}_{\ell}\gamma^{4}-k^{3}\gamma^{3}+m)(1-i
\gamma^{1}\gamma^{2}\tanh(|eB|s))-\gamma^{\bot}\cdot
\mathbf{k}_{\bot}(1-\tanh^{2}(|eB|s))].
\end{eqnarray}
In the following, we indicate the Matsubara frequencies in the
fermionic case with $\hat{\omega}_{\ell}\equiv (2\ell+1) \pi T$ and
those in the bosonic case by $\omega_{n}\equiv 2 n \pi T$. Next, the
free photon propagator $D_{\mu\nu}^{(0)}(k)$ and the full photon
propagator ${\cal{D}}_{\mu\nu}(k)$ in a constant magnetic field and
at finite temperature are given by \cite{shabad}
\begin{eqnarray}\label{Z19}
D_{\mu\nu}^{(0)}\left(k_{0},{\mathbf{k}}\right)&=&-\sum\limits_{i=1}^{4}\frac{1}{k^{2}_{E}}\
\frac{b_{\mu}^{(i)}b_{\nu}^{\star
(i)}}{\left(b_{\rho}^{(i)}b^{\star\rho(i)}\right)},\nonumber\\
{\cal{D}}_{\mu\nu}\left(k_{0},{\mathbf{k}}\right)&=&-\sum\limits_{i=1}^{4}\frac{1}{\left(k_{E}^{2}+\kappa_{i}(k_{0},{\mathbf{k}})\right)}\
\frac{b_{\mu}^{(i)}b_{\nu}^{\star
(i)}}{\left(b_{\rho}^{(i)}b^{\star\rho(i)}\right)},
\end{eqnarray}
where $\kappa_{i}$ and $b_{\mu}^{(i)}$ are eigenvalue and
eigenfunctions of the vacuum polarization tensor $\Pi_{\mu\nu}$,
\textit{i.e.}
\begin{eqnarray}\label{Z20}
\Pi_{\mu\nu}(k)b_{\nu}^{(i)}=\kappa_{i}(k)b_{\mu}^{(i)}.
\end{eqnarray}
In (\ref{Z19}) $k_{E}$ is the Euclidean four-momentum and
$k^{2}_{E}\equiv 4\pi^{2} n^{2}T^{2}+{\mathbf{k}}^{2}$.  In the next
paragraph, we will determine $\kappa_{i}(k)$ in the IR limit
\textit{i.e.} for $k_{0}\to 0$ ($n=0$) but finite ${\mathbf{k}}$.
Then, using the eigenfunctions $b_{\mu}^{(i)}$, QED vacuum
polarization tensor $\Pi_{\mu\nu}(k_{0},{\mathbf{k}})$ will be
diagonalized and eventually determined in the IR limit.
\subsection{QED vacuum polarization tensor in a constant magnetic
field at finite temperature}
In \cite{shabad} it is shown that the vacuum polarization tensor
$\Pi_{\mu\nu}$ in the presence of external magnetic field and at
finite temperature can be diagonalized as
\begin{eqnarray}\label{Z21}
\Pi_{\mu\nu}(k_{0},{\mathbf{k}})=\sum\limits_{i=1}^{4}\kappa_{i}(k_{0},{\mathbf{k}})\frac{b_{\mu}^{(i)}b_{\nu}^{\star
(i)}}{\left(b_{\rho}^{(i)}b^{\star\rho(i)}\right)},
\end{eqnarray}
where $\kappa_{i}$ and $b_{\mu}^{(i)}$ are defined in (\ref{Z20}).
This relation can be proved by plugging it back in (\ref{Z20}) and
using the property $b_{\mu}^{(i)}b^{\star\mu(j)}=0,\ \forall i\neq
j$. According to the results in \cite{shabad}, the eigenvalues
$\kappa_{i}$ in the \textit{Minkowski} space are given by
\begin{eqnarray}\label{Z22}
\kappa_{1,2}(k_{0},{\mathbf{k}})&=&\frac{1}{2}\bigg\{P+S\pm\big[\left(P-S\right)^{2}-4Q^{2}\big]^{1/2}\bigg\},\nonumber\\
\kappa_{3}(k_{0},{\mathbf{k}})&=&R,\qquad\qquad
\kappa_{4}(k_{0},{\mathbf{k}})=0,
\end{eqnarray}
with
\begin{eqnarray}\label{Z23}
P(k_{0},{\mathbf{k}})&\equiv& k^{2}\pi_{1}+\left(k\cdot
\tilde{F}^{2}\cdot k\right)\pi_{3}-\frac{(u\cdot k)^{2}\left(k\cdot
F^{2}\cdot k\right)}
{\left(k\cdot \tilde{F}^{2}\cdot k\right)}\pi_{4} \nonumber\\
Q(k_{0},{\mathbf{k}})&\equiv&\frac{(u\cdot k)(u\cdot \tilde{F}\cdot k)}{(k\cdot \tilde{F}^{2}\cdot k)}\left(-\frac{k\cdot F^{2}\cdot k}{k^{2}}\right)^{1/2}\pi_{4}\nonumber\\
S(k_{0},{\mathbf{k}})&\equiv& k^{2}\pi_{1}-\frac{(u\cdot
\tilde{F}\cdot k)^{2}}{(k\cdot
\tilde{F}^{2}\cdot k)} \pi_{4}\nonumber\\
R(k_{0},{\mathbf{k}})&\equiv & k^{2}\pi_{1}-(k\cdot F^{2}\cdot
k)\pi_{2}+2\ \mathcal{F}\ k^{2}\pi_{3},
\end{eqnarray}
where the notation $a\cdot b\equiv a_{\mu}b^{\mu}$ is used. Here,
$F_{\mu\nu}$ is the field strength tensor, $\tilde{F}_{\mu\nu}$ the
dual tensor $\tilde{F}_{\mu\nu}\equiv
\frac{1}{2}\epsilon_{\mu\nu\rho\sigma}F^{\rho\sigma}$, $u\equiv
\left(1,0,0,0\right)$  the rest frame vector in the Minkowski space,
$k^{2}\equiv k_{0}^{2}-{\mathbf{k}}^{2}$ and ${\cal{F}}\equiv
-\frac{1}{4}F_{\mu\rho}F^{\rho\mu}$. Assuming that there exists only
a magnetic field $B_{\ell}=\frac{1}{2}\epsilon_{\ell mn}F^{mn}$
directed along the 3-axis (\textit{i.e.}
${\mathbf{B}}=B{\mathbf{e}}_{3}$), we get $F_{12}=-F_{21}=B$. The
other components of $F_{\mu\nu}$ vanish. For the dual tensor
$\tilde{F}_{\mu\nu}$ only the components
$\tilde{F}_{03}=-\tilde{F}_{30}=B$ survive.
\par
Considering again (\ref{Z23}), $\pi_{i}, i=1,2,3,4,$ are the
coefficients of the expansion of $\Pi_{\mu\nu}$ as a second rank
tensor in certain basis $\Psi^{(i)}_{\mu\nu}$ \cite{shabad}
\begin{eqnarray}\label{Z24}
\Pi_{\mu\nu}(k_{0},{\mathbf{k}})=\sum\limits_{i=1}^{4}\pi_{i}\left((u\cdot
k)^{2},k\cdot F^{2}\cdot k,{\mathcal{F}},
k^{2}\right)\Psi_{\mu\nu}^{(i)},
\end{eqnarray}
where the basis $\Psi^{(i)}_{\mu\nu}$ are second rank tensors, that
are built up from four vectors $k_{\mu},F_{\mu\rho}k^{\rho},
F_{\rho\mu}F^{\mu\nu}k_{\nu}$, and $u_{\mu}$.\footnote{In
Minkowskian space there are indeed 16 independent tensors
$\Psi_{\mu\nu}^{(i)}$. But, as it is shown in \cite{shabad}, for
zero chemical potential and due to symmetry properties only
$\Psi_{\mu\nu}^{(i)}, i=1,\cdots,4$ from (\ref{Z25}) are
nonvanishing.} They are given by
\begin{eqnarray}\label{Z25}
\Psi_{\mu\nu}^{(1)}(k_{0},{\mathbf{k}})&=&k^{2}g_{\mu\nu}-k_{\mu}k_{\nu}, \nonumber\\
\Psi_{\mu\nu}^{(2)}(k_{0},{\mathbf{k}})&=&F_{\mu\lambda}k^{\lambda}F_{\nu\sigma}k^{\sigma},\nonumber\\
\Psi_{\mu\nu}^{(3)}(k_{0},{\mathbf{k}})&=&-k^{2}\left(
g_{\mu\lambda}-\frac{k_{\mu}k_{\lambda}}{k^{2}}\right)F^{\lambda}_{\rho}F^{\rho\eta}\left(g_{\eta\nu}-\frac{k_{\eta}k_{\nu}}{k^{2}}\right),\nonumber\\
\Psi_{\mu\nu}^{(4)}(k_{0},{\mathbf{k}})&=&\left(u_{\mu}-\frac{k_{\mu}(u\cdot
k)}{k^{2}}\right)\left(u_{\nu}-\frac{k_{\nu}(u\cdot
k)}{k^{2}}\right),
\end{eqnarray}
where $\Psi_{\mu\nu}^{(i)}, i=1,\cdots,4$ satisfy
$\Psi_{\mu\nu}^{(i)}=\Psi_{\nu\mu}^{(i)}, \forall i$. It is the
purpose of this section to determine $P,Q,S$ and $R$ from
(\ref{Z23}) and eventually $\kappa_{i},i=1,\cdots,4$ from
(\ref{Z22}) in the IR limit where $k_{0}$ is taken to zero but
${\mathbf{k}}$ is nonvanishing. This will enable us to determine the
ring contribution of the effective potential in the IR limit. To do
this we have to determine $\pi_{i}, i=1,\cdots, 4$ from (\ref{Z24})
explicitly. Multiplying (\ref{Z24}) with $\Psi_{\nu\sigma}^{(j)}$
and adding over $\nu$, we arrive at
\begin{eqnarray}\label{Z26}
{\cal{P}}^{j}=\sum_{i=1}^{4}{\mathcal{A}}^{ji}\pi^{i},
\end{eqnarray}
where ${\cal{P}}^{j}\equiv \Pi^{\mu\nu}\Psi_{\nu\mu}^{(j)}$ and
${\mathcal{A}}^{ij}\equiv \Psi^{(i)\mu\nu}\Psi_{\nu\mu}^{(j)}$. To
calculate $\pi_{i}$, we consider (\ref{Z26}) first as a generic
system of equations in terms of generic ${\cal{P}}^{j}$ and
${\cal{A}}^{ji}$. Solving this system of equations $\pi_{i}$ are
given by {\small{
\begin{eqnarray}\label{Z27}
\pi_{1}&=&\frac{{\cal{P}}_{1}}{Y}\left(A_{22}A_{34}^2+A_{23}^2A_{44}-A_{22}A_{33}A_{44}\right)+\frac{{\cal{P}}_{2}}{Y}\left(
A_{14}A_{23}A_{34}-A_{12}A_{34}^{2}-A_{13}A_{23}A_{44}+A_{12}A_{33}A_{44}
\right)
\nonumber\\
&&+
\frac{{\cal{P}}_{3}}{Y}\left(-A_{22}\left(A_{14}A_{34}-A_{13}A_{44}\right)-A_{12}A_{23}A_{44}\right)
+\frac{{\cal{P}}_{4}}{Y}\left(-A_{14}A_{23}^{2}+A_{22}\left(A_{14}A_{33}-A_{13}A_{34}\right)+A_{12}A_{23}A_{34}\right),\nonumber\\
\pi_{2}&=&\frac{{\cal{P}}_{1}}{Y}\left(-A_{14}A_{23}A_{34}+A_{13}A_{23}A_{44}+A_{12}\left(A_{34}^{2}-A_{33}A_{44}\right)\right)+
\frac{{\cal{P}}_{2}}{Y}\left(
-A_{14}\left(A_{14}A_{33}-2A_{13}A_{34}\right)-A_{13}^{2}A_{44}\right.\nonumber\\
&&\left.-A_{11}\left(A_{34}^{2}-A_{33}A_{44}\right)
\right)+\frac{{\cal{P}}_{3}}{Y}\left(
A_{14}^{2}A_{23}-A_{12}\left(A_{14}A_{34}-A_{13}A_{44}\right)-A_{11}A_{23}A_{44}
\right)\nonumber\\
&&+\frac{{\cal{P}}_{4}}{Y}\left(A_{12}A_{14}A_{33}+A_{11}A_{23}A_{34}-A_{13}\left(A_{14}A_{23}+A_{12}A_{34}\right)\right),\nonumber\\
\pi_{3}&=&\frac{{\cal{P}}_{1}}{Y}\left(A_{22}\left(A_{14}A_{34}-A_{13}A_{44}\right)+A_{12}A_{23}A_{44}\right)
+\frac{{\cal{P}}_{2}}{Y}\left(A_{14}\left(A_{14}A_{23}-A_{12}A_{34}\right)+A_{44}\left(A_{12}A_{13}-A_{11}A_{23}\right)\right)\nonumber\\
&&-\frac{{\cal{P}}_{3}}{Y}\left(A_{14}^{2}A_{22}+A_{44}\left(A_{12}^{2}-A_{11}A_{22}\right)\right)
+\frac{{\cal{P}}_{4}}{Y}\left(A_{14}\left(A_{13}A_{22}-A_{12}A_{23}\right)+A_{34}\left(A_{12}^{2}-A_{11}A_{22}\right)\right),\nonumber\\
\pi_{4}&=&\frac{{\cal{P}}_{1}}{Y}\left(A_{14}\left(-A_{23}^{2}+A_{22}A_{33}\right)+A_{34}\left(-A_{13}A_{22}+A_{12}A_{23}\right)\right)
+\frac{{\cal{P}}_{2}}{Y}\left(A_{14}\left(A_{13}A_{23}-A_{12}A_{33}\right)+\right.\nonumber\\
&&\left.A_{34}\left(A_{12}A_{13}-A_{11}A_{23}\right)\right) +
\frac{{\cal{P}}_{3}}{Y}\left(A_{14}\left(-A_{13}A_{22}+A_{12}A_{23}\right)+A_{34}\left(-A_{12}^{2}+A_{11}A_{22}\right)\right)\nonumber\\
&&+ \frac{{\cal{P}}_{4}}{Y}\left(
A_{13}^{2}A_{22}+A_{12}\left(-2A_{13}A_{23}+A_{12}A_{33}\right)+A_{11}\left(A_{23}^{2}-A_{22}A_{33}\right)
\right),
\end{eqnarray}
}} with the denominator
\begin{eqnarray*}
Y&=&-\bigg[A_{14}^{2}\left(A_{23}^{2}-A_{22}A_{33}\right)+2A_{14}A_{34}\left(A_{13}A_{22}-A_{12}A_{23}\right)-A_{11}A_{22}A_{34}^{2}-A_{13}^{2}A_{22}A_{44}
\nonumber\\
&&+2A_{12}A_{13}A_{23}A_{44}-A_{11}A_{23}^{2}A_{44}
+A_{11}A_{22}A_{33}A_{44}+A_{12}^{2}\left(A_{34}^{2}-A_{33}A_{44}\right)\bigg].
\end{eqnarray*}
As next we have to calculate ${\cal{A}}^{ij}$ and ${\cal{P}}^{j}$
from (\ref{Z26}). Here, ${\cal{A}}^{ij}$ can be determined using the
$\Psi_{\mu\nu}^{(i)}$ from (\ref{Z25}). They are given by
\begin{eqnarray}\label{Z28}
{\cal{A}}^{11}&=& 3 (k^{2})^{2},\nonumber\\
{\cal{A}}^{12}&=&-k^{2}\left(k\cdot F^{2}\cdot k\right)=-B^2 k^{2}{\mathbf{k}}_{\perp}^{2},\nonumber\\
{\cal{A}}^{13}&=&-\left(k^{2}\right)^{2}\mbox{tr}(F^{2})+k^{2}\left(k\cdot
F^{2}\cdot k\right)=B^{2}k^{2}\left(2k^{2}+{\mathbf{k}}_{\perp}^{2}
\right),\nonumber\\
{\cal{A}}^{14}&=&k^{2}u^{2}-(u\cdot k)^{2}=-{\mathbf{k}}^{2},\nonumber\\
{\cal{A}}^{22}&=&\left(k\cdot F^{2}\cdot k\right)^{2}={B}^{4}(\mathbf{k}_{\bot}^{2})^{2},\nonumber\\
{\cal{A}}^{23}&=&k^{2}\left(k\cdot F^{4}\cdot k\right)=-{B}^{4}k^{2}{\mathbf{k}}_{\perp}^{2},\nonumber\\
{\cal{A}}^{24}&=&0,\nonumber\\
{\cal{A}}^{33}&=&\left(k^{2}\right)^{2}\mbox{tr}(F^{4})-2k^{2}\left(k\cdot
F^{4}\cdot k\right)+\left(k\cdot F^{2}\cdot
k\right)^{2}=B^{4}\left(2(k^{2})^{2}+2k^{2}{\mathbf{k}}_{\perp}^{2}+({\mathbf{k}}_{\perp}^{2})^{2}\right),\nonumber\\
A^{34}&=&-\frac{\left(k\cdot F^{2}\cdot k\right)\left(u\cdot k\right)^{2}}{k^{2}}=-\frac{B^{2}k_{0}^{2}{\mathbf{k}}^{2}_{\perp}}{k^{2}}\nonumber\\
{\cal{A}}^{44}&=&1-\frac{2(u\cdot k)^{2}}{k^{2}}+\frac{(u\cdot
k)^{4}}{(k^{2})^{2}}=1-\frac{2k_{0}^{2}}{k^{2}}+\frac{k_{0}^{4}}{(k^{2})^{2}},
\end{eqnarray}
where the rest frame constraint $u_{\mu}F^{\mu\nu}=0$ and the
following relations are used
\begin{eqnarray*}
k\cdot F\cdot k=0,\qquad k\cdot F^{2}\cdot
k=B^{2}{\mathbf{k}}_{\perp}^{2},\qquad k\cdot F^{4}\cdot
k=-B^{4}{\mathbf{k}}_{\perp}^{2},\qquad
\mbox{tr}(F^{2})=-2B^{2},\qquad \mbox{tr}(F^{4})=2B^{4}.
\end{eqnarray*}
Other components of ${\cal{A}}^{ji}$ are determined by the symmetry
property ${\cal{A}}^{ji}={\cal{A}}^{ij}$. To determine
${\cal{P}}^{j}$ from (\ref{Z26}), we use $\Psi_{\mu\nu}^{(i)}$ from
(\ref{Z25}) and $k_{\mu}\Pi^{\mu\nu}=0$ to get
\begin{eqnarray}\label{Z29-a}
{\cal{P}}_{1}&=&k^{2}\mbox{tr}\left(\Pi\right)=k^{2}\left(\Pi_{00}-\sum_{i=1}^{3}\Pi_{ii}\right),\nonumber\\
{\cal{P}}_{2}&=& - k\cdot F\cdot \Pi\cdot F\cdot
k=B^{2}\left(-k_{1}k_{2}\Pi_{21}+k_{1}^{2}\Pi_{22}+k_{2}^{2}\Pi_{11}-k_{1}k_{2}\Pi_{12}\right),\nonumber\\
{\cal{P}}_{3}&=& -k^{2}\mbox{tr}\left(\Pi\cdot  F^{2}\right)=-B^{2}k^{2}\left(\Pi_{11}+\Pi_{22}\right),\nonumber\\
{\cal{P}}_{4}&=& u\cdot \Pi\cdot u=\Pi_{00}.
\end{eqnarray}
The components of QED vacuum polarization tensor $\Pi_{\mu\nu}$ in a
magnetic field at finite temperature are explicitly calculated in
\cite{alexandre} using the Schwinger proper-time formalism
\cite{schwinger}.\footnote{To determine the vacuum polarization
tensor $\Pi_{\mu\nu}$ in the Schwinger proper-time formalism, it is
enough to replace the free fermion propagator in ordinary
$\Pi_{\mu\nu}$ by the free fermion propagator in the constant
background magnetic field arising from Schwinger proper-time
formalism. At finite temperature, this is done in \cite{alexandre}
to determine $\Pi_{\mu\nu}$.} To determine ${\cal{P}}^{j}$ we need
only the components $\Pi_{\mu\mu}\left(n,{\mathbf{k}}\right)$ with
$\mu=1,\cdots, 4$, and $\Pi_{12}\left(n,{\mathbf{k}}\right)$ of the
vacuum polarization tensor. In the Euclidean space, they are given
by
\begin{eqnarray}\label{Z29}
\Pi_{ii}\left(n,{\mathbf{k}}\right)&=&\frac{-\alpha T
eB}{\sqrt{\pi}}\int_{\frac{1}{\Lambda^{2}}}^{\infty}du\sqrt{u}\int_{-1}^{+1}dv\sum_{\ell=-\infty}^{\infty}e^{\phi_{n}(u,v;\ell)}
\bigg[\omega_{n}W_{\ell n}\ \coth\bar{u}\
\frac{\sinh\bar{u}v}{\sinh\bar{u}}
\nonumber\\
&&+({\mathbf{k}}_{\bot}^{2}-k_{i}^{2})\frac{\left(\cosh\bar{u}-\cosh\bar{u}v\right)}{\sinh^{3}\bar{u}}
+\frac{(\omega_{n}^{2}+k_{3}^{2})}{2}\frac{\left(\cosh\bar{u}v-v\coth\bar{u}\sinh\bar{u}v\right)}{\sinh\bar{u}}\bigg],
\qquad i=1,2,\nonumber\\
\Pi_{33}\left(n,{\mathbf{k}}\right)&=&\frac{-\alpha T
e{B}}{\sqrt{\pi}}\int_{\frac{1}{\Lambda^{2}}}^{\infty} du\sqrt{u}\int_{-1}^{+1}dv\sum\limits_{\ell=-\infty}^{\infty}e^{\phi_{n}(u,v;\ell)}\nonumber\\
&&\times\bigg[v\ \omega_{n}W_{\ell
n}\coth\bar{u}+\frac{{\mathbf{k}}_{\bot}^{2}}{2}\frac{\left(\cosh\bar{u}v-v\coth\bar{u}\sinh\bar{u}v\right)}{\sinh\bar{u}}
+\omega_{n}^{2}\ \frac{\left(1-v^{2}\right)}{2}\coth\bar{u}\bigg],\nonumber\\
\Pi_{44}\left(n,{\mathbf{k}}\right)&=& -\frac{\alpha T
eB}{\sqrt{\pi}}\int_{\frac{1}{\Lambda^{2}}}^{\infty}
du\sqrt{u}\int_{-1}^{+1}dv\sum\limits_{\ell=-\infty}^{+\infty}e^{\phi_{n}(u,v;\ell)}\nonumber\\
&&\times\bigg[\frac{\mathbf{k}_{\bot}^{2}}{2}\frac{\left(\cosh\bar{u}v-v\coth\bar{u}\sinh\bar{u}v\right)}{\sinh\bar{u}}-\coth{\bar{u}}\left(\frac{1}{u}-
2W_{\ell n}^{2}+v\ \omega_{n}
W_{\ell n}-\frac{\left(1-v^{2}\right)}{2}k_{3}^{2}\right)\bigg],\nonumber\\
\Pi_{12}\left(n,{\mathbf{k}}\right)&=&+\frac{\alpha T
eB}{\sqrt{\pi}}\int_{\frac{1}{\Lambda^{2}}}^{\infty}du
\sqrt{u}\int_{-1}^{+1}dv
\sum_{\ell=
-\infty}^{+\infty}e^{\phi_{n}(u,v;\ell)}k_{1}k_{2}\frac{\left(\cosh\bar{u}-\cosh\bar{u}v\right)}{\sinh^{3}\bar{u}},
\end{eqnarray}
where ${\mathbf{k}}_{\bot}\equiv (k_{1},k_{2})$, $\bar{u}\equiv u
eB$ and $\phi_{n}(u,v;\ell)$ is defined by
\begin{eqnarray*}
\phi_{n}(u,v;\ell)\equiv
-\frac{\mathbf{k}_{\bot}^{2}}{eB}\frac{\left(\cosh\bar{u}-\cosh\bar{u}v\right)}{2\sinh\bar{u}}-u\big[m^{2}+W_{\ell
n}^{2}+\frac{(1-v^{2})}{4}
\left(\omega_{n}^{2}+k_{3}^{2}\right)\big].
\end{eqnarray*}
In all the above expressions $W_{\ell n}\equiv
\hat{\omega}_{\ell}-\frac{1-v}{2}\omega_{n}$, where
$\hat{\omega}_{\ell}$ and $\omega_{n}$ are the Matsubara frequencies
of the fermionic and bosonic fields, respectively.\footnote{Note
that compared to the results in \cite{alexandre}, there are some
temperature independent contact terms missing in the above
expression. We will omit them here, keeping in mind that they are
relevant only to cancel temperature independent imaginary terms in
Sect. III.A.} In the IR limit $n=0$ (or equivalently $k_{0}\to 0$)
they are given by
\begin{eqnarray}\label{Z30}
\Pi^{11}\left(n=0,{\mathbf{k}}\right)&=&2k_{2}^{2}I_{4}+k_{3}^{2}I_{1},\nonumber\\
\Pi^{22}\left(n=0,{\mathbf{k}}\right)&=&2k_{1}^{2}I_{4}+k_{3}^{2}I_{1},\nonumber\\
\Pi^{33}\left(n=0,{\mathbf{k}}\right)&=&\mathbf{k}_{\perp}^{2}I_{1},\nonumber\\
\Pi^{44}\left(n=0,{\mathbf{k}}\right)&=&\mathbf{k}_{\perp}^{2}I_{1}+k_{3}^{2}I_{3}-I_{2},\nonumber\\
\Pi^{12}\left(n=0,{\mathbf{k}}\right)&=&-2k_{1}k_{2}I_{4},
\end{eqnarray}
where the integrals $I_{i},i=1,\cdots,4$ can be determined using
(\ref{Z29}) with $n=0$. For future purposes, we will separate
$I_{i},i=1,\cdots,4$ in a temperature independent part $I_{i}^{0}$
and a temperature dependent part $I_{i}^{T}$. This can be done using
the Possion resummation formula
\begin{eqnarray*}
\sum\limits_{\ell=-\infty}^{\infty}e^{-a(\ell-z)^{2}}=\left(\frac{\pi}{a}\right)^{1/2}
\sum\limits_{\ell=-\infty}^{\infty}\exp\left(-\frac{\pi^{2}\ell^{2}}{a}-2i\pi
z\ell\right),
\end{eqnarray*}
that leads to
\begin{eqnarray*}
\sum\limits_{\ell=-\infty}^{+\infty}e^{-uW_{\ell
0}^{2}}=\frac{1}{T\sqrt{\pi u}}\sum\limits_{\ell\geq
1}(-1)^{\ell}e^{-\frac{\ell^{2}}{4uT^{2}}}+\frac{1}{2T\sqrt{\pi u}},
\end{eqnarray*}
and
\begin{eqnarray*}
\sum\limits_{\ell=-\infty}^{+\infty}\left(-2W_{\ell
0}^{2}+\frac{1}{u}\right)e^{-uW_{\ell 0}^{2}}=\frac{1}{2T\sqrt{\pi
}}\sum\limits_{\ell\geq 1}(-1)^{\ell}\frac{1}{u^{5/2}}\
\frac{\ell^{2}}{T^{2}}e^{-\frac{\ell^{2}}{4uT^{2}}}.
\end{eqnarray*}
We arrive therefore at $I_{i}=I_{i}^{0}+I_{i}^{T},i=1,\cdots,4$
with\footnote{In \cite{alexandre}, a similar method is used to
separate $\Pi_{44}$ into a temperature dependent and a temperature
independent part.}
\begin{eqnarray}\label{Z31}
I_{1}^{0}&=&-\frac{\alpha eB}{4
\pi}\int_{\frac{1}{\Lambda^{2}}}^{\infty}du\int_{-1}^{+1}dv\
e^{\phi(u,v)}\frac{\left(\cosh\bar{u}v-v\coth\bar{u}\sinh\bar{u}v\right)}{\sinh\bar{u}},\nonumber\\
I_{1}^{T}&=&-\frac{\alpha eB}{2
\pi}\int_{0}^{\infty}du\int_{-1}^{+1}dv\
e^{\phi(u,v)}\sum_{\ell=1}^{\infty}(-1)^{\ell}e^{-\frac{\ell^{2}}{4uT^{2}}}\frac{\left(\cosh\bar{u}v-v\coth\bar{u}\sinh\bar{u}v\right)}{\sinh\bar{u}},\nonumber\\
I^{0}_{2}&=&0,\nonumber\\
I_{2}^{T}&=&-\frac{\alpha eB}{2
\pi}\int_{0}^{\infty}du\int_{-1}^{+1}dv\
e^{\phi(u,v)}\sum_{\ell=1}^{\infty}(-1)^{\ell}\frac{\ell^{2}}{T^{2}}\ e^{-\frac{\ell^{2}}{4uT^{2}}}\ \frac{\coth\bar{u}}{u^{2}},\nonumber\\
I_{3}^{0}&=&-\frac{\alpha eB}{4
\pi}\int_{\frac{1}{\Lambda^{2}}}^{\infty}du\int_{-1}^{+1}dv\
e^{\phi(u,v)}(1-v^{2})\coth\bar{u},\nonumber\\
I_{3}^{T}&=&-\frac{\alpha eB}{2
\pi}\int_{0}^{\infty}du\int_{-1}^{+1}dv\
e^{\phi(u,v)}\sum_{\ell=1}^{\infty}(-1)^{\ell}e^{-\frac{\ell^{2}}{4uT^{2}}}(1-v^{2})\coth\bar{u},\nonumber\\
I_{4}^{0}&=&-\frac{\alpha eB}{4
\pi}\int_{\frac{1}{\Lambda^{2}}}^{\infty}du\int_{-1}^{+1}dv\
e^{\phi(u,v)}\ \frac{\left(\cosh\bar{u}-\cosh\bar{u}v\right)}{\sinh^{3}\bar{u}},\nonumber\\
I_{4}^{T}&=&-\frac{\alpha eB}{2
\pi}\int_{0}^{\infty}du\int_{-1}^{+1}dv\
e^{\phi(u,v)}\sum_{\ell=1}^{\infty}(-1)^{\ell}e^{-\frac{\ell^{2}}{4uT^{2}}}\
\frac{\left(\cosh\bar{u}-\cosh\bar{u}v\right)}{\sinh^{3}\bar{u}},
\end{eqnarray}
where $\phi(u,v)\equiv \phi_{0}(u,v;\ell)+uW^{2}_{\ell 0}$. Plugging
now  $\Pi^{\mu\nu}\left(n=0,{\mathbf{k}}\right)$ from (\ref{Z30}) in
(\ref{Z29-a}), ${\cal{P}}_{i}, i=1,\cdots,4$ in the IR limit are
given by
\begin{eqnarray}\label{Z32}
{{\cal{P}}}_{1}\left(k_{0}\to 0,{\mathbf{k}}\right)&=&{\mathbf{k}}^{2}(2\mathbf{k}^{2}I_{1}+2\mathbf{k}_{\bot}^{2}I_{4}+k_{3}^{2}I_{3}-I_{2}),\nonumber\\
{{\cal{P}}}_{2}\left(k_{0}\to 0,{\mathbf{k}}\right)&=&B^{2}\mathbf{k}_{\bot}^{2}\left(2\mathbf{k}_{\bot}^{2}I_{4}+k_{3}^{2}I_{1}\right),\nonumber\\
{{\cal{P}}}_{3}\left(k_{0}\to 0,{\mathbf{k}}\right)&=&+B^{2}\mathbf{k}^{2}(2k_{3}^{2}I_{1}+2\mathbf{k}_{\bot}^{2}I_{4}),\nonumber\\
{{\cal{P}}}_{4}\left(k_{0}\to
0,{\mathbf{k}}\right)&=&-\mathbf{k}_{\bot}^{2}I_{1}-k_{3}^{2}I_{3}+I_{2}.
\end{eqnarray}
Plugging further ${\cal{A}}^{ji}$ from (\ref{Z28}) and
${\cal{P}}^{j}$ from (\ref{Z32}) in (\ref{Z27}) and taking carefully
the limit $k_{0}\to 0$, $\pi_{i},i=1,\cdots,4$ can be determined in
the IR limit. They are given by
\newcommand{\kt}{${\mathbf{k}}_{\perp}^{2}$}
\begin{eqnarray}\label{Z34}
\pi_{1}&=&\frac{I_{1}\left(k_{3}^{2}{\mathbf{k}}_{\perp}^{2}+3{\mathbf{k}}_{\perp}^{4}\right)+I_{2}\left(2k_{3}^{2}-{\mathbf{k}}_{\perp}^{2}\right)
+I_{3}\left(k_{3}^{2}{\mathbf{k}}_{\perp}^{2}-2k_{3}^{4}\right)+2I_{4}{\mathbf{k}}_{\perp}^{4}
}{2{\mathbf{k}}_{\perp}^{2}{\mathbf{k}}^{2}},\nonumber\\
\pi_{2}&=&\frac{
I_{1}\left({\mathbf{k}}_{\perp}^{4}-3k_{3}^{2}{\mathbf{k}}_{\perp}^{2}\right)+I_{2}\left(2k_{3}^{2}-{\mathbf{k}}_{\perp}^{2}\right)
+I_{3}\left(k_{3}^{2}{\mathbf{k}}_{\perp}^{2}-2k_{3}^{4}\right)+I_{4}\left(4k_{3}^{2}{\mathbf{k}}_{\perp}^{2}+2{\mathbf{k}}_{\perp}^{4}\right)
}{2B^{2}k_{3}^{2}{\mathbf{k}}_{\perp}^{2}},\nonumber\\
\pi_{3}&=&\frac{
I_{1}\left(k_{3}^{2}{\mathbf{k}}_{\perp}^{2}-{\mathbf{k}}_{\perp}^{4}\right)+I_{2}\left({\mathbf{k}}_{\perp}^{2}-2k_{3}^{2}\right)
+I_{3}\left(2k_{3}^{4}-k_{3}^{2}{\mathbf{k}}_{\perp}^{2}\right)
-2I_{4}{\mathbf{k}}_{\perp}^{4}
}{2B^{2}k_{3}^{2}{\mathbf{k}}_{\perp}^{2}},\nonumber\\
\pi_{4}&=&\frac{
I_{1}\left(k_{3}^{2}{\mathbf{k}}_{\perp}^{2}+{\mathbf{k}}_{\perp}^{4}\right)+I_{2}\left(2k_{3}^{2}+{\mathbf{k}}_{\perp}^{2}\right)
-I_{3}\left(2k_{3}^{4}+k_{3}^{2}{\mathbf{k}}_{\perp}^{2}\right)
+2I_{4}{\mathbf{k}}_{\perp}^{4} }{2{\mathbf{k}}_{\perp}^{2}},
\end{eqnarray}
where $I_{i}=I_{i}^{0}+I_{i}^{T}, i=1,\cdots, 4$ are given in
(\ref{Z31}). The above information are enough to determine
$\kappa_{i}, i=1,\cdots,4$ from (\ref{Z22}) in the IR limit. To do
this, let us replace $\pi_{i},\ i=1,\cdots,4$ from (\ref{Z34}) in
(\ref{Z23}) to get
\begin{eqnarray}\label{Z35}
P(k_{0}\to 0,{\mathbf{k}})&=&
-\mathbf{k}^{2}\pi_{1}-B^{2}k_{3}^{2}\pi_{3}=-{\mathbf{k}}^{2}I_{1}\nonumber\\
Q(k_{0}\to 0,{\mathbf{k}})&=& 0\nonumber\\
S(k_{0}\to 0,{\mathbf{k}})&=&
-\mathbf{k}^{2}\pi_{1}+\pi_{4}=-\mathbf{k}_{\perp}^{2}I_{1}+I_{2}-k_{3}^{2}I_{3}
\nonumber\\
R(k_{0}\to
0,{\mathbf{k}})&=&-{\mathbf{k}}^{2}\pi_{1}-B^{2}{\mathbf{k}}^{2}\pi_{3}-B^{2}{\mathbf{k}}_{\perp}^{2}\pi_{2}=
-k_{3}^{2}I_{1}-2{\mathbf{k}}_{\perp}^{2}I_{4}.
\end{eqnarray}
Thus, in the basis of $b_{\mu}^{(i)}$ from (\ref{Z20}), QED vacuum
polarization tensor in the IR limit $k_{0}\to 0$ reads
\begin{eqnarray}\label{Z36}
\Pi_{\mu\nu}(k_{0}\to
0,{\mathbf{k}})=\sum\limits_{i=1}^{4}\kappa_{i}(k_{0}\to
0,{\mathbf{k}})\frac{b_{\mu}^{(i)}b_{\nu}^{\star
(i)}}{\left(b_{\mu}^{(i)}b_{\mu}^{\star(i)}\right)},
\end{eqnarray}
where $\kappa_{i}(k_{0}\to 0,{\mathbf{k}})$ are determined by
plugging (\ref{Z35}) in (\ref{Z22}). They are given by
\begin{eqnarray}\label{Z37}
{\kappa}_{1}(k_{0}\to 0,{\mathbf{k}})&=&P(k_{0}\to
0,{\mathbf{k}}),\nonumber\\
{\kappa}_{2}(k_{0}\to 0,{\mathbf{k}})&=&S(k_{0}\to
0,{\mathbf{k}}),\nonumber\\
{\kappa}_{3}(k_{0}\to 0,{\mathbf{k}})&=&R(k_{0}\to
0,{\mathbf{k}}),\nonumber\\
{\kappa}_{4}(k_{0}\to 0,{\mathbf{k}})&=&0.
\end{eqnarray}
In the next section (\ref{Z35})-(\ref{Z37}) will be used to
determine the ring potential of QED in a constant magnetic field at
finite temperature.
\section{QED effective potential for $T\neq 0$ and
$B\neq 0$ beyond the static limit}
\setcounter{equation}{0} In this section QED effective potential in
a constant magnetic field at finite temperature will be determined
in an approximation beyond the static limit. It receives
contributions from the one-loop and ring (plasmon) potentials. Let
us first look at the one-loop part of the effective potential,
$V^{(1)}$, which is calculated in \cite{sato} for constant magnetic
field and zero chemical potential.\footnote{There are different
equivalent methods to determine the one-loop effective potential in
the presence of constant magnetic field. One of these methods is to
use the Schwinger proper-time formalism \cite{miransky-2}, where the
one-loop effective potential is defined by
$V^{(1)}=-i\Omega^{-1}\mbox{Tr}\ln S^{-1}$. Here, $\Omega$ is the
4-dimensional space-tine volume, and ${S}$ is the free propagator of
massive fermions in a constant magnetic field with $m$
(\ref{Z2})-(\ref{Z3}) [for a definition of one-loop effective
potential in LLL approximation see (\ref{AB19}) in Appendix A]. The
same one-loop effective potential is calculated in \cite{sato} using
the worldline formalism \cite{worldline}. Although this method is
different from the well-known Schwinger proper-time formalism, but
the final result for the one-loop effective potential is the same as
in the Schwinger's method (see \cite{miransky-nc} and
\cite{jafari-sadooghi} for recent example on the equivalence between
these two methods).} It is given by the following integral over the
Schwinger parameter $s$
\begin{eqnarray}\label{A1}
V^{(1)}(m,eB;T)=-\frac{2eB}{\beta}\int_{0}^{\infty}ds\
\frac{\Theta_{2}(0|is\frac{4\pi}{\beta^{2}})}{(4\pi
s)^{\frac{3}{2}}}\coth(seB)e^{-sm^{2}}.
\end{eqnarray}
Here, $\beta$ is the inverse of temperature $\beta\equiv
\frac{1}{T}$ and
\begin{eqnarray}\label{A2}
\Theta_{2}(u|\tau)\equiv
2\sum\limits_{n=0}^{\infty}e^{i\pi\tau\left(n+\frac{1}{2}\right)^{2}}\cos\left(\left(2n+1\right)u\right),
\end{eqnarray}
is the elliptic $\Theta$-function of second kind. The above
potential can be separated into a temperature independent part,
$V_{0}^{(1)}$, and a temperature dependent part, $V_{T}^{(1)}$,
\begin{eqnarray}\label{A3}
V^{(1)}(m,eB;T)=V^{(1)}_{0}(m,eB;\Lambda)+V_{T}^{(1)}(m,eB).
\end{eqnarray}
To do this, we use the identity \cite{miransky-1}
\begin{eqnarray}\label{A4}
\Theta_{2}(u|\tau)=\left(\frac{i}{\tau}\right)^{1/2}e^{-\frac{iu^{2}}{\pi\tau}}\Theta_{4}\left(\frac{u}{\tau}|
-\frac{1}{\tau}\right),
\end{eqnarray}
where
\begin{eqnarray}\label{A5}
\Theta_{4}(u|\tau)=1+2\sum\limits_{n=1}^{\infty}(-1)^{n}\ e^{i\pi
n^{2}\tau}\cos\left(2nu\right),
\end{eqnarray}
is the fourth Jacobian $\Theta$-function. Using the above identities
and the series expansion
$$\coth t=1+2\sum\limits_{m=1}^{\infty}e^{-2mt},$$
the temperature independent part is given by
\begin{eqnarray}\label{A6}
V_{0}^{(1)}\left(m,eB;\Lambda\right)&=&-\frac{eB}{8\pi^{2}}\int_{\frac{1}{\Lambda^{2}}}^{\infty}\frac{ds}{s^{2}}\bigg[e^{-sm^{2}}+2\sum\limits_{\ell=1}^{\infty}
e^{-s(m^2+2eB\ell)}\bigg],\nonumber\\
&=&-\frac{eB}{8\pi^2}\bigg[m^2\Gamma\left(-1,\frac{m^2}{\Lambda^2}\right)
+2\sum_{\ell=1}^{\infty}\left(m^2+2eB\ell\right)\
\Gamma\left(-1,\frac{\left(m^2+2eB\ell\right)}{\Lambda^2}\right)\bigg].
\end{eqnarray}
with $\Lambda$ the ultraviolet (UV) cutoff, and
$\Gamma\left(n,z\right)\equiv \int_{z}^{\infty}dt\ t^{n-1}e^{-t}$
the incomplete $\Gamma$-function. The temperature dependent part of
the one-loop effective potential (\ref{A1}) reads
\begin{eqnarray}\label{A7}
V_{T}^{(1)}(m,eB)&=&
-\frac{eB}{4\pi^{2}}\sum\limits_{n=1}^{\infty}(-1)^{n}\bigg[\int
\limits_{0}^{\infty}\frac{ds}{s^{2}}e^{-\left(sm^{2}+\frac{n^{2}\beta^{2}}{4s}\right)}
+2\sum\limits_{\ell=1}^{\infty}\int\limits_{0}^{\infty}\frac{ds}{s^{2}}e^{-\left(s(m^{2}+2eB\ell)+\frac{n^{2}\beta^{2}}{4s}\right)}\bigg]\nonumber\\
&=&-\frac{eB}{\pi^{2}}\sum\limits_{n=1}^{\infty}(-1)^{n}\bigg[\frac{m}{n\beta}
K_{1}\left(n\beta
m\right)+2\sum\limits_{\ell=1}^{\infty}\frac{\sqrt{\left(m^{2}+2eB\ell\right)}}{n
\beta}K_{1}\left(n\beta\sqrt{\left(m^{2}
+2eB\ell\right)}\right)\bigg],\nonumber\\
\end{eqnarray}
where $K_{n}(z)$ is the modified Bessel-function of the second kind
defined by
\begin{eqnarray}\label{A8}
\int\limits_{0}^{\infty}dx
x^{\nu-1}\exp\left(-\frac{\beta}{x}-\gamma
x\right)=2\left(\frac{\beta}{\gamma}\right)^{\frac{\nu}{2}}K_{\nu}\left(2\sqrt{\beta\gamma}\right).
\end{eqnarray}
Next, we will focus on the ring contribution to QED effective
potential, that will be determined in the IR limit. The general
structure of the ring diagram is given by
\begin{eqnarray}\label{A9}
V_{ring}&=&\frac{T}{2}\sum\limits_{n=-\infty}^{+\infty}\int\frac{d^{3}k}{(2\pi)^{3}}\sum\limits_{N=1}^{\infty}\frac{(-1)^{N}}{N}\bigg[
D^{(0)}_{\mu\rho}\left(n,{\mathbf{k}}\right)
\Pi^{\rho\mu}\left(n,{\mathbf{k}}\right) \bigg]^{N}\nonumber\\
&=&-\frac{T}{2}\sum\limits_{n=-\infty}^{+\infty}
\int\frac{d^{3}k}{(2\pi)^{3}}\ln[1+D^{(0)}_{\mu\rho}(n,{\mathbf{k}})\Pi^{\rho\mu}(n,{\mathbf{k}})].
\end{eqnarray}
To simplify this potential we use the definition of the free photon
propagator $D_{\mu\nu}^{(0)}(k)$ from (\ref{Z19}) and the vacuum
polarization tensor $\Pi_{\mu\nu}(k)$ from (\ref{Z21}), and arrive
at\footnote{To build the ring potential, we have used the free
(bare) photon propagator $D_{\mu\nu}^{(0)}(k)$ to be consistent with
the result from our one-loop effective potential throughout this
paper. Note that the fermion propagator that are used to determine
the polarization tensor $\Pi_{\mu\nu}$ in the ring potential, are
free propagator of massive fermions in the LLL approximation. This
is also consistent with the approximations used in this paper.}
\begin{eqnarray}\label{A10}
V_{ring}(m,eB;T)&=&-\frac{T}{2}\sum\limits_{n=-\infty}^{+\infty}\int\frac{d^{3}k}{(2\pi)^{3}}\sum\limits_{i=1}^{4}
\ln\left(1-\frac{\kappa_{i}(k_{0},{\mathbf{k}})}{k_{E}^{2}}\right),
\end{eqnarray}
where for Euclidean four-momentum $k_{E}$, we have
$k_{E}^{2}={\mathbf{k}}^{2}+4\pi^{2}n^{2}T^{2}$. Here, the
orthogonality of the eigenfunctions $b_{\mu}^{(i)}$ from (\ref{Z20})
and the relation $b_{\mu}^{(i)}b_{\mu}^{\star(j)}=0, \forall i\neq
j$ are used. To take the IR limit of this potential, we set $n=0$ or
equivalently $k_{0}\to 0$ in
$\kappa_{i}\left(k_{0},{\mathbf{k}}\right)$ as well as in
$k_{E}^{2}$. We arrive therefore at
\begin{eqnarray}\label{A11}
V_{ring}^{\mbox{\tiny{IR
limit}}}(m,eB;T)=-\frac{T}{2}\int\frac{d^{3}k}{(2\pi)^{3}}\sum\limits_{i=1}^{4}\ln\left(1-\frac{\kappa_{i}(k_{0}\to
0,{\mathbf{k}})}{{\mathbf{k}}^{2}}\right),
\end{eqnarray}
where $\kappa_{i}\left(k_{0}\to 0,{\mathbf{k}}\right)$ are given in
(\ref{Z37}). To compare this result  with the result (\ref{XX1})
from \cite{bordag, demchik}, let us consider the static (zero
momentum) limit ${\mathbf{k}}\to 0$ in (\ref{A11}). Using the
(\ref{Z37}) and (\ref{Z35}) and taking ${\mathbf{k}}\to 0$, we have
\begin{eqnarray}\label{A12}
\kappa_{i}(0,{\mathbf{0}})=0,\qquad\mbox{for}\qquad i=1,3,4
\qquad\mbox{and}\qquad\kappa_{2}(0,{\mathbf{0}})=S(0,{\mathbf{0}})=I_{2}.
\end{eqnarray}
Further, using (\ref{Z30}),
$\kappa_{2}(0,{\mathbf{0}})=I_{2}=-\Pi_{44}(0,{\mathbf{0}})$.
Continuing into the Minkowski space we have
$\kappa_{2}^{\mbox{\tiny{Mink.}}}\equiv
-\Pi_{00}=\Pi_{44}=-\kappa_{2}$. Plugging this result in
(\ref{A11}), the ring contribution to QED effective potential in the
\textit{static limit} reads
\begin{eqnarray}\label{A13}
V_{ring}^{\mbox{\tiny{static
limit}}}(m,eB;T)&=&-\frac{T}{2}\int\frac{d^{3}k}{(2\pi)^{3}}\sum\limits_{i=1}^{4}
\ln\left(1-\frac{\kappa_{i}^{\mbox{\tiny{Mink.}}}\left(0,{\mathbf{0}}\right)}{{\mathbf{k}}^{2}}\right)=-\frac{T}{2}\int\frac{d^{3}k}{(2\pi)^{3}}
\ln\left(1+\frac{\Pi_{00}(0,{\mathbf{0}})}{{\mathbf{k}}^{2}}\right)\nonumber\\
&=&-\frac{T}{4\pi^{2}}\int\limits_{0}^{\Lambda}
{\mathbf{k}}^{2}d{\mathbf{k}}
\ln\left(1+\frac{\Pi_{00}(0,{\mathbf{0}})}{{\mathbf{k}}^{2}}\right)=\frac{T}{12\pi}\bigg[\Pi_{00}\left(0,{\mathbf{0}}\right)\bigg]^{3/2}
+\mbox{$\Lambda$
dependent terms}.\nonumber\\
\end{eqnarray}
Taking the Higgs mass $m(v)=0$, this result indeed coincides with
(\ref{XX1}) from \cite{bordag,demchik}.
\par
The ring improved effective potential for QED in a constant magnetic
field at finite temperature is therefore given by adding the
one-loop effective potential (\ref{A6})-(\ref{A7}) and the ring
(plasmon) potential (\ref{A10})
\begin{eqnarray}\label{A14}
V_{\mbox{\tiny{eff}}}(m,eB;
T,\Lambda)&=&-\frac{eB}{8\pi^2}\bigg[m^2\Gamma\left(-1,\frac{m^2}{\Lambda^2}\right)
+2\sum_{\ell=1}^{\infty}\left(m^2+2eB\ell\right)\
\Gamma\left(-1,\frac{\left(m^2+2eB\ell\right)}{\Lambda^2}\right)\bigg]\nonumber\\
&&-\frac{eB}{\pi^{2}}\sum\limits_{n=1}^{\infty}(-1)^{n}\bigg[\frac{m}{n\beta}
K_{1}\left(n\beta
m\right)+2\sum\limits_{\ell=1}^{\infty}\frac{\sqrt{\left(m^{2}+2eB\ell\right)}}{n
\beta}K_{1}\left(n\beta\sqrt{\left(m^{2}
+2eB\ell\right)}\right)\bigg]\nonumber\\
&&-\frac{T}{2}\sum\limits_{n=-\infty}^{+\infty}\int\frac{d^{3}k}{(2\pi)^{3}}\sum\limits_{i=1}^{4}\ln\left(1-\frac{\kappa_{i}\left(k_{0},{\mathbf{k}}\right)}
{{\mathbf{k}}^{2}}\right).
\end{eqnarray}
In the following two paragraphs, we will determine QED effective
potential $V_{\mbox{\tiny{eff}}}$ in the limit of weak and strong
magnetic field.
\subsection{QED effective potential in the limit of weak magnetic field}
The weak magnetic field limit is characterized by $eB\ll m^{2}\ll
T^{2}$. To determine the effective potential in this limit, let us
first consider the one-loop effective potential (\ref{A1}).
Expanding $\coth(eBs)$ on the right hand side (r.h.s.) of (\ref{A1})
in the orders of $eB$ up to second order, we get
\begin{eqnarray}\label{A15}
V^{(1)}(m,eB;T)=-\frac{2}{\beta}\int_{{\cal{S}}}\frac{ds}{(4\pi
s)^{3/2}}\Theta_{2}\left(0|is\frac{4\pi
}{\beta^{2}}\right)\left(\frac{1}{s}+\frac{s(eB)^{2}}{3}+\cdots\right)e^{-sm^{2}}.
\end{eqnarray}
To separate (\ref{A15}) into a temperature independent and a
temperature dependent part, we use (\ref{A4}) and (\ref{A5}) and
arrive first at
\begin{eqnarray}\label{A16}
V^{(1)}(m,eB;T)=-\frac{1}{8\pi^{2}}\int_{{\cal{S}}}\frac{ds}{s^{2}}\left(\frac{1}{s}+\frac{s
(eB)^{2}}{3}\right)e^{-sm^{2}}\left(1+2\sum_{n=1}^{\infty}(-1)^{n}\
e^{-\frac{n^{2}\beta^{2}}{4s}}\right)+\cdots.
\end{eqnarray}
Here, the integration region ${\cal{S}}$ spans over
$s\in[\frac{1}{\Lambda^{2}},\infty[$ for the temperature independent
part, and over $s\in[0,\infty[$ for the temperature dependent part.
Using the definition of the incomplete $\Gamma$-function,
$\Gamma(n,z)=\int_{z}^{\infty}dt\ t^{n-1}e^{-t}$, as well as
(\ref{A8}), the one-loop effective potential can be determined in
the limit of weak magnetic field up to second order in $eB$
\begin{eqnarray}\label{A17}
V^{(1)/\mbox{\tiny{weak}}}\left(m,eB;T\right)&=&-\frac{1}{8\pi^{2}}\left\{m^{4}\Gamma\left(-2,\frac{m^{2}}{\Lambda^{2}}\right)+\frac{(eB)^{2}
}{3}\Gamma\left(0,\frac{m^{2}}{\Lambda^{2}}\right)\right.\nonumber\\
&&\left.+4\sum_{n=1}^{\infty}(-1)^{n}\left(\frac{4m^{2}}{n^{2}\beta^{2}}K_{2}(n
m \beta)+\frac{(eB)^{2}}{3}K_{0}(n m \beta)\right)\right\}.
\end{eqnarray}
To determine the ring contribution to the effective potential in the
limit of weak magnetic field, let us consider (\ref{A11}), where
$\kappa_{i}\left(0,{\mathbf{k}}\right), i=1,\cdots,4$ are given in
(\ref{Z37}). To determine $\kappa_{i}$ in limit of weak magnetic
field, we have to evaluate $P,S$ and $R$ from (\ref{Z35}), and
consequently the functions $I_{i}, i=1,\cdots,4$ from (\ref{Z31}) in
this limit. To do this we expand $I_{i}$ up to second order in $eB$.
Assuming ${\mathbf{k}}_{\perp}^{2}\ll eB\ll k_{3}^{2}$ and
neglecting therefore the terms proportional to
${\mathbf{k}}_{\perp}^{2}(eB)^{2}$ \cite{huang}, we arrive first
at\footnote{In the following, $I_{4}$ will be skipped since, as it
turns out, the ring potential in the limit of weak magnetic field is
determined only by $I_{i},i=1,2,3$ [See (\ref{A22})].}
\begin{eqnarray}\label{A18}
\tilde{I}_{1}^{0}&=&-\frac{\alpha}{4\pi}\int_{\frac{1}{\Lambda^{2}}}^{\infty}du\int\limits_{-1}^{+1}dv\
e^{-u\left(m^{2}+\frac{1-v^{2}}{4}\mathbf{k}^{2}\right)}\bigg[\frac{1-v^{2}}{u}-\frac{(eB)^{2}}{6}
u\left(1-v^{2}\right)^{2}\bigg],\nonumber\\
\tilde{I}_{1}^{T}&=&-\frac{\alpha}{2\pi}\int_{0}^{\infty}du\int\limits_{-1}^{+1}dv\
e^{-u\left(m^{2}+\frac{1-v^{2}}{4}\mathbf{k}^{2}\right)}
\sum_{\ell=1}^{\infty}\left(-1\right)^{\ell}\
e^{-\frac{\ell^{2}}{4uT^{2}}}\bigg[\frac{1-v^{2}}{u}-\frac{(eB)^{2}}{6}
u\left(1-v^{2}\right)^{2}\bigg],\nonumber\\
\tilde{I}_{2}^{0}&=&0\nonumber\\
\tilde{I}_{2}^{T}&=&-\frac{\alpha}{2\pi}\int_{0}^{\infty}du\int\limits_{-1}^{+1}dv\
e^{-u\left(m^{2}+\frac{1-v^{2}}{4}\mathbf{k}^{2}\right)}
\sum_{\ell=1}^{\infty}\left(-1\right)^{\ell}\
e^{-\frac{\ell^{2}}{4uT^{2}}}\frac{\ell^{2}}{T^{2}}
\bigg[\frac{1}{u^{3}}+\frac{(eB)^{2}}{3u}\bigg],\nonumber\\
\nonumber\\
\tilde{I}_{3}^{0}&=&-\frac{\alpha}{4\pi}\int_{\frac{1}{\Lambda^{2}}}^{\infty}du\int\limits_{-1}^{+1}dv\
e^{-u\left(m^{2}+\frac{1-v^{2}}{4}\mathbf{k}^{2}\right)}\bigg[\frac{1-v^{2}}{u}+\frac{(eB)^{2}}{3}u(1-v^{2})\bigg],\nonumber\\
\nonumber\\
\tilde{I}_{3}^{T}&=&\frac{-\alpha}{2\pi}\int_{0}^{\infty}du\int\limits_{-1}^{+1}dv\
e^{-u\left(m^{2}+\frac{1-v^{2}}{4}\mathbf{k}^{2}\right)}
\sum_{\ell=1}^{\infty}\left(-1\right)^{\ell}\
e^{-\frac{\ell^{2}}{4uT^{2}}}
\bigg[\frac{1-v^{2}}{u}+\frac{(eB)^{2}}{3}u(1-v^{2})\bigg].
\end{eqnarray}
To perform then the integrations over $u$ and $v$, we expand the
above expressions in the order of
$\frac{{\mathbf{k}}^{2}}{m^{2}}$.We get the following general
structure
\begin{eqnarray}\label{A19}
\tilde{I}_{i}^{0}=a_{i}^{0}+\frac{{\mathbf{k}}^{2}}{m^{2}}
b_{i}^{0}\qquad \mbox{for}\qquad i=1,3,\qquad\mbox{as well as}\qquad
\tilde{I}_{i}^{T}=a_{i}^{T}+\frac{{\mathbf{k}}^{2}}{m^{2}}
b_{i}^{T}, \qquad
\mbox{for}\qquad i=1,2,3,\nonumber\\
\end{eqnarray}
where, the temperature independent parts are\footnote{Note that the
temperature independent part consists of imaginary terms. These
terms cancel the contact terms in (\ref{Z29}) [see footnote 11].}
\begin{eqnarray}\label{A20}
a_{1}^{0}&=&+\frac{2\alpha}{45\pi}\frac{(eB)^{2}}{m^{4}},\qquad\
b_{1}^{0}=+\frac{\alpha}{15\pi}-\frac{2\alpha}{105\pi}\frac{(eB)^{2}}{m^{4}},\nonumber\\
a_{3}^{0}&=&-\frac{\alpha}{9\pi}\frac{(eB)^{2}}{m^{4}},\qquad\ \
b_{3}^{0}=+\frac{\alpha}{15\pi}+\frac{2\alpha}{45\pi}\frac{(eB)^{2}}{m^{4}},
\end{eqnarray}
and the temperature dependent parts are
\begin{eqnarray}\label{A21}
a_{1}^{T}&=&\sum\limits_{\ell=1}^{\infty}
(-1)^{\ell}\bigg[-\frac{4\alpha}{3\pi}K_{0}\left(\ell
m\beta\right)+\frac{2\alpha}{45\pi}\frac{(eB)^{2}}{m^{4}}(\ell
m\beta)^{2}K_{2}\left(\ell m\beta\right)\bigg],\nonumber\\
b_{1}^{T}&=&\sum\limits_{\ell=1}^{\infty}(-1)^{\ell}\bigg[\frac{2\alpha}{15\pi}(\ell
m\beta)K_{1}\left(\ell
m\beta\right)-\frac{\alpha}{210\pi}\frac{(eB)^{2}}{m^{4}}(\ell
m\beta)^{3}K_{3}\left(\ell m\beta\right)\bigg],\nonumber\\
a_{2}^{T}&=&\sum\limits_{\ell=1}^{\infty}(-1)^{\ell}\bigg[-\frac{8\alpha}{\pi}m^{2}K_{2}\left(\ell
m\beta\right)-\frac{2\alpha}{3\pi}\frac{(eB)^{2}}{m^{2}}(\ell
m\beta)^{2}K_{0}\left(\ell m\beta\right)\bigg],\nonumber\\
b_{2}^{T}&=&\sum\limits_{\ell=1}^{\infty}(-1)^{\ell}\bigg[\frac{2\alpha}{3\pi}m^2(\ell
m\beta)K_{1}\left(\ell
m\beta\right)+\frac{\alpha}{18\pi}\frac{(eB)^{2}}{m^{2}}(\ell
m\beta)^{3}K_{1}\left(\ell m\beta\right)\bigg],\nonumber\\
a_{3}^{T}&=&\sum\limits_{\ell=1}^{\infty}
(-1)^{\ell}\bigg[-\frac{4\alpha}{3\pi}K_{0}\left(\ell
m\beta\right)-\frac{\alpha}{9\pi}\frac{(eB)^{2}}{m^{4}}(\ell
m\beta)^{2}K_{2}\left(\ell m\beta\right)\bigg],\nonumber\\
b_{3}^{T}&=&\sum\limits_{\ell=1}^{\infty}(-1)^{\ell}\bigg[\frac{2\alpha}{15\pi}(\ell
m\beta)K_{1}\left(\ell
m\beta\right)+\frac{\alpha}{90\pi}\frac{(eB)^{2}}{m^{4}}(\ell
m\beta)^{3}K_{3}\left(\ell m\beta\right)\bigg].
\end{eqnarray}
To evaluate $P,S$ and $R$ from (\ref{Z35}) in the limit of weak
$eB$, we use again ${\mathbf{k}}_{\perp}^{2}\ll eB\ll k_{3}^{2}$
\cite{huang}. The most dominant terms in
$\kappa_{i}(0,{\mathbf{k}})$ are therefore given by
\begin{eqnarray}\label{A22}
\kappa_{1}\left(k_{0}\to 0,{\mathbf{k}}\right)&=&-k_{3}^{2}\tilde{I}_{1}+{\cal{O}}\left(\frac{k_{\bot}^{2}}{eB}\right)\nonumber\\
{\kappa}_{2}\left(k_{0}\to 0,{\mathbf{k}}\right)&=&-k_{3}^{2}\tilde{I}_{3}+\tilde{I}_{2}+{\cal{O}}\left(\frac{k_{\bot}^{2}}{eB}\right)\nonumber\\
{\kappa}_{3}\left(k_{0}\to 0,{\mathbf{k}}\right)&=&-k_{3}^{2}\tilde{I}_{1}+{\cal{O}}\left(\frac{k_{\bot}^{2}}{eB}\right)\nonumber\\
{\kappa}_{4}\left(k_{0}\to 0,{\mathbf{k}}\right)&=&0.
\end{eqnarray}
Plugging as next these expressions in (\ref{A11}), the ring
potential in the limit of weak magnetic field is given by
\begin{eqnarray}\label{A23}
V_{ring}^{\mbox{\tiny{IR limit/weak}}}&\approx&-\frac{T}{2}\int
\frac{d^{3}k}{(2\pi)^{3}}\bigg[
2\ln\left(1+\frac{k_{3}^{2}}{{\mathbf{k}}^{2}}\left(a_{1}^{0}+a_{1}^{T}\right)+\frac{k_{3}^{2}}{m^{2}}\left(b_{1}^{0}+b_{1}^{T}\right)\right)\nonumber\\
&&
+\ln\left(1+\frac{k_{3}^{2}}{{\mathbf{k}}^{2}}\left(a_{3}^{0}+a_{3}^{T}\right)+\frac{k_{3}^{2}}{m^{2}}\left(b_{3}^{0}+b_{3}^{T}\right)-
\left(\frac{a_{2}^{T}}{{\mathbf{k}}^{2}}+\frac{b_{2}^{T}}{m^{2}}\right)\right)\bigg].
\end{eqnarray}
To perform the integration over three-momentum $\mathbf{k}$, we will
use the same procedure as was discussed in part B of the
Introduction. Adding and substracting an appropriate integral to the
ring potential (\ref{A23}), whose integrand is independent of
$k_{3}^{2}$, we arrive at
\begin{eqnarray}\label{A24}
V_{ring}^{\mbox{\tiny{IR
limit/weak}}}=V^{(f)}_{ring}+V^{\Lambda}_{ring},
\end{eqnarray}
where the finite part is
\begin{eqnarray}\label{A25}
V_{ring}^{(f)}=-\frac{T}{2}\int \frac{d^{3}k}{(2\pi)^{3}}\
\ln\left(1-\left(\frac{a_{2}^{T}}{{\mathbf{k}}^{2}}+\frac{b_{2}^{T}}{m^{2}}\right)\right),
\end{eqnarray}
and the cutoff ($\Lambda$) dependent part
\begin{eqnarray}\label{A26}
\lefteqn{V^{\Lambda}_{ring}=-\frac{T}{2}\int
\frac{d^{3}k}{(2\pi)^{3}}\bigg[
2\ln\left(1+\frac{k_{3}^{2}}{{\mathbf{k}}^{2}}\left(a_{1}^{0}+a_{1}^{T}\right)+\frac{k_{3}^{2}}{m^{2}}\left(b_{1}^{0}+b_{1}^{T}\right)\right)}\nonumber\\
&&
+\ln\left(1+\frac{k_{3}^{2}}{{\mathbf{k}}^{2}}\left(a_{3}^{0}+a_{3}^{T}\right)+\frac{k_{3}^{2}}{m^{2}}\left(b_{3}^{0}+b_{3}^{T}\right)-
\left(\frac{a_{2}^{T}}{{\mathbf{k}}^{2}}+\frac{b_{2}^{T}}{m^{2}}\right)\right)\bigg]-\frac{T}{2}\int
\frac{d^{3}k}{(2\pi)^{3}}\
\ln\left(1-\left(\frac{a_{2}^{T}}{{\mathbf{k}}^{2}}+\frac{b_{2}^{T}}{m^{2}}\right)\right).\nonumber\\
\end{eqnarray}
Performing the integration over $\mathbf{k}$ in $V_{ring}^{(f)}$ we
get
\begin{eqnarray}\label{A27}
V_{ring}^{(f)}\approx {T}
\frac{({a}_{2}^{T})^{3/2}}{\left(1-\frac{{b}_{2}^{T}}{m^{2}}\right)^{3/2}}+\mbox{Cutoff
dependent terms},
\end{eqnarray}
whereas for $V_{ring}^{\Lambda}$ we have
\begin{eqnarray}\label{A28}
V_{ring}^{\Lambda}\approx\alpha T\
{\cal{O}}\left(\Lambda^{3}\right).
\end{eqnarray}
Note that $V_{ring}^{\Lambda}$ can be derived by expanding the
logarithms in (\ref{A26}) and performing the three dimensional
integration over $k$ using a momentum cutoff $\Lambda$. Neglecting
now the cutoff dependent terms, we arrive at
\begin{eqnarray}\label{A29}
V_{ring}^{\mbox{\tiny{IR limit/weak}}}&=& {\cal{C}} T
\frac{({a}_{2}^{T})^{3/2}}{\left(1-\frac{{b}_{2}^{T}}{m^{2}}\right)^{3/2}}\nonumber\\
&=&{\cal{C}}  {m^3 T}\left\{
\frac{\frac{8\alpha}{\pi}\sum\limits_{\ell=1}^{\infty}(-1)^{\ell+1}K_{2}\left(\ell
m\beta\right)}
{1-\frac{2\alpha}{3\pi}\sum\limits_{\ell=1}^{\infty}(-1)^{\ell}(\ell
m\beta)K_{1}\left(\ell
m\beta\right)}\right\}^{3/2}+{\cal{O}}\left(\left(\frac{eB}{m^{2}}\right)^{2}\right).
\end{eqnarray}
Here, the proportionality constant ${\cal{C}}={\cal{O}}(1)$. To
compare this result with the ring potential in the leading
\textit{static limit}, (\ref{A29}) will be evaluated in the high
temperature expansion $m\beta\to 0$. This can be determined from the
behavior of Bessel functions in this limit
\begin{eqnarray}\label{A30}
K_{\nu}(x)\stackrel{x\to
0}{\longrightarrow}\frac{1}{2}\Gamma\left(\nu\right)\left(\frac{2}{x}\right)^{\nu},
\end{eqnarray}
and the Bessel function identities \cite{ogilvie, gradshteyn}
\begin{eqnarray}\label{A31}
\sum\limits_{\ell=1}^{\infty}K_{0}(\ell
z)\cos\left(\ell\phi\right)=\frac{1}{2}\left(\gamma+\ln\frac{z}{4\pi}\right)+C_{0}(z,\phi).
\end{eqnarray}
Here, $\gamma\simeq 0.577$ is the Euler-Mascheroni constant and
$C_{0}(z,\phi)$ is given by
\begin{eqnarray}\label{A32}
C_{0}(z,\phi)\equiv\frac{\pi}{2}\sum_{\ell}^{'}
\left(\frac{1}{\sqrt{z^2+\left(\phi-2\pi\ell\right)^2}}-\frac{1}{2\pi|\ell|}\right),
\end{eqnarray}
where the notation $\sum_{\ell}^{'}$ indicates that singular terms
are omitted when $\ell=0$ \cite{ogilvie}. Deriving (\ref{A31}) with
respect to $\ln z$ and reminding the fact that $\frac{\partial
K_{0}(z)}{\partial \ln z}=-zK_{1}(z)$, we get
\begin{eqnarray}\label{A33}
\sum\limits_{\ell=1}^{\infty}(\ell z) K_{1}\left(\ell
z\right)\cos\left(\ell\phi\right)=-\frac{1}{2}+C_{1}(z,\phi),
\end{eqnarray}
with
\begin{eqnarray}\label{A34}
C_{1}(z,\phi)\equiv -\frac{\partial C_{0}(z,\phi)}{\partial \ln
z}=\frac{\pi}{2}\sum\limits_{\ell=-\infty}^{\infty}\frac{z^2}{\left(z^2+\left(\phi-2\pi\ell\right)^2\right)^{3/2}}.
\end{eqnarray}
Choosing now $\nu=2$ as well as $x\equiv \ell m\beta$ in (\ref{A30})
and $z\equiv m\beta$ as well as $\phi\equiv \pi$ in
(\ref{A31})-(\ref{A34}), and plugging (\ref{A30}) in the numerator
and (\ref{A33}) in the denominator of (\ref{A29}), we arrive first
at
\begin{eqnarray}\label{A35}
V_{ring}^{\mbox{\tiny{IR
limit/weak}}}&\longrightarrow&{\cal{C}}\frac{8\pi}{3}\sqrt{\frac{\pi}{3}}\frac{T^4\alpha^{3/2}}
{\left(1+\frac{\alpha}{3\pi}-\frac{2\alpha}{3\pi}C_{1}\left(m\beta,\pi\right)\right)^{3/2}}.
\end{eqnarray}
Here, we have used
$\sum\limits_{\ell=1}^{\infty}\frac{(-1)^{\ell+1}}{\ell^2}=\frac{\pi^{2}}{12}$.
Expanding now $C_{1}(m\beta,\pi)$ in the denominator of (\ref{A35})
in the orders of $m\beta$ and using
$C_{1}(m\beta,\pi)=\frac{7(m\beta)^{2}\zeta(3)}{8\pi^{2}}+{\cal{O}}((m\beta)^{3})$,
we arrive for $\alpha\to 0$ at
\begin{eqnarray}\label{A36}
V_{ring}^{\mbox{\tiny{IR
limit/weak}}}&=&\frac{8\pi}{3}\sqrt{\frac{\pi}{3}}{\cal{C}}T^{4}\left(\frac{\alpha}{1+\frac{\alpha}{2\pi}}\right)^{3/2}
\bigg[1-
\frac{7\zeta(3)}{8\pi^{3}}\frac{\alpha}{(1+\frac{\alpha}{2\pi})}(m\beta)^{2}\bigg]+{\cal{O}}\left((m\beta)^{3}\right),
\nonumber\\
&=&\frac{8\pi}{3}\sqrt{\frac{\pi}{3}}{\cal{C}}T^{4}\alpha^{3/2}\bigg[1-\frac{\alpha}{2\pi}
\left(1-\frac{7\zeta(3)}{4\pi^{2}}\left(m\beta\right)^{2}\right)\bigg]
+{\cal{O}}\left(\alpha^{7/2},(m\beta)^{3}\right).
\end{eqnarray}
The first term is the usual $\alpha^{3/2}$ contribution to the ring
potential from the \textit{static limit} \cite{kapusta, bordag,
demchik} [see also part B of the Introduction and in particular
(\ref{X11})]. The second term, however, arises only in the
\textit{IR limit}. It is a consequence of the additional $b_{2}^{T}$
term in the denominator of $V_{ring}$ from (\ref{A29}). The above
result (\ref{A29}) can be viewed as a nonperturbative correction of
QED effective potential in addition to the perturbative loop
corrections to this potential. Note that in QCD at finite
temperature and zero magnetic fields $\alpha_{s}^{3/2}$ and
$\alpha_{s}^{5/2}$ terms are calculated using Hard Thermal Loop
expansion (see \cite{braaten, alpha5-2} and references therein). The
above result are relevant in studying the standard electroweak phase
transition in the presence of weak external magnetic field
\cite{ayala}.
\subsection{QED effective potential in the limit of strong magnetic field}
The strong magnetic field is characterized by $m^{2}\ll T^{2}\ll
eB$. To determine QED effective potential in the limit of strong
magnetic field, let us consider first the one-loop effective
potential (\ref{A1}). For $eB\to \infty$ (\ref{A1}) is given by
\begin{eqnarray}\label{A37}
V^{(1)}(m,eB;T)=-\frac{2eB}{\beta}\int_{0}^{\infty} ds\
\frac{\Theta_{2}\left(0|is\frac{4\pi}{\beta^{2}}\right)}{\left(4\pi
s\right)^{\frac{3}{2}}}e^{-sm^{2}},
\end{eqnarray}
where $\coth\left(eBs\right)\approx 1$ is used. To separate
(\ref{A37}) into a temperature dependent and a temperature
independent part, we use (\ref{A4}) and (\ref{A5}) and arrive first
at
\begin{eqnarray}\label{A38}
V^{(1)}(m,eB;T)=-\frac{eB}{8\pi^{2}}\int_{{\cal{S}}}\frac{ds}{s^{2}}\left(1+2\sum\limits_{n=1}^{\infty}(-1)^{n}
e^{-\frac{n^{2}\beta^{2}}{4s}}\right)e^{-sm^{2}},
\end{eqnarray}
where the integration region ${\cal{S}}$ spans over
$s\in[\frac{1}{\Lambda^{2}},\infty[$ for the temperature independent
part, and over $s\in[0,\infty[$ for the temperature dependent part.
Using the definition of the incomplete $\Gamma$-function as well as
(\ref{A8}), the one-loop effective potential in the limit of strong
magnetic field is given by
\begin{eqnarray}\label{A39}
V^{(1)/\mbox{\tiny{strong}}}(m,eB;\Lambda,T)=-\frac{
eB}{8\pi^{2}}\left\{m^{2}\Gamma\left(-1,\frac{m^{2}}{\Lambda^{2}}\right)+\frac{8m}{\beta}\sum\limits_{n=1}^{\infty}
\frac{(-1)^{n}}{n}K_{-1}\left(nm\beta\right)\right\}.
\end{eqnarray}
To determine the ring contribution to QED effective potential in the
limit of strong magnetic field, let us consider (\ref{A11}) with
$\kappa_{i}(0,{\mathbf{k}}), i=0,\cdots,4$ from (\ref{Z37}). To
determine $\kappa_{i}, i=1,\cdots,4$ in the limit of strong magnetic
field, we have to evaluate $P,S$ and $R$ from (\ref{Z35}), and
consequently the functions $I_{i},i=1,\cdots,4$ from (\ref{Z31}) in
this limit. Note that in a strong magnetic field at finite
temperature, as in the zero temperature case, QED dynamics is
dominated by LLL, where the chiral symmetry is dynamically broken as
a consequence of the external magnetic field. As we have mentioned
in Sect. II.A, the LLL is characterized by
$k_{3}^{2},{\mathbf{k}}_{\perp}^{2}\ll eB$ and a small dynamical
mass $m^{2}\ll eB$ \cite{miransky-2, miransky1-5}. Keeping these
facts in mind, it is easy to determine the most dominant $I_{i}$
among $I_{i}, i=1,\cdots,4$ in the limit of strong magnetic field. A
simple calculation shows that in the limit $eB\to \infty$ only
$I_{2}$ and $I_{3}$ survive. They are given by
\begin{eqnarray}\label{A40}
{I}_{2}^{T}&\approx&-\frac{\alpha
eB}{2\pi}\int_{0}^{\infty}\frac{du}{u^{2}}\int_{-1}^{1}dv\
e^{{\phi}}
\sum\limits_{\ell=1}^{\infty}(-1)^{\ell}\frac{\ell^{2}}{T^{2}}e^{-\frac{\ell^{2}}{4uT^2}},\nonumber\\
{I}_{3}^{0}&\approx&-\frac{\alpha
eB}{4\pi}\int_{0}^{\infty}du\int_{-1}^{1}dv\
e^{{\phi}}\left(1-v^{2}\right),\nonumber\\
{I}_{3}^{T}&\approx&-\frac{\alpha
eB}{2\pi}\int_{0}^{\infty}du\int_{-1}^{1}dv\ e^{{\phi}}
\sum\limits_{\ell=1}^{\infty}(-1)^{\ell}e^{-\frac{\ell^{2}}{4uT^2}}\left(1-v^{2}\right),
\end{eqnarray}
where
\begin{eqnarray}\label{A41}
{{\phi}}\approx
{-\frac{{\mathbf{k}}_{\perp}^{2}}{2eB}-u\big[m^{2}+\frac{\left(1-v^2\right)}{4}k_{3}^{2}\big]}.
\end{eqnarray}
Plugging (\ref{A40}) in (\ref{Z35}), we get
\begin{eqnarray}\label{A42}
P,R\stackrel{eB\to\infty}{\longrightarrow} 0, \qquad\mbox{and}\qquad
S\stackrel{eB\to
\infty}{\longrightarrow}-k_{3}^{2}\left({I}_{3}^{0}+{I}_{3}^{T}\right)+{I}_{2}^{T}.
\end{eqnarray}
Using now the relations (\ref{Z37}), only $\kappa_{2}(k_{0}\to
0,{\mathbf{k}})=S$ survives in (\ref{A11}). The ring potential is
thus given by
\begin{eqnarray}\label{A43}
V_{ring}^{\mbox{\tiny{IR
limit/strong}}}&=&-\frac{T}{2}\int\frac{d^{3}k}{(2\pi)^{3}}\ln\left(1-\frac{\kappa_2\left(k_{0}\to 0,{\mathbf{k}}\right)}{{\mathbf{k}^{2}}}\right)\nonumber\\
&\approx&-\frac{T}{8\pi^{2}}\int_{0}^{\infty}d({\mathbf{k}}^{2}_{\perp})\int_{0}^{\infty}
dk_{3}\ln\left(1+\frac{k_{3}^{2}\left({I}_{3}^{0}+{I}_{3}^{T}\right)-{I}_{2}^{T}}{\left(k_{3}^{2}
+{\mathbf{k}}_{\perp}^{2}\right)}\right).
\end{eqnarray}
As next the integration over $k_{3}$ will be evaluated separately in
two different regimes of dynamical mass in the LLL. These two
regimes will be indicated by $k_{3}^{2}\ll m^{2}\ll eB$ and
$m^{2}\ll k_{3}^{2}\ll eB$ [see Sect. II.A and in particular
(\ref{Z13}) and (\ref{Z14})]. To do this we use the
relation\footnote{The same method is also used in \cite{miransky-2}.
Here, we have matched the asymptotics at $k_{3}=m$.}
$$\int_{0}^{\infty}dk_{3}=\int_{0}^{m}dk_{3}+\int_{m}^{\infty}dk_{3},$$
where the first integral $\int_{0}^{m}dk_{3}$ corresponds to the
first regime $k_{3}^{2}\ll m^{2}\ll eB$ and the second integral
$\int_{m}^{\infty}dk_{3}$ to the second regime $m^{2}\ll
k_{3}^{2}\ll eB$ in the LLL. As for the integration (\ref{A43}) only
the phase $\phi$ from (\ref{A41}) is different in these two regimes.
Thus taking
\begin{eqnarray}
\phi&\approx& {-\frac{{\mathbf{k}}_{\perp}^{2}}{2eB}-u
m^{2}}\qquad\qquad\ \ \ \ \mbox{for}\qquad k_{3}^{2}\ll m^{2}\ll
eB,\label{A44}\\
\phi&\approx&
{-\frac{{\mathbf{k}}_{\perp}^{2}}{2eB}-\frac{\left(1-v^2\right)}{4}uk_{3}^{2}}\qquad\mbox{for}\qquad
m^{2}\ll k_{3}^{2}\ll eB,\label{A45}
\end{eqnarray}
in $I_{2}^{T},I_{3}^{0}$ and $I_{3}^{T}$ the integration over $u$
and $v$ can be easily performed. As next, we will determine the
corresponding ring contribution to the effective potential for these
two regimes separately. The results will be added eventually.
\subsubsection*{i) Ring potential in the first regime $k_{3}^{2}\ll
m^{2}\ll eB$ of LLL} To determine the ring potential in the
$k_{3}^{2}\ll m^{2}\ll eB$ regime in the LLL, we have to calculate
first the integrals $I_{2}^{T}$, $I_{3}^{0}$ and $I_{3}^{T}$ from
(\ref{A40}) in this regime. Using the phase $\phi$ from (\ref{A44})
we get
\begin{eqnarray}\label{A46}
{I}_{2}^{T}&\approx&-\frac{\alpha
eB}{2\pi}e^{-\frac{{\mathbf{k}}_{\perp}^{2}}{2eB}}\int_{-1}^{1}dv\
\sum\limits_{\ell=1}^{\infty}(-1)^{\ell}\left(\ell\beta\right)^{2}\int_{0}^{\infty}\frac{du}{u^{2}}\ e^{-um^2-\frac{\ell^{2}}{4uT^2}}\equiv
e^{-\frac{{\mathbf{k}}_{\perp}^{2}}{2eB}} A_{2}^{T},\nonumber\\
{I}_{3}^{0}&\approx&-\frac{\alpha
eB}{4\pi}e^{-\frac{{\mathbf{k}}_{\perp}^{2}}{2eB}}\int_{-1}^{1}dv\
\left(1-v^{2}\right)
\int_{0}^{\infty}du\ e^{-um^2}\equiv e^{-\frac{{\mathbf{k}}_{\perp}^{2}}{2eB}}A_{3}^{0},\nonumber\\
{I}_{3}^{T}&\approx&-\frac{\alpha
eB}{2\pi}e^{-\frac{{\mathbf{k}}_{\perp}^{2}}{2eB}}\int_{-1}^{1}dv\
\left(1-v^{2}\right)
\sum\limits_{\ell=1}^{\infty}(-1)^{\ell}\int_{0}^{\infty}du\
e^{-um^{2}-\frac{\ell^{2}}{4uT^2}}\equiv
e^{-\frac{{\mathbf{k}}_{\perp}^{2}}{2eB}} A_{3}^{T}.
\end{eqnarray}
Here, using the notation $M_{\gamma}^{2}\equiv \frac{2\alpha
eB}{\pi}$, we have $A_{3}^{0}\equiv -\frac{M_{\gamma}^{2}}{6m^2}$
and
\begin{eqnarray}\label{A47}
A_{2}^{T}\equiv
-2M_{\gamma}^{2}\sum\limits_{\ell=1}^{\infty}(-1)^{\ell}(\ell
m\beta)K_{1}(\ell m\beta),\qquad A_{3}^{T}\equiv
-\frac{M_{\gamma}^{2}}{3m^{2}}\sum\limits_{\ell=1}^{\infty}(-1)^{\ell}(\ell
m\beta)K_{1}(\ell m\beta).
\end{eqnarray}
The ring potential (\ref{A43}) corresponding to the first regime
$k_{3}^{2}\ll m^{2}\ll eB$ in LLL reads therefore
\begin{eqnarray}\label{A48}
\hspace{-0.5cm}V_{ring}^{\mbox{\tiny{IR
limit/strong}}}\bigg|_{k_{3}^{2}\ll m^{2}\ll
eB}\approx-\frac{T}{8\pi^{2}}\int_{0}^{\infty}d({\mathbf{k}}^{2}_{\perp})\int_{0}^{m}
dk_{3}\ln\left(1+\frac{e^{-\frac{{\mathbf{k}}_{\perp}^{2}}{2eB}}\left(k_{3}^{2}A_{3}-A_{2}^{T}\right)}
{\left(k_{3}^{2}+{\mathbf{k}}_{\perp}^{2}\right)}\right),
\end{eqnarray}
where $A_{3}\equiv A_{3}^{0}+A_{3}^{T}$. Using the expression
(\ref{A33}), $A_{3}$ and $A_{2}^{T}$ can be simplified
\begin{eqnarray}\label{A49}
A_{3}=-\frac{M_{\gamma}^{2}}{3m^{2}}C_{1}(m\beta,\pi),\qquad\mbox{and}\qquad
A_{2}^{T}=M_{\gamma}^{2}\left(1-2C_{1}(m\beta,\pi)\right),
\end{eqnarray}
where $C_{1}(z,\phi)$ is defined in (\ref{A34}). Performing now the
integration over $k_{3}\in [0,m]$, we get first
\begin{eqnarray}\label{A50}
V_{ring}^{\mbox{\tiny{IR limit/strong}}}\bigg|_{k_{3}^{2}\ll
m^{2}\ll eB}\approx
-\frac{mT}{8\pi^{2}}\int_{0}^{\infty}d({\mathbf{k}}^{2}_{\perp})\ln\left(1-\frac{e^{-\frac{{\mathbf{k}}_{\perp}^{2}}{2eB}}\left(A_{2}^{T}-m^{2}A_{3}\right)}
{{\mathbf{k}}_{\perp}^{2}+m^{2}}\right)+J,
\end{eqnarray}
where
\begin{eqnarray}\label{A51}
J\equiv
+\frac{T}{4\pi^{2}}\int_{0}^{\infty}d({\mathbf{k}}^{2}_{\perp})
\sqrt{{\mathbf{k}}_{\perp}^{2}}\mbox{arctan}\left(\frac{m}{\sqrt{{\mathbf{k}}_{\perp}^{2}}}\right)
+{\cal{O}}\left(\frac{m}{\sqrt{eB}}\right).
\end{eqnarray}
Here, an expansion in the orders of $\frac{m}{\sqrt{eB}}$ is
performed, as we are in a regime where $m^{2}\ll eB$. To perform the
integration over the first term in (\ref{A50}), we use the identity
\begin{eqnarray}\label{A52}
\int\limits_{0}^{\infty} dy
\ln\left(1-\frac{e^{-\frac{y}{x}}}{y+z}\right)=x\
\mbox{Li}_{2}\left(-\frac{1}{z}\right),
\end{eqnarray}
where the dilogarithm is defined by
\begin{eqnarray*}
\mbox{Li}_{2}(z)\equiv
-\int\limits_{0}^{z}\frac{\ln\left(1-z\right)}{z}dz=
-\int\limits_{0}^{z}\ln\left(1-z\right)\frac{d}{dz}\ln z dz.
\end{eqnarray*}
Choosing now $y \left(A_{2}^{T}-m^{2}A_{3}\right)\equiv
{\mathbf{k}}_{\perp}^{2}$, $x
\left(A_{2}^{T}-m^{2}A_{3}\right)\equiv 2eB$ and
$z\left(A_{2}^{T}-m^{2}A_{3}\right)\equiv m^{2}$ in (\ref{A52}), the
ring contribution in the first regime $k_{3}^{2}\ll m^{2}\ll eB$ in
the LLL reads
\begin{eqnarray}\label{A53}
V_{ring}^{\mbox{\tiny{IR limit/strong}}}\bigg|_{k_{3}^{2}\ll
m^{2}\ll eB}\approx -\frac{mT
eB}{4\pi^{2}}\mbox{Li}_{2}\left(-\frac{\left(A_{2}^{T}-m^{2}A_{3}\right)}{m^{2}}\right)+J,
\end{eqnarray}
where $A_{2}^{T}-m^{2}A_{3}$ can be simplified using (\ref{A49}) and
reads
\begin{eqnarray}\label{A54}
A_{2}^{T}-m^{2}A_{3}=M_{\gamma}^{2}\left(1-\frac{5}{3}C_{1}(m\beta,\pi)\right).
\end{eqnarray}
As it turns out, the second term on the r.h.s. of (\ref{A53})
vanishes with the ring potential corresponding to the second regime
$m^{2}\ll k_{3}^{2}\ll eB$ in the LLL.
\subsubsection*{ii) Ring potential in the second regime $m^{2}\ll
k_{3}^{2}\ll eB$ of LLL} As for the second regime $ m^{2}\ll
k_{3}^{2}\ll eB$, we have to determine $I_{2}^{T}, I_{3}^{0}$ and
$I_{3}^{T}$ from (\ref{A40}). Taking $\phi$ from (\ref{A45}), we get
\begin{eqnarray}\label{A55}
{I}_{2}^{T}&\approx&-\frac{\alpha
eB}{2\pi}e^{-\frac{{\mathbf{k}}_{\perp}^{2}}{2eB}}\int_{-1}^{1}dv\
\sum\limits_{\ell=1}^{\infty}(-1)^{\ell}\left(\ell\beta\right)^{2}\int_{0}^{\infty}\frac{du}{u^{2}}\ e^{-\frac{(1-v^{2})}{4}
uk_{3}^{2}-\frac{\ell^{2}}{4uT^2}}\equiv e^{-\frac{{\mathbf{k}}_{\perp}^{2}}{2eB}} B_{2}^{T}\nonumber\\
{I}_{3}^{0}&\approx&-\frac{\alpha
eB}{4\pi}e^{-\frac{{\mathbf{k}}_{\perp}^{2}}{2eB}}\int_{-1}^{1}dv\
\left(1-v^{2}\right)
\int_{0}^{\infty}du\ e^{-\frac{(1-v^{2})}{4}uk_{3}^{2}}=-\frac{2\alpha eB}{\pi k_{3}^{2}}e^{-\frac{{\mathbf{k}}_{\perp}^{2}}{2eB}}\equiv
-e^{-\frac{{\mathbf{k}}_{\perp}^{2}}{2eB}}\frac{M_{\gamma}^{2}}{k_{3}^{2}},\nonumber\\
{I}_{3}^{T}&\approx&-\frac{\alpha
eB}{2\pi}e^{-\frac{{\mathbf{k}}_{\perp}^{2}}{2eB}}\int_{-1}^{1}dv\
\left(1-v^{2}\right)
\sum\limits_{\ell=1}^{\infty}(-1)^{\ell}\int_{0}^{\infty}du\
e^{-\frac{(1-v^2)}{4}uk_{3}^{2}-\frac{\ell^{2}}{4uT^2}}\equiv
e^{-\frac{{\mathbf{k}}_{\perp}^{2}}{2eB}}
\frac{B_{3}^{T}}{k_{3}^{2}},
\end{eqnarray}
where $M_{\gamma}^{2}\equiv \frac{2\alpha eB}{\pi}$, and
\begin{eqnarray}\label{A56}
B_{2}^{T}&=&B_{3}^{T}\equiv -M_{\gamma}^{2}\int\limits_{-1}^{+1}dv\
\sum\limits_{\ell=1}^{\infty}(-1)^{\ell}\left(\frac{\ell\beta
k_{3}}{2}\sqrt{1-v^2}\right)K_{1}\left(\frac{\ell \beta
k_{3}}{2}\sqrt{1-v^2}\right).
\end{eqnarray}
Plugging (\ref{A55}) in (\ref{A43}), the contribution from
$B_{2}^{T}$ and $B_{3}^{T}$ cancel and we are left with
\begin{eqnarray}\label{A57}
V_{ring}^{\mbox{\tiny{IR limit/strong}}}\bigg|_{m^{2}\ll
k_{3}^{2}\ll eB}\approx -
\frac{T}{8\pi^{2}}\int_{0}^{\infty}d({\mathbf{k}}^{2}_{\perp})\int_{m}^{\infty}
dk_{3}\ln\left(1-\frac{e^{-\frac{{\mathbf{k}}_{\perp}^{2}}{2eB}}{M}_{\gamma}^{2}}
{\left(k_{3}^{2}+{\mathbf{k}}_{\perp}^{2}\right)}\right).
\end{eqnarray}
Here, the integration over $k_{3}$ can be performed and we arrive
first at
\begin{eqnarray}\label{A58}
V_{ring}^{\mbox{\tiny{IR limit/strong}}}\bigg|_{m^{2}\ll
k_{3}^{2}\ll eB}\approx W_{ring}^{\mbox{\tiny{IR/nonpert.}}}+
V_{ring}^{\mbox{\tiny{IR/pert.}}}-J,
\end{eqnarray}
where $W_{ring}^{\mbox{\tiny{IR/nonpert.}}}$ is the nonperturbative,
$V_{ring}^{\mbox{\tiny{IR/pert.}}}$ is the perturbative part of
$V_{ring}^{\mbox{\tiny{IR limit/strong}}}$ and $J$ is given in
(\ref{A50}).  The nonperturbative part of the ring potential
$W_{ring}^{\mbox{\tiny{IR/nonpert.}}}$ is given by
\begin{eqnarray}\label{A59}
W_{ring}^{\mbox{\tiny{IR/nonpert.}}}&=&\frac{mT}{8\pi^{2}}\int_{0}^{\infty}d({\mathbf{k}}^{2}_{\perp})
\ln\left(1-\frac{e^{-\frac{{\mathbf{k}}_{\perp}^{2}}{2eB}}{M}_{\gamma}^{2}}{\left({\mathbf{k}}_{\perp}^{2}+m^{2}\right)}\right)=
\frac{mT
eB}{4\pi^{2}}\mbox{Li}_{2}\left(-\frac{M_{\gamma}^{2}}{m^{2}}\right).
\end{eqnarray}
To evaluate this integral we have used (\ref{A52}) with $y
M_{\gamma}^{2}\equiv {\mathbf{k}}_{\perp}^{2}$, $x
M_{\gamma}^{2}\equiv 2eB$ and $zM_{\gamma}^{2}\equiv m^{2}$. As for
the perturbative part of the ring potential,
$V_{ring}^{\mbox{\tiny{IR/pert.}}}$, it is given by the substitution
$z\equiv \frac{{\mathbf{k}}_{\perp}^{2}}{eB}$. It reads
\begin{eqnarray}\label{A60}
V_{ring}^{\mbox{\tiny{IR/pert.}}}\equiv
\frac{TeB\sqrt{eB}}{8\pi}\int\limits_{0}^{\infty}\left(\sqrt{z}-\sqrt{z-\frac{2\alpha}{\pi}e^{-\frac{z}{2}}}\right)\
dz=\frac{\alpha
TeB\sqrt{2eB}}{8\pi^{3/2}}+{\cal{O}}\left(\alpha^{2}\right).
\end{eqnarray}
Here, we have expanded the integrand in the orders of $\alpha$ and
performed eventually the integration over $z$.
\subsubsection*{iii) QED Ring potential in the LLL; the IR limit}
At this stage we are able to give the ring potential in the limit of
strong magnetic field at finite temperature. It is determined by
adding the contribution from the first regime (\ref{A53}) with the
contribution from the second regime (\ref{A58}). It consists of a
perturbative and a nonperturbative part
\begin{eqnarray}\label{A61}
V_{ring}^{\mbox{\tiny{IR
limit/strong}}}=V_{ring}^{\mbox{\tiny{IR/pert.}}}+V_{ring}^{\mbox{\tiny{IR/nonpert.}}}.
\end{eqnarray}
The perturbative part, $V_{ring}^{\mbox{\tiny{IR/pert.}}}$, is given
by (\ref{A60}) and the nonperturbative part,
$V_{ring}^{\mbox{\tiny{IR/nonpert.}}}$, is given by adding up the
ring contribution (\ref{A53}) and (\ref{A59}) from the first and
second regime of LLL, respectively. It is given by
\begin{eqnarray}\label{A62}
V_{ring}^{\mbox{\tiny{IR/nonpert.}}}= -\frac{mT
eB}{4\pi^{2}}\left(\mbox{Li}_{2}\left(-\frac{M_{\gamma}^{2}}{m^{2}}\left(1-\frac{5}{3}C_{1}(m\beta,\pi)\right)\right)-
\mbox{Li}_{2}\left(-\frac{M_{\gamma}^{2}}{m^{2}}\right)\right).
\end{eqnarray}
Here, we have  replaced $A_{2}^{T}-m^{2}A_{3}$ in (\ref{A53}) by its
value from (\ref{A54}). It is interesting to examine the behavior of
the ring potential in the high temperature limit. To do this we use
the asymptotic expansion of the dilogarithm
\begin{eqnarray}\label{A63}
\mbox{Li}_{2}\left(-x\right)\stackrel{x\to\infty}{\longrightarrow}-\frac{\pi^{2}}{6}-
\frac{1}{2}\left(\ln\left(x\right)\right)^{2}+\frac{1}{x}+{\cal{O}}\left(\frac{1}{x^2}\right),
\end{eqnarray}
and expand (\ref{A62}) in the orders of $x\equiv\frac{eB}{m^{2}}$
and then in the orders of $t\equiv m\beta$ to get
\begin{eqnarray}\label{A64}
V_{ring}^{\mbox{\tiny{IR/nonpert.}}}\bigg|_{\frac{eB}{m^{2}}\to\infty,
m\beta\to 0}\approx
-\frac{35m^{4}\zeta(3)}{192\pi^{3}\alpha}\bigg[1+\frac{2\alpha}{\pi}\frac{eB
}{m^{2}}\ln\left(\frac{2\alpha}{\pi}\frac{eB}{m^{2}}\right)\bigg](m\beta)
+{\cal{O}}\left(\frac{m^{4}}{(eB)^2},
\left(m\beta\right)^{2}\right),
\end{eqnarray}
where
$C_{1}\left(t,\pi\right)=\frac{7t^2\zeta(3)}{8\pi^{2}}+{\cal{O}}(t^4)$
is also used. Together with the perturbative contribution to the
effective potential, (\ref{A60}), the most dominant part of the ring
potential in the limit $\frac{eB}{m^{2}}\to \infty$ is given by
\begin{eqnarray}\label{A65}
V_{ring}^{\mbox{\tiny{IR limit/strong}}}\approx
-\frac{35m^{4}\zeta(3)}{192\pi^{3}\alpha}\bigg[1+\frac{2\alpha}{\pi}\frac{eB
}{m^{2}}\ln\left(\frac{2\alpha}{\pi}\frac{eB}{m^{2}}\right)\bigg](m\beta)+\frac{\alpha
TeB\sqrt{2eB}}{8\pi^{3/2}}+{\cal{O}}\left(\frac{m^{4}}{(eB)^2},\alpha^{2},(m\beta)^{2}\right).\nonumber\\
\end{eqnarray}
The same result will arise when we keep $\frac{eB}{m^{2}}$ in
(\ref{A62}) fixed and after replacing $m\to \frac{t}{\beta}$ expand
the resulting expression in the orders of $t$. This means that the
two limits $eB\to \infty, m\beta\to 0$ and $m\beta\to 0$ and
$eB\to\infty$ yield the same result.
\subsection{Ring potential of QED in the LLL
in the static limit; A comparison with the IR limit}
\par\noindent
Let us now compare the above results in the IR limit ($k_{0}\to 0$)
with the ring potential in the static limit ($k_{0}\to 0$ and
${\mathbf{k}}\to {\mathbf{0}}$). In this case, the ring potential
(\ref{A43}) is only determined by
$\kappa_{2}(0,{\mathbf{0}})=I_{2}^{T}(0,{\mathbf{0}})$
\begin{eqnarray}\label{A66}
V_{ring}^{\mbox{\tiny{static
limit}}}=-\frac{T}{2}\int\frac{d^{3}k}{(2\pi)^{3}}
\ln\left(1-\frac{\kappa_{2}(0,{\mathbf{0}})}{k_{3}^{2}+{\mathbf{k}}_{\perp}^{2}}\right)=
-\frac{T}{8\pi^{2}}\int\limits_{0}^{\infty}d({\mathbf{k}}_{\perp}^{2})\int\limits_{0}^{\infty}dk_{3}
\ln\left(1-\frac{I_{2}^{T}(0,{\mathbf{0}})}{k_{3}^{2}+{\mathbf{k}}_{\perp}^{2}}\right).
\end{eqnarray}
As we have seen in the previous paragraphs, in the limit of strong
magnetic field the integration over $k_{3}$ must be separated into a
regime where $k_{3}^{2}\ll m^{2}\ll eB$ and a regime with $m^{2}\ll
k_{3}^2\ll eB$. As for $I_{2}^{T}$ in the first regime $k_{3}^{2}\ll
m^{2}\ll eB$, it is given in (\ref{A46}) with $A_{2}^{T}$ from
(\ref{A49})
\begin{eqnarray}\label{A67}
I_{2}^{T}(0,{\mathbf{0}})\bigg|_{k_{3}^{2}\ll m^{2}\ll
eB}&=&-\frac{\alpha
eB}{2\pi}\int\limits_{-1}^{1}dv\sum\limits_{\ell=1}^{\infty}(-1)^{\ell}\left(\ell\beta\right)^{2}\int\limits_{0}^{\infty}
\frac{du}{u^{2}}e^{-um^{2}-\frac{\ell^{2}}{4uT^2}}\nonumber\\
&=& A_{2}^{T}=M_{\gamma}^{2}\left(1-2C_{1}(m\beta,\pi)\right).
\end{eqnarray}
Note that the only difference between the static and the previous IR
limit is a factor $e^{-\frac{{\mathbf{k}}_{\perp}^{2}}{2eB}}$ that
appears in $I_{2}^{T}(0,{\mathbf{k}}\neq {\mathbf{0}})$ in
(\ref{A46}). This factor vanishes in the static limit where we take
the limit $k_{0}\to 0$ \textit{and} ${\mathbf{k}}\to {\mathbf{0}}$.
In the second regime $m^{2}\ll k_{3}^2\ll eB$,
$I_{2}^{T}(0,{\mathbf{k}}\neq {\mathbf{0}})$ is given in (\ref{A55})
with $B_{2}^{T}$ from (\ref{A56}). In the static limit for
${\mathbf{k}}\to {\mathbf{0}}$, it is, however,  given by
\begin{eqnarray}\label{A68}
I_{2}^{T}(0,{\mathbf{0}})\bigg|_{ m^{2}\ll k_{3}^{2}\ll
eB}=-\lim\limits_{\varepsilon\to 0}\frac{\alpha
eB}{2\pi}\int\limits_{-1}^{1}dv\sum\limits_{\ell=1}^{\infty}(-1)^{\ell}\left(\ell\beta\right)^{2}\int\limits_{0}^{\infty}
\frac{du}{u^{2}}e^{-u\varepsilon-\frac{\ell^{2}}{4uT^2}},
\end{eqnarray}
where $\varepsilon$ is an IR cutoff. Using now the definition of the
Bessel function (\ref{A8}), we get
\begin{eqnarray}\label{A69}
I_{2}^{T}(0,{\mathbf{0}})\bigg|_{ m^{2}\ll k_{3}^{2}\ll
eB}=-\lim\limits_{\varepsilon\to
0}2M_{\gamma}^{2}\sum\limits_{\ell=1}^{\infty}(-1)^{\ell} (\ell
\sqrt{\varepsilon}\beta)K_{1}\left(\ell
\sqrt{\varepsilon}\beta\right)= M_{\gamma}^{2}.
\end{eqnarray}
Here, we have used (\ref{A33}) and the fact that $C_{1}(
\sqrt{\varepsilon},\pi)$ from (\ref{A34}) vanishes for
$\varepsilon\to 0$. Plugging now (\ref{A67}) and (\ref{A69}) in
(\ref{A66}), the ring potential in the static limit and in the
presence of strong magnetic field is given by
\begin{eqnarray}\label{A70}
\lefteqn{V_{ring}^{\mbox{\tiny{static limit/strong}}}=}\nonumber\\
&=&-\frac{T}{8\pi^{2}}\int\limits_{0}^{\infty}d({\mathbf{k}}_{\perp}^{2})\bigg[\int\limits_{0}^{m}dk_{3}
\ln\left(1-\frac{I_{2}^{T}(0,{\mathbf{0}})\big|_{k_{3}^{2}\ll
m^{2}\ll
eB}}{k_{3}^{2}+{\mathbf{k}}_{\perp}^{2}}\right)+\int\limits_{m}^{\infty}dk_{3}
\ln\left(1-\frac{I_{2}^{T}(0,{\mathbf{0}})\big|_{ m^{2}\ll
k_{3}^{2}\ll
eB}}{k_{3}^{2}+{\mathbf{k}}_{\perp}^{2}}\right)\bigg]\nonumber\\
&=&-\frac{T}{8\pi^{2}}\int\limits_{0}^{\infty}d({\mathbf{k}}_{\perp}^{2})
\bigg[\int\limits_{0}^{m}dk_{3}\ln\left(1-\frac{M_{\gamma}^{2}\left(1-2C_{1}(m\beta,\pi)\right)}{k_{3}^{2}+{\mathbf{k}}_{\perp}^{2}}\right)+
\int\limits_{m}^{\infty}dk_{3}\ln\left(1-\frac{M_{\gamma}^{2}}{k_{3}^{2}+{\mathbf{k}}_{\perp}^{2}}\right)
 \bigg].
\end{eqnarray}
Following now the same steps as in the previous paragraph, the ring
potential in the static limit can be decomposed into a perturbative
and a nonperturbative part
\begin{eqnarray}\label{A71}
V_{ring}^{\mbox{\tiny{static
limit/strong}}}=V_{ring}^{\mbox{\tiny{static/pert.}}}+V_{ring}^{\mbox{\tiny{static/nonpert.}}}+\mbox{cutoff
dependent terms}.
\end{eqnarray}
The perturbative part is given, as in the previous case, by the
substitution $z=\frac{{\mathbf{k}}_{\perp}^{2}}{eB}$. It reads
\begin{eqnarray}\label{A72}
V_{ring}^{\mbox{\tiny{static/pert.}}}=\frac{TeB\sqrt{eB}}{8\pi}\int\limits_{0}^{\Lambda}dz
\left(\sqrt{z}-\sqrt{z-\frac{2\alpha}{\pi}}\right)= \frac{\alpha
TeB\sqrt{eB}}{4\pi}\sqrt{\Lambda}+{\cal{O}}\left(\alpha^{2}\right).
\end{eqnarray}
Here, we have first expanded the integrand in the orders of $\alpha$
and then performed the integration over $z$. Comparing to the
perturbative part in the IR limit (\ref{A60}),
$V_{ring}^{\mbox{\tiny{static/pert.}}}$ diverges for $\Lambda\to
\infty$. This is due to the lack of a factor $e^{-\frac{z}{2}}$ in
the second term of the integrand. This factor arises only in the IR
limit where ${\mathbf{k}}\neq {\mathbf{0}}$ and damps the integral.
The perturbative part (\ref{A60}) in the IR approximation remains
therefore convergent and yields a finite contribution to the
perturbative loop potential.
\par
As for the nonperturbative part of
$V_{ring}^{\mbox{\tiny{static/strong}}}$ (\ref{A71}), it is given by
\begin{eqnarray}\label{A73}
V_{ring}^{\mbox{\tiny{static/nonpert.}}}=\frac{mT}{8\pi^{2}}\bigg[(m^{2}-M_{\gamma}^{2})\ln\left(1+\frac{2M_{\gamma}^{2}C_{1}}{m^{2}-M_{\gamma}^{2}}\right)-2C_{1}M_{\gamma}^{2}\bigg]+\mbox{cutoff
dependent terms}.
\end{eqnarray}
In the high temperature expansion $m\beta\to 0$,  the most dominant
term of the potential (\ref{A73}) for $\frac{eB}{m^{2}}\to \infty$
is given by
\begin{eqnarray}\label{A74}
V_{ring}^{\mbox{\tiny{static/nonpert.}}}\bigg|_{ m\beta\to 0,
\frac{eB}{m^{2}}\to\infty}=\frac{49m^4
(\zeta(3))^{2}}{256\pi^{6}}\left(1+\frac{2\alpha}{\pi}\frac{eB}{m^{2}}\right)(m\beta)^{3}
+{\cal{O}}\left(\frac{m^{4}}{(eB)^{2}},(m\beta)^4\right).
\end{eqnarray}
The same result will arise if we expand (\ref{A73}) first in the
orders of $\frac{eB}{m^{2}}\to \infty$ for fixed $m\beta$ and then
take the limit $m\beta\to 0$. Together with the perturbative
contribution to the effective potential, (\ref{A72}), the most
dominant part of the ring potential in the limit
$\frac{eB}{m^{2}}\to \infty$ is given by
\begin{eqnarray}\label{A75}
V_{ring}^{\mbox{\tiny{static limit/strong}}}\approx \frac{49m^4
(\zeta(3))^{2}}{256\pi^{6}}\left(1+\frac{2\alpha}{\pi}\frac{eB}{m^{2}}\right)(m\beta)^{3}+\mbox{cutoff
dependent terms}.
\end{eqnarray}
This result can be compared with (\ref{A65}) where, in the order
$m\beta$, a novel term proportional to
$\frac{2}{\pi}\frac{eB}{m^{2}}\ln(\frac{2\alpha}{\pi}\frac{eB}{m^{2}})$
appears.
\subsection{A second possibility to determine the ring
potential in the static limit; The strong limit}
\par\noindent
In III.C, we have determined the ring potential in the static limit
using $I_{2}$ from (\ref{A40}), where we have first taken the limit
$eB\to\infty$. This means that a transition to the LLL is occurred
before the ring potential is calculated. Once we are in the LLL, it
is necessary to distinguish between two different dynamical regimes
in the LLL: $k_{3}^{2}\ll m^{2}\ll eB$ and $m^{2}\ll k_{3}^{2}\ll
eB$, and separate consequently the $k_{3}$ integration interval
$k_{3}\in [0,\infty[$ into two parts, $k_{3}\in [0,m]$ and $k_{3}\in
[m,\infty]$. As we have seen in (\ref{A70}) the integrands are also
different in these two dynamical regimes. Mathematically, in the
LLL,  the three-dimensional integration over ${\mathbf{k}}=(k_{1},
k_{2},k_{3})$ in (\ref{A66}) and (\ref{A70}) are separated into two
integrals over $k_{3}$ and a symmetric integral over $k_{1}$-$k_{2}$
plane, which is perpendicular to the direction of external magnetic
field.\footnote{In other words, the integrations over
${\mathbf{k}}=(k_{1},k_{2},k_{3})$ in (\ref{A66}) and (\ref{A70})
are to be performed over a cylinder with the basis in
$k_{1}$-$k_{2}$ plane and the height in the $k_{3}$ direction.}
Physically, this can be viewed as a consequence of the dimensional
reduction which is one of the well-known properties of the LLL
dynamics \cite{miransky-2}.
\par
In this section, we will point out that a second approach is also
possible to determine the static ring potential in the limit $eB\to
\infty$. In this approach, one starts directly from the ring
potential (\ref{A13}) and take the limit $eB\to\infty$
afterwards.\footnote{Note that mathematically in (\ref{A13}), there
is no difference between $k_{3}$ and
${\mathbf{k}}_{\perp}=(k_{1},k_{2})$ integration. Both the integrand
and the integration interval are spherical symmetric over a
three-sphere, once we introduce a sharp momentum cutoff $\Lambda$ to
calculate the integral.} Doing this, one arrives at
\begin{eqnarray}\label{A76}
V_{ring}^{\mbox{\tiny{strong limit}}}=
\frac{T}{12\pi}\left(\Pi_{00}\left(0,{\mathbf{0}}\right)\right)^{3/2}
&=&\frac{T}{12\pi}\left(I_{2}^{T}\left(0,{\mathbf{0}}\right)\right)^{3/2}\nonumber\\
&\stackrel{eB\to
\infty}{\longrightarrow}&\frac{T}{12\pi}\bigg[M_{\gamma}^{2}\left(1-2C_{1}\left(m\beta,\pi\right)\right)\bigg]^{3/2},
\end{eqnarray}
where
$\Pi_{00}\left(0,{\mathbf{0}}\right)=-\Pi_{44}\left(0,{\mathbf{0}}\right)=I_{2}^{T}(0,{\mathbf{0}})$
from (\ref{Z31}) and
\begin{eqnarray*}
I_{2}^{T}\left(0,{\mathbf{0}}\right)&=&A_{2}^{T}=-2M_{\gamma}^{2}\sum\limits_{\ell=1}^{\infty}(-1)^{\ell}
\left(\ell m\beta\right)K_{1}\left(\ell m\beta\right)\nonumber\\
&=&M_{\gamma}^{2}\left(1-2C_{1}\left(m\beta,\pi\right)\right),
\end{eqnarray*}
from (\ref{A47}) are used. In the next section, the ring potential
(\ref{A62}), (\ref{A73}) and (\ref{A76}) will be used to study the
dynamical chiral symmetry breaking of QED in the LLL.
\section{Dynamical chiral symmetry breaking of QED in the LLL}
\setcounter{equation}{0}
\par\noindent For a given one-loop effective
potential $V^{(1)}$, the gap equation is given by [see Appendix B
for the derivation of the gap equation at $T=0$ and Appendix C for a
generalization to $T\neq 0$ case]
\begin{eqnarray}\label{S1}
\frac{\partial V^{(1)}\left(m,eB;T\right)}{\partial\langle
\bar{\psi}\psi\rangle}=Gm,
\end{eqnarray}
where $G$ is an appropriate coupling and $m$ is the dynamical
mass.\footnote{In this section, we will omit the subscript $dyn$ in
$m_{dyn.}$ from Appendices A-C.} Using the identity $\langle
\bar{\psi}\psi\rangle\equiv\frac{\partial V^{(1)}}{\partial m}$, the
above equation is given by
\begin{eqnarray}\label{S2}
\frac{\partial V^{(1)}\left(m,eB;T\right)}{\partial
m}=Gm\frac{\partial^{2}V^{(1)}(m,eB;T)}{\partial m^{2}}.
\end{eqnarray}
In this section, we will use (\ref{S2}) to determine the dynamical
mass and critical temperature of the dynamical chiral symmetry
breaking of QED at finite temperature and in the presence of
\textit{strong} magnetic field. To determine the gap equation of
this theory, the ring improved effective potential including the
one-loop and the ring contributions will be considered. To fix our
notations and at the same time to check our procedure, we will first
determine in Sect. IV.A, the dynamical mass $m(T)$ and the critical
temperature $T_{c}$ arising from (\ref{A39}), the one-loop effective
potential of the theory.\footnote{Note that (\ref{A39}) is
determined using the worldline formalism \cite{worldline}, which is
supposed to lead to the same one-loop effective potential arising
from the well-known Schwinger proper-time formalism \cite{schwinger,
miransky1-5, miransky-2}. For the exact definition of the one-loop
effective potential in the LLL approximation see Appendix A.} In the
limit of strong magnetic field $eB\gg m^{2}$ and in the high
temperature limit $m\beta\to 0$, our results indeed coincide with
the results from \cite{shovkovy,ebert}.\footnote{In \cite{ebert} the
dynamical mass and the critical temperature of an effective
Nambu-Jona Lasinio (NJL) model at finite temperature are determined
using the one-loop effective potential of the theory in the presence
of an external chromomagnetic field. In \cite{shovkovy}, the same
quantities are determined by solving the SD equation of QED in the
presence of a strong magnetic field using a ladder approximation.
The common result in these two papers is the well-known relation
$T^{(1)}_{c}= {\cal{C}}m^{(1)}(T=0)$, where $T_{c}^{(1)}$ is the
critical temperature arising from one-loop effective
potential/solution of SD equation in the ladder approximation and
$m^{(1)}(T=0)$ is the corresponding dynamical mass at zero
temperature, and ${\cal{C}}$ is a numerical factor. This relation
seems to be model independent. In this section, using the one-loop
effective potential of QED in the presence of strong magnetic field,
(\ref{A39}), and following the procedure described in Sect. IV.A, we
arrive at the same result $T^{(1)}_{c}\approx m^{(1)}(T=0)$ [see
(\ref{SS11})].}
\par
In the rest of this section, we will examine the possible effects of
the ring contribution to the ring improved (one-loop) effective
potential on the dynamical mass and critical temperature of QED at
finite temperature and in the LLL. To this purpose, we will use the
ring potentials in the IR, static and strong magnetic field limits
that are calculated in the previous sections. As it is mentioned
before, the ring potential arises only in the finite temperature
field theory and reflects the infrared behavior of the theory at
$T\neq 0$ [see Sect. I.A]. Its contribution to the one-loop
effective potential has various effects. In \cite{takahashi}, for
instance, the phase transition of a simple scalar field theory is
considered using the \textit{ring improved} effective potential,
including the one-loop \textit{and} the ring contributions. It is
shown that the addition of the ring to the one-loop effective
potential has indeed two effects: First, in the ring contributions
there are terms that cancel certain imaginary terms arising from the
one-loop effective potential, and second, the order of phase
transition changes from second to first order. The same effect
happens also in \cite{carrington}, where in particular, it is shown
that after adding the ring contribution to the one-loop effective
potential of the electroweak Standard Model (SM), the critical
temperature of electroweak symmetry breaking decreases from its
value arising from the one-loop effective potential of the theory.
It is the purpose of this section to show that the critical
temperature of the dynamical chiral symmetry breaking of QED at
finite temperature and in the LLL approximation is indeed affected
by the contribution from the ring potential. To show this, we will
consider in Sect. IV.B the ring potential in the improved IR limit,
(\ref{A62}), and calculate the dynamical mass and the critical
temperature in the strong magnetic field limit $eB\gg m^{2}$ and
high temperature limit $m\beta\to 0$. Then, in Sect. IV.C and IV.D,
the ring potential in the static limit, (\ref{A73}), and the strong
limit, (\ref{A76}), will be considered separately and the
corresponding dynamical mass and critical temperature will be
determined. Whereas adding the ring potential in the improved IR and
strong limits to the one-loop effective potential decreases the
critical temperature arising from one-loop effective potential, the
ring potential in the static limit does not change $T_{c}^{(1)}$. In
Sect. IV.E, we will finally consider the ratio
$T_{c}^{(1)}/{\cal{T}}_{c}$, where ${\cal{T}}_{c}$ is the ring
improved critical temperature arising from the \textit{ring
improved} effective potential including the one-loop and the ring
potential. Further, we will define an efficiency factor
$\eta=1-u^{-1}$. In this way we will be able to compare numerically
the effect of the ring potential in the IR limit with effect of the
ring potential in the strong limit in changing $T_{c}^{(1)}$. As it
turns out, compared to the strong limit, the IR limit is more
efficient in decreasing the critical temperature arising from
one-loop effective potential. This is indeed a promising result in
view of the baryogenesis problem in the electroweak SM, when the
ring potential in the improved IR limit is considered to calculate
the critical temperature of the electroweak SM in the presence of
strong magnetic field.\footnote{In \cite{bordag, demchik} the
standard electroweak symmetry breaking is considered in the presence
of a strong hypermagnetic field. To determine the ring potential in
the strong magnetic field limit, first the ring integral is
calculated and then $eB\to \infty$ limit is taken, as in
(\ref{A76}). Our calculation shows that the results in \cite{bordag,
demchik} may be improved, if in place of the strong limit, the
improved IR limit is used \cite{sadooghi}.}
\subsection{Dynamical mass and critical temperature arising from QED
one-loop effective potential} Let us start from QED one-loop
effective potential $V^{(1)/\mbox{\tiny{strong}}}(m,eB;\Lambda,T)$
in the limit of strong magnetic field (\ref{A39}). It consists of a
temperature dependent and a temperature independent part. Whereas
the temperature dependent part is renormalization free, the
temperature independent part depends explicitly on a sharp momentum
cutoff $\Lambda$.\footnote{The temperature independent part of the
one-loop effective potential is also called ''the renormalized
effective potential'' \cite{sato}.}
 In the limit of strong magnetic field, where QED dynamics
is described by an effective field theory in the LLL, the momentum
cutoff $\Lambda$ can be replaced by $\Lambda_{B}\equiv \sqrt{eB}$.
In this section, using (\ref{A39}) with $\Lambda\to \Lambda_{B}$, we
will determine the dynamical mass and the critical temperature of
the dynamical chiral phase transition arising from QED one-loop
effective potential. The results will then be compared with the
corresponding results in \cite{shovkovy} and \cite{ebert}. We will
show that in the limit of strong magnetic field, $m^{2}\ll eB$, and
in the high temperature limit, $m\beta\to 0$, our results coincides
with the results from \cite{shovkovy,ebert}.
\par
To start, let us replace the potential $V$ in (\ref{S2}) by
(\ref{A39}) which is given by
\begin{eqnarray*}
V^{(1)/\mbox{\tiny{strong}}}(m,eB;T)=-\frac{eB}{8\pi^{2}}\bigg\{m^{2}\Gamma\left(-1,\frac{m^{2}}{\Lambda_{B}^{2}}\right)+\frac{8m}{\beta}
\sum\limits_{n=1}^{\infty}\frac{(-1)^{n}}{n}K_{-1}\left(nm\beta\right)\bigg\}.
\end{eqnarray*}
We arrive first at
\begin{eqnarray}\label{SS3}
\lefteqn{\hspace{-1cm} \frac{\partial
V^{(1)/\mbox{\tiny{strong}}}}{\partial m}-Gm\frac{\partial^{2}
V^{(1)/\mbox{\tiny{strong}}}}{\partial m^{2}}=\frac{eBm}{4\pi^{2}}
\bigg[(1-G)\Gamma\left(0,\frac{m^{2}}{\Lambda_{B}^{2}}\right)+2G
\Gamma\left(1,\frac{m^{2}}{\Lambda_{B}^{2}}\right)
}\nonumber\\
&& +4(1-G)\sum\limits_{n=0}^{\infty}(-1)^{n}K_{0}(nm\beta) +4G
\sum\limits_{n=1}^{\infty}(-1)^n\left(nm\beta\right)K_{1}\left(nm\beta\right)\bigg]=0.
\end{eqnarray}
The above equation will be simplified as follows. First, we take the
high temperature limit  $m\beta\to 0$ in the temperature dependent
part of (\ref{SS3}) which is given in the second line of the above
expression. Using (\ref{A31}) and (\ref{A33}), it turns out that the
term proportional to $(nm\beta)K_{1}(nm\beta)$ behaves as $\approx
{\cal{O}}\left(1\right)$, whereas the term proportional to
$K_{0}(nm\beta)$ is proportional to $\sim \ln(m\beta)$. Keeping only
the logarithmic divergent terms in in the limit $m\beta\to 0$, the
term proportional to $(nm\beta) K_{1}(nm\beta)$ can therefore be
neglected in this limit. Next, in the temperature independent part
of (\ref{SS3}) which is given in the first line of the above
expression, we take the strong magnetic field limit, $z\equiv
\frac{m^{2}}{eB}\ll 1$. Using the relation
\begin{eqnarray*}
\Gamma(0,z)\stackrel{z\to 0}{\longrightarrow} -\gamma-\ln
z,\qquad\mbox{and}\qquad \Gamma(1,z)\stackrel{z\to
0}{\longrightarrow} 1,
\end{eqnarray*}
where $\gamma$ is the Euler-Mascheroni constant, we arrive therefore
at the gap equation
\begin{eqnarray}\label{SS4}
\ln\frac{m}{\Lambda_{B}}=-\frac{\gamma}{2}+\frac{G}{1-G}+2\sum\limits_{n=1}^{\infty}(-1)^{n}K_{0}(nm\beta),
\end{eqnarray}
where $m\equiv m(G;T)$ is the temperature dependent dynamical mass,
which is given by
\begin{eqnarray}\label{SS5}
m(G;T)=\Lambda_{B}\exp\left(-\frac{\gamma}{2}+\frac{G}{1-G}+2\sum\limits_{n=1}^{\infty}(-1)^{n}K_{0}(nm\beta)\right).
\end{eqnarray}
At $T=0$ we have
\begin{eqnarray}\label{SS6}
m(G_{0};T=0)=\Lambda_{B}\exp\left(-\frac{\gamma}{2}+\frac{G_{0}}{1-G_{0}}\right).
\end{eqnarray}
This is a general structure for the dynamical mass as a function of
the effective coupling $G_{0}$ at zero temperature (see Appendices B
and C for an exact definition of $G_{0}$ at zero temperature as well
as $G$ at finite temperature). As it is shown in (\ref{AB28}), in
the lowest order of $\alpha$ correction, $G^{(1)}_{0}$ receives
contribution from diagrams shown in Fig. 3 (Appendix A) of order
$\alpha$ (only the $T=0$ part of these diagrams are relevant for
$G_{0}^{(1)}$). In higher orders of $\alpha$ expansion $G_{0}$
receives contribution from the temperature independents parts of all
diagrams shown in Fig. 1 and (\ref{X6}) (ring diagrams) at zero
temperature.
\par
As we have mentioned in Sect. II.A, the dynamically generated
fermion mass in the lowest order of $\alpha$ correction at zero
temperature is calculated in \cite{miransky-2} and \cite{shovkovy}
in the ladder LLL approximation.\footnote{Note that in
\cite{shovkovy} apart from the ladder approximation, a constant mass
approximation (CMA) is also used. In this approximation, one
neglects the fermion mass structure in the solution of the
corresponding SD equation \cite{cma}. In other words momentum
dependence of self-energy in the gap equation are neglected in this
approximation.  As it is shown in \cite{ferrer-ward} this turns out
to be a reliable approximation in QED (with only one coupling
constant) in the presence of a strong magnetic field limit, although
there is no general principle that guarantees the validity of this
approximation for the whole range of physical coupling. In other
theories with more than one coupling constant, due to the richness
of parameter space, the reliability of CMA is questionable and
should be investigated in detail (see Elizalde \textit{et al.} in
\cite{cma}).} It is given by (\ref{Z16}). Comparing to the exact
result, $m(G_{0},T=0)$ from (\ref{SS6}), this first correction to
the dynamical mass can be indicated as: $m_{dyn.}^{(1)}\equiv
m^{(1)}(G_{0}^{(1)};T=0)$. As it is shown in Appendix B, the result
from (\ref{SS6}) is indeed comparable with the dynamical mass
$m^{(1)}(G_{0}^{(1)};T=0)$ (\ref{Z16}), provided $G^{(1)}_{0}$ in
this lowest order of $\alpha$ correction is fixed as in (\ref{AB36})
\textit{i.e.} by $G_{0}^{(1)}=
1/(1-\sqrt{\frac{\alpha}{\pi}})\approx 1+\sqrt{\frac{\alpha}{\pi}}$.
Plugging this result in (\ref{SS6}), we get
\begin{eqnarray}\label{SS7}
m^{(1)}(G_{0}^{(1)};T=0)={\cal{C}}_{m}\Lambda_{B}\exp\left(-\sqrt{\frac{\pi}{\alpha}}\right),\qquad
{\cal{C}}_{m}=e^{-\gamma/2}=0.749306.
\end{eqnarray}
As next, let us determine the critical temperature of chiral
symmetry breaking of QED in the LLL approximation. To do this we use
the gap equation (\ref{SS4}), and  replace [see (\ref{A31})]
\begin{eqnarray*}
\sum\limits_{n=1}^{\infty}(-1)^{n}K_{0}(nm\beta)\to
\frac{1}{2}\left(\gamma+\ln\frac{m\beta}{4\pi}\right)+C_{0}(m\beta,\pi).
\end{eqnarray*}
The critical temperature arising from one-loop effective potential
(\ref{A39}), $T_{c}(G;T)$, can now be determined from the condition
$m(T_{c})=0$. It is given as a function of $G$ by [see Appendix C
for an exact definition of $G$]
\begin{eqnarray}\label{SS8}
T_{c}(G;T)=\Lambda_{B}\exp\left(\frac{\gamma}{2}+\frac{G}{1-G}+2\ln
2\right).
\end{eqnarray}
Here, we have used the expansion of $C_{0}(t,\pi)$
\begin{eqnarray}\label{SS9}
C_{0}(t,\pi)=\ln
2-\frac{7t^{2}\zeta(3)}{16\pi^{2}}+{\cal{O}}\left(t^{4}\right),
\end{eqnarray}
at $t=0$.  The critical temperature as it is given in (\ref{SS8})
includes all orders of $\alpha$ correction through definition of $G$
from (\ref{D6}). All diagrams contributing to $G$ are to be
considered at finite temperature. In the lowest order of $\alpha$
correction, $G^{(1)}$ receives contributions from diagrams shown in
Fig. 3 of order $\alpha$. This fact allows us to compare (\ref{SS8})
in the lowest order of $\alpha$ correction, \textit{i.e.}
$T_{c}^{(1)}$, with the critical temperature arising from the SD
equation including the contributions from the same two diagrams in
Fig. 3, that are relevant in determining $m^{(1)}(G_{0}^{(1)};T=0)$
\cite{shovkovy}. Doing this $G^{(1)}$ is again fixed to be
$G^{(1)}=G_{0}^{(1)}= 1/(1-\sqrt{\frac{\alpha}{\pi}})\approx
1+\sqrt{\frac{\alpha}{\pi}}$. Plugging this expression in
(\ref{SS8}), we arrive at
\begin{eqnarray}\label{SS10}
T_{c}^{(1)}(G^{(1)};T)={\cal{C}}_{T}\Lambda_{B}\exp\left(-\sqrt{\frac{\pi}{\alpha}}\right),
\qquad {\cal{C}}_{T}=\pi^{-1}e^{\gamma/2}= 0.156277,
\end{eqnarray}
which is comparable with the result from \cite{shovkovy}. Comparing
to $m^{(1)}(0)$ from (\ref{SS7}), we arrive at the well-known
relation
\begin{eqnarray}\label{SS11}
T_{c}^{(1)}=\pi^{-1}e^{\gamma}m^{(1)}(0)=0.424806\ m^{(1)}(0).
\end{eqnarray}
The same relation arises in \cite{ebert} between the critical
temperature $T_{c}^{(1)}$ and the dynamical mass $m^{(1)}(T=0)$ in
the effective NJL model in the presence of constant chromomagnetic
field. Note that in higher orders of $\alpha$ correction, the
critical temperature $T_{c}(G;T)$ receive contribution from higher
loop diagrams through the definition of $G$ from (\ref{D6}).
\par
As next, we will determine the gap equation, the dynamical mass and
the critical temperature for the \textit{ring improved} effective
potential including the one-loop effective potential (\ref{A39})
\textit{and} the ring potential in the IR, static and strong limits
[see (\ref{A62}), (\ref{A73}) and (\ref{A76}), respectively].
\subsection{Full dynamical mass and critical temperature of QED in the IR limit}
\subsubsection*{Full dynamical mass in the IR limit}
\par\noindent
As in (\ref{S2}), the general structure of the gap equation
corresponding to the \textit{ring improved} effective potential (see
Appendix C for more details), $\tilde{V}\equiv V^{(1)}+V_{ring}$, is
given by
\begin{eqnarray}\label{S12}
\frac{\partial \tilde{V}}{\partial
\tilde{m}}=\tilde{G}\tilde{m}\frac{\partial^{2}\tilde{V}}{\partial
\tilde{m}^{2}}.
\end{eqnarray}
In $\tilde{V}$, $V^{(1)}$ and $V_{ring}$ denote the one-loop
effective potential and the ring contribution to the effective
potential, respectively. Here, comparing to the effective coupling
$G$ in (\ref{S1}), the modified coupling $\tilde{G}$ is given
by\footnote{See Appendix C.3 for an exact definition of
$\tilde{G}$.}
\begin{eqnarray*}
G=\tilde{G}-\frac{1}{\tilde{m}}\frac{\partial
V_{ring}(\tilde{m},eB;T)}{\partial\langle\bar{\psi}\psi\rangle}.
\end{eqnarray*}
Using the gap equation arising from ring improved effective
potential including the one-loop effective potential and the ring
potential and following the same procedure as was described in Sect.
IV.A, the \textit{full} dynamical mass $\tilde{m}(\tilde{G},T)$
reads
\begin{eqnarray}\label{S13}
\tilde{m}(\tilde{G};T)=m(\tilde{G};T)\exp\left(+\frac{2\pi^{2}}{eB
\tilde{m}(1-\tilde{G})}\bigg[\frac{\partial V_{ring}}{\partial
\tilde{m}}-\tilde{G}\tilde{m}\frac{\partial^{2}V_{ring}}{\partial
\tilde{m}^{2}}\bigg]\right).
\end{eqnarray}
On the r.h.s. of this expression $m(G;T)$ is given by the dynamical
mass arising from one-loop effective potential (\ref{SS5}) with $G$
replaced by $\tilde{G}$. Here, to determine the full dynamical mass
in the IR limit, $\tilde{m}^{\mbox{\tiny{IR}}}$,  we will replace
$V_{ring}$ on the r.h.s. of (\ref{S13}) by the ring contribution of
the LLL ring potential in the IR limit (\ref{A62}),\footnote{ The
perturbative part of the ring potential in the IR limit from
(\ref{A60}) is mass independent and has therefore no contribution to
the gap equation, the dynamical mass and the critical temperature.
We will therefore omit this mass independent contribution in this
section.} and arrive first at
\begin{eqnarray}\label{SS14}
\lefteqn{\frac{2\pi^{2}}{eB
\tilde{m}^{\mbox{\tiny{}}}(1-\tilde{G}^{\mbox{\tiny{}}})}\bigg[\frac{\partial
V_{ring}^{\mbox{\tiny{IR/nonpert.}}}}{\partial
\tilde{m}^{\mbox{\tiny{}}}}-\tilde{G}^{\mbox{\tiny{}}}\tilde{m}^{\mbox{\tiny{}}}\frac{\partial^{2}V_{ring}^{\mbox{\tiny{IR/nonpert.}}}}{\partial
\tilde{m}^{\mbox{\tiny{}}2}}\bigg]=}\nonumber\\
&=&+\frac{2}{\tilde{m}^{\mbox{\tiny{}}}\beta\left(1-\tilde{G}^{\mbox{\tiny{}}}\right)}
\left\{\frac{\tilde{G}^{\mbox{\tiny{}}}z^{2}}{\left(1+z^{2}\right)}
-\frac{\tilde{G}^{\mbox{\tiny{}}}z^{2}\left(6-10C_1+5
\tilde{m}^{\mbox{\tiny{}}}\beta
C_{1}^{'}\right)^2}{36\left(1-\frac{5}{3}C_{1}\right)
\left(1+z^2\left(1-\frac{5}{3}C_{1}\right)\right)}\right.\nonumber\\
&&\left.+\frac{\left(1-\tilde{G}^{\mbox{\tiny{}}}\right)}{2}\bigg[\ln\left(1+z^{2}\right)-
\ln\left(1+z^2\left(1-\frac{5}{3}C_{1}\right)\right)\bigg]+\frac{1}{4}\bigg[\mbox{Li}_{2}\left(-z^{2}\right)-
\mbox{Li}_{2}\left(-z^2\left(1-\frac{5}{3}C_{1}\right)\right)\bigg]\right.\nonumber\\
&&\left. -\frac{5
\tilde{m}^{\mbox{\tiny{}}}\beta\left(\left(1-2\tilde{G}^{\mbox{\tiny{}}}\right)\left(3-5C_{1}\right)C_{1}^{'}-5\tilde{G}^{\mbox{\tiny{}}}\tilde{m}^{\mbox{\tiny{}}}\beta
C_{1}^{'2}-\tilde{G}^{\mbox{\tiny{}}}\tilde{m}^{\mbox{\tiny{}}}\beta\left(3-5C_{1}\right)C_{1}^{''}\right)}{36\left(1-\frac{5}{3}C_{1}\right)^{2}}\ln\left(1+z^{2}\left(1-\frac{5}{3}C_{1}\right)\right)
\right\}.\nonumber\\
\end{eqnarray}
Here, $\tilde{G}=\tilde{G}^{\mbox{\tiny{IR}}}$ and
$\tilde{m}=\tilde{m}^{\mbox{\tiny{IR}}}$.\footnote{See (\ref{DA}) in
Appendix C for the difference between $\tilde{G}$ and
$\tilde{G}^{\mbox{\tiny{IR}}}$.} Further, $z^2(\tilde{m})\equiv
\frac{M^2_{\gamma}}{\tilde{m}^2}=\frac{2\alpha}{\pi}\frac{eB}{\tilde{m}^{2}}$.
Using (\ref{SS9}),  the expansion of $C_{0}(t,\pi)$ in the orders of
$t\equiv \tilde{m}\beta$, and the relations
\begin{eqnarray}\label{SS15}
C_{1}\left(t,\pi\right)&=&\frac{7t^{2}\zeta(3)}{4\pi^{2}}+{\cal{O}}\left(t^{3}\right),\nonumber\\
C_{1}^{'}\left(t,\pi\right)&=&\frac{7t\zeta(3)}{4\pi^{2}}+{\cal{O}}\left(t^{3}\right),\nonumber\\
C_{1}^{''}\left(t,\pi\right)&=&\frac{7\zeta(3)}{4\pi^{2}}-
\frac{279t^{2}\zeta(5)}{16\pi^{4}}+{\cal{O}}\left(t^{3}\right),
\end{eqnarray}
the full dynamical mass in the IR limit is given by
\begin{eqnarray}\label{SS16}
\tilde{m}^{\mbox{\tiny{IR}}}(\tilde{G}^{\mbox{\tiny{IR}}};T)\approx
m(\tilde{G}^{\mbox{\tiny{IR}}};T)\left(1+(\tilde{m}^{\mbox{\tiny{IR}}}\beta){\cal{E}}^{\mbox{\tiny{IR}}}\right),
\end{eqnarray}
where
\begin{eqnarray}\label{SS17}
{\cal{E}}^{\mbox{\tiny{IR}}}\equiv
+\frac{35\zeta(3)}{24\pi^{2}}\frac{\left(1-5\tilde{G}^{\mbox{\tiny{IR}}}\right)}{\left(1-\tilde{G}^{\mbox{\tiny{IR}}}\right)}
\left(1-\frac{3}{2}\frac{\left(1-2\tilde{G}^{\mbox{\tiny{IR}}}\right)}{\left(1-5\tilde{G}^{\mbox{\tiny{IR}}}\right)}\ln\left(\frac{2\alpha}{\pi}\frac{eB}{\tilde{m}_{\mbox{\tiny{IR}}}^{2}}\right)\right)
+{\cal{O}}\left(\frac{\tilde{m}_{\mbox{\tiny{IR}}}^{4}}{(eB)^{2}}\right).
\end{eqnarray}
\subsubsection*{Full critical temperature in the IR limit}
\par\noindent
Using the gap equation (\ref{S12}) with the ring improved effective
potential $\tilde{V}$, and following the same procedure described in
Sect. IV.A to determine the critical temperature arising from the
one-loop effective potential, we arrive first at the following
general expression for the \textit{full} critical temperature
\begin{eqnarray}\label{SS18}
\tilde{T}_{c}(\tilde{G};T)=T_{c}(\tilde{G};T)\exp\left(+\frac{2\pi^{2}}{eB
\tilde{m}^{\mbox{\tiny{}}}(1-\tilde{G})}\bigg[\frac{\partial
V_{ring}}{\partial
\tilde{m}^{\mbox{\tiny{}}}}-\tilde{G}\tilde{m}^{\mbox{\tiny{}}}\frac{\partial^{2}V_{ring}}{\partial
\tilde{m}^{\mbox{\tiny{}}2}}\bigg]\right)\bigg|_{\tilde{m}^{\mbox{\tiny{}}}(T_{c})=0}.
\end{eqnarray}
Here, $T_{c}(\tilde{G};T)$ is given in (\ref{SS8}) by replacing $G$
by $\tilde{G}$. To determine the full critical temperature
$\tilde{T}_{c}$ of dynamical chiral symmetry restoration in the IR
limit, we have to recalculate the ring potential, (\ref{A43}), for a
fixed, temperature independent mass cutoff $m_{0}$\footnote{$m_{0}$
plays the role of an IR regulator.}. To do this we separate the
integral over $k_{3}\in [0,\infty]$ in (\ref{A43}) into two regimes
$[0,m_{0}]$ and $[m_{0},\infty]$ and follow the same procedure which
led from (\ref{A43}) to (\ref{A62}). We arrive at the relevant
nonpertubative part of ring potential in the IR limit
\begin{eqnarray}\label{SS19}
V_{ring}^{\mbox{\tiny{IR}}}(eB,m_{0};T)=-\frac{m_{0}T
eB}{4\pi^{2}}\left\{\mbox{Li}_{2}\left(-\frac{M_{\gamma}^{2}}{m_{0}^{2}}\left(1-\frac{5}{3}C_{1}(m\beta,\pi)\right)\right)-
\mbox{Li}_{2}\left(-\frac{M_{\gamma}^{2}}{m_{0}^{2}}\right)\right\}.
\end{eqnarray}
Replacing (\ref{SS19}) on the r.h.s. of (\ref{SS18}), we get first
\begin{eqnarray}\label{SS20}
\lefteqn{\frac{2\pi^{2}}{eB
\tilde{m}^{\mbox{\tiny{}}}(1-\tilde{G}^{\mbox{\tiny{}}})}\bigg[\frac{\partial
V_{ring}^{\mbox{\tiny{}}}}{\partial
\tilde{m}^{\mbox{\tiny{}}}}-\tilde{G}^{\mbox{\tiny{}}}\tilde{m}^{\mbox{\tiny{}}}\frac{\partial^{2}V_{ring}^{\mbox{\tiny{}}}}{\partial
\tilde{m}^{\mbox{\tiny{}}2}}\bigg]=}\nonumber\\
&=& -\frac{5m^{\mbox{\tiny{}}}_{0}\beta
}{6\left(1-\frac{5}{3}C_{1}\right)\left(1-\tilde{G}^{\mbox{\tiny{}}}\right)}\left\{\frac{5}{3}C_{1}^{'2}\tilde{G}^{\mbox{\tiny{}}}\bigg[\frac{z_{0}^2}{\left(1+z_{0}^{2}\left(1-\frac{5}{3}C_{1}\right)\right)}
-\frac{\ln\left(1+z_{0}^{2}\left(1-\frac{5}{3}C_{1}\right)\right)}{1-\frac{5}{3}C_{1}}
\bigg]\right.
\nonumber\\
&&+\left.\bigg[\left(\frac{C_{1}^{'}}{\tilde{m}^{\mbox{\tiny{}}}\beta}-
C_{1}^{''}\tilde{G}^{\mbox{\tiny{}}}\right)\ln\left(1+z_{0}^{2}\left(1-\frac{5}{3}C_{1}\right)\right)\bigg]\right\}.
\end{eqnarray}
Here, $\tilde{G}=\tilde{G}^{\mbox{\tiny{IR}}}$ and
$\tilde{m}=\tilde{m}^{\mbox{\tiny{IR}}}$. Further, $z_{0}^{2}\equiv
z^{2}(m_{0})=
\frac{M_{\gamma}^{2}}{m_{0}^{2}}=\frac{2\alpha}{\pi}\frac{eB}{m_{0}^{2}}$.
Using now the definition $m(T_{c})=0$ and using (\ref{SS9}) as well
as (\ref{SS15}) to determine $C_{i}, i=0,1$, $C'_{1}$ and $C''_{1}$
at $m=0$, we arrive at the full critical temperature of QED in the
IR limit
\begin{eqnarray}\label{S21}
\tilde{T}_{c}^{\mbox{\tiny{IR}}}(\tilde{G}^{\mbox{\tiny{IR}}};T)=T_{c}(\tilde{G}^{\mbox{\tiny{IR}}};T)\exp\left(-\frac{35\zeta(3)}
{24\pi^{2}}(m_{0}\tilde{\beta}_{c}^{\mbox{\tiny{IR}}})\ln\left(1+z_{0}^{2}\right)\right).
\end{eqnarray}
Here, $T_{c}^{(1)}$ is given in (\ref{SS8}) and
$\tilde{\beta}_{c}^{\mbox{\tiny{IR}}}\equiv
1/\tilde{T}_{c}^{\mbox{\tiny{IR}}}$.
\subsection{Full dynamical mass and critical temperature of QED in the static limit}
\par\noindent
In Sect. III.C and III.D, we have presented two different approaches
leading to the ring contribution to the effective potential in the
static limit $(k_{0},\mathbf{k})=(0,\mathbf{0})$ for strong magnetic
field. First using the method presented in III.C, we arrived at the
nonperturbative part of the ring potential (\ref{A73}). In a second
approach in III.D, we just started from the ring potential
(\ref{A13}) and took the limit $eB\to \infty$. This leads to the
ring potential (\ref{A76}).  Although it seems that these two
approaches are physically equivalent, but according to our arguments
in Sect. III.D, they are indeed different.\footnote{In the first
approach, we take first the limit $eB\to\infty$ and then calculate
the ring integral. In the second approach, however, we calculate
first the ring integral and then take the limit $eB\to \infty$. As
it turns out, the limit $eB\to \infty$ and the integration over
$k_{3}$ in the LLL are not commutative. Physically, this can viewed
as a direct consequence of dimensional reduction, which is one of
the well-known properties of the LLL dynamics.} In this section, we
will determine the full dynamical mass and  critical temperature
using the ring potential in the static limit. In the next section,
these quantities are calculated using the ring potential in the
strong limit.\footnote{Although we believe that in the LLL the first
approach (static limit) is more reliable, we will present the
results corresponding to the strong limit in Sect. IV.D. This is
just to compare them with the result from IR limit in Set. IV.B.}
\subsubsection*{Full dynamical mass in the static limit}
\par\noindent
Let us consider first the nonperturbative part of the LLL ring
potential in the static limit (\ref{A73})
\begin{eqnarray}\label{SS22}
V_{ring}^{\mbox{\tiny{st./nonpert.}}}&=&\frac{mT}{8\pi^{2}}\bigg[(m^{2}-M_{\gamma}^{2})\ln\left(1+\frac{2M_{\gamma}^{2}C_{1}}
{m^{2}-M_{\gamma}^{2}}\right)-2C_{1}M_{\gamma}^{2}\bigg],
\end{eqnarray}
where we have omitted the irrelevant mass independent terms. The
\textit{full} dynamical mass $\tilde{m}(\tilde{G};T)$, arising from
one-loop and ring contribution to the effective potential is given
in (\ref{S13}), where $V_{ring}$ is to be replaced by (\ref{SS22}).
Doing this, we arrive first at
\begin{eqnarray}\label{SS23}
\lefteqn{\frac{2\pi^{2}}{eB
\tilde{m}(1-\tilde{G})}\bigg[\frac{\partial
V_{ring}^{\mbox{\tiny{static/nonpert.}}}}{\partial
\tilde{m}}-\tilde{G}\tilde{m}\frac{\partial^{2}V_{ring}^{\mbox{\tiny{static/nonpert.}}}}{\partial
\tilde{m}^{2}}\bigg]=}\nonumber\\
&=&-\frac{\tilde{m}T}{4eB\left(1-\tilde{G}\right)}\left\{2z^2\bigg[\tilde{m}\beta
\left((1-2\tilde{G})C_{1}^{'}-\tilde{m}\beta \tilde{G}
C_{1}^{''}\right)+C_{1}\bigg]-
\frac{4z^{4}\left(2C_{1}-\tilde{m}\beta(1-z^{2})C_{1}^{'}\right)^{2}}{(1-z^{2})\left(1-z^2+2z^{2}C_{1}\right)^{2}}
\right.\nonumber\\
&&
\left.-\big[3(1-2\tilde{G})-z^{2}\big]\ln\left(1+\frac{2z^{2}C_{1}}{1-z^{2}}\right)\right.\nonumber\\
&&\left.+\frac{2
z^{2}\bigg[2(1-3\tilde{G})C_{1}-\tilde{m}\beta(1-z^{2})\left((1-2\tilde{G})C_{1}^{'}-\tilde{m}\beta
\tilde{G} C_{1}^{''}\right)\bigg]}{(1-z^{2}+2z^{2}C_{1})} \right\}.
\end{eqnarray}
Here, $\tilde{G}=\tilde{G}^{\mbox{\tiny{st.}}}$ and
$\tilde{m}=\tilde{m}^{\mbox{\tiny{st.}}}$. Using (\ref{SS9}) and
(\ref{SS15}) to expand $C_{i}, i=0,1$ as well as $C'_{1}$ and
$C''_{1}$ in the order of $m\beta$, the full dynamical mass in the
static limit is given by
\begin{eqnarray}\label{SS24}
\tilde{m}^{\mbox{\tiny{st.}}}(\tilde{G}^{\mbox{\tiny{st.}}};T)\approx
m(\tilde{G}^{\mbox{\tiny{st.}}};T)\left(1+(\tilde{m}^{\mbox{\tiny{st.}}}\beta)^{3}{\cal{E}}^{\mbox{\tiny{st.}}}\right),
\end{eqnarray}
where $m^{(1)}(T)$ is given in (\ref{SS5}) and
\begin{eqnarray}\label{SS25}
{\cal{E}}^{\mbox{\tiny{st.}}}\equiv +\frac{245\alpha(\zeta(3))^{2}}
{64\pi^{5}}\frac{(1-4\tilde{G}^{\mbox{\tiny{st.}}})}{(1-\tilde{G}^{\mbox{\tiny{st.}}})}+
{\cal{O}}\left(\frac{\left(\tilde{m}^{\mbox{\tiny{st.}}}\right)^{4}}{(eB)^{2}}\right).
\end{eqnarray}
This result can be compared with (\ref{SS16})-(\ref{SS17}) from the
improved IR limit. Whereas $\tilde{m}^{\mbox{\tiny{IR}}}$ in
(\ref{SS17}) consists of a $\ln\alpha$ in the leading order of
$eB\to \infty$ and $m\beta\to 0$ limits,
$\tilde{m}^{\mbox{\tiny{st.}}}$ in (\ref{SS24})-(\ref{SS25}) has no
contribution in the order $\tilde{m}\beta$. Thus, in the high
temperature limit $m\beta\to 0$, we have practically
$\tilde{m}^{\mbox{\tiny{st.}}}\approx \tilde{m}$. As it will be
shown below, the ring potential in the static limit does not change
the full critical temperature in this limit too.
\subsubsection*{Full critical temperature in the static limit}
\par\noindent
Here, as in the previous part, the critical temperature of dynamical
chiral symmetry breaking can be determined only after recalculating
the ring potential (\ref{A66}) for a fixed, temperature independent
mass cutoff $m_{0}$. We separate the interval $[0,\infty]$ of the
integration over $k_{3}$ in (\ref{A66}) into two intervals
$[0,m_{0}]$ and $[m_{0},\infty]$ and follow the same steps leading
from (\ref{A66}) to (\ref{A73}) as the relevant nonperturbative part
of the ring potential. We arrive therefore at
\begin{eqnarray}\label{SS26}
V_{ring}^{\mbox{\tiny{static
limit}}}(eB,m_{0};T)=\frac{m_{0}^{3}T(1-z_{0}^{2})}{8\pi^2}\ln\left(1+\frac{2C_{1}z_{0}^{2}}{1-z_{0}^{2}}\right)
-\frac{m_{0}^{3}z_{0}^{2}TC_{1}}{4\pi^{2}}.
\end{eqnarray}
Replacing (\ref{SS26}) on the r.h.s. of  (\ref{SS18}) we get first
\begin{eqnarray}\label{SS27}
\lefteqn{\frac{2\pi^{2}}{eB
\tilde{m}(1-\tilde{G})}\bigg[\frac{\partial
V_{ring}^{\mbox{\tiny{IR}}}}{\partial
\tilde{m}}-\tilde{G}\tilde{m}\frac{\partial^{2}V_{ring}^{\mbox{\tiny{IR}}}}{\partial
\tilde{m}^{2}}\bigg]=}\nonumber\\
&=&
-\frac{M_{\gamma}^{2}m_{0}\beta}{2eB(1-\tilde{G})}\left\{\left(\frac{C_{1}^{'}}{\tilde{m}\beta}-C_{1}^{''}\tilde{G}\right)\left(1-\frac{(1-z_{0}^{2})}{\left(1-z_{0}^{2}(1-2C_{1})\right)}
\right)-2\tilde{G}C_{1}^{'2}z_{0}^{2}\left(\frac{(1-z_{0}^{2})}{\left(1-z_{0}^{2}(1-2C_{1})\right)^{2}}
\right)\right\}.\nonumber\\
\end{eqnarray}
Here, $\tilde{G}=\tilde{G}^{\mbox{\tiny{st.}}}$ and
$\tilde{m}=\tilde{m}^{\mbox{\tiny{st.}}}$. Using now the definition
$m(T_{c})=0$ and (\ref{SS9}) as well as (\ref{SS15}) to determine
$C_{i}, i=0,1$, $C'_{1}$ and $C''_{1}$ at $m=0$, it turns out that
full critical temperature of QED receives no contribution from the
ring potential in the static limit, \textit{i.e.}
\begin{eqnarray}\label{S28}
\tilde{T}_{c}^{\mbox{\tiny{st.}}}(\tilde{G}^{\mbox{\tiny{st.}}};T)=T_{c}(\tilde{G}^{\mbox{\tiny{st.}}};T),
\end{eqnarray}
where $T_{c}(\tilde{G}^{\mbox{\tiny{st.}}};T)$ can be read from
(\ref{SS8}) by replacing $G$ with $\tilde{G}^{\mbox{\tiny{st.}}}$.
\subsection{Full dynamical mass and critical temperature of QED in the strong limit}
\subsubsection*{Full dynamical mass in the strong limit}
\par\noindent
To determine the \textit{full} dynamical mass in the strong limit,
we use (\ref{S13}) and replace $V_{ring}$ by (\ref{A76})
\begin{eqnarray}\label{SS29}
V^{\mbox{\tiny{strong}}}_{ring}=\frac{T}{12\pi}\bigg[M_{\gamma}^{2}\left(1-2C_{1}\left(m\beta,\pi\right)\right)\bigg]^{3/2}.
\end{eqnarray}
Here, we have neglected the cutoff dependent terms. The full
dynamical mass $\tilde{m}(T)$, arising from one-loop effective
potential (\ref{A39}) and the ring contribution to the effective
potential is given in (\ref{S13}), where $V_{ring}$ is to be
replaced by (\ref{SS29}). Doing this, we arrive first at
\begin{eqnarray}\label{SS30}
\frac{2\pi^{2}}{eB \tilde{m}(1-\tilde{G})}\bigg[\frac{\partial
V_{ring}^{\mbox{\tiny{strong}}}}{\partial
\tilde{m}}-\tilde{G}\tilde{m}\frac{\partial^{2}V_{ring}^{\mbox{\tiny{strong}}}}{\partial
\tilde{m}^{2}}\bigg]=-
\frac{M_{\gamma}^{3}\pi\beta\left(1-2C_{1}\right)^{1/2}}{2eB(1-\tilde{G})}
\left(\frac{C_{1}^{'}}{\tilde{m}\beta}-C_{1}^{''}\tilde{G}+\frac{\tilde{G}C_{1}^{2}}{(1-2C_{1})}\right).
\end{eqnarray}
Here, $\tilde{G}=\tilde{G}^{\mbox{\tiny{str.}}}$ and
$\tilde{m}=\tilde{m}^{\mbox{\tiny{str.}}}$. In the high temperature
limit $m\beta\to 0$ and for the strong magnetic field, the dynamical
mass behaves as
\begin{eqnarray}\label{SS31}
\tilde{m}^{\mbox{\tiny{str.}}}(T)\approx
m(1+(\tilde{m}^{\mbox{\tiny{str.}}}\beta){\cal{E}}^{\mbox{\tiny{strong}}}),\qquad\mbox{with}\qquad
{\cal{E}}^{\mbox{\tiny{strong}}}\equiv
-\frac{7\zeta(3)\alpha}{2\pi^{2}}\left(\frac{2\alpha}{\pi}\frac{eB}{(\tilde{m}^{\mbox{\tiny{str.}}})^{2}}\right)^{1/2}.
\end{eqnarray}
This result is in contrast to ${\cal{E}}^{\mbox{\tiny{IR}}}$ from
(\ref{SS16})-(\ref{SS17}), where a novel term proportional to
$\ln\alpha$ appears. Besides it is in contrast to
(\ref{SS24})-(\ref{SS25}), where the first nonvanishing coefficient
in the $m\beta\to 0$ expansion is of order $(m\beta)^{3}$ and can be
practically neglected in the very high temperature.
\subsubsection*{Full critical temperature in the strong limit}
\par\noindent
To determine the critical temperature corresponding to the ring
improved effective potential, (\ref{SS18}) has to be used. Replacing
(\ref{SS29}) in the exponent of (\ref{SS18}), we arrive, as it is
shown above, at (\ref{SS30}).\footnote{Note that here, in contrast
to the previous two cases, no constant mass cutoff $m_{0}$ is
necessary to calculate the ring potential leading to the critical
temperature.} Setting $m=0$ in (\ref{SS30}) and using (\ref{SS9}) as
well as (\ref{SS15}), we arrive at
\begin{eqnarray}\label{SS32}
\tilde{T}_{c}^{\mbox{\tiny{strong}}}(\tilde{G}^{\mbox{\tiny{strong}}};T)=T_{c}(\tilde{G}^{\mbox{\tiny{strong}}};T)\exp\left(-\frac{7\zeta(3)\alpha}{4\pi^{2}}(m_{0}\tilde{\beta}_{c}^{\mbox{\tiny{strong}}})z_{0}\right).
\end{eqnarray}
Here, the temperature independent mass $m_{0}$ is introduced by
hand. This enables us to compare this result with the previous
results from the IR limit (\ref{S21}) and the static limit
(\ref{S28}).
\par
In the next section, we study the effect of ring contribution to the
effective potential in decreasing the critical temperature arising
only from the one-loop effective potential.\footnote{As it is known
from \cite{carrington}, the ring contribution to the effective
potential of Standard Model without magnetic field decreases the
critical temperature arising from one-loop effective potential. The
same phenomenon is shown to be true in the presence of magnetic
field \cite{demchik, bordag} in the static limit
$(k_{0},\mathbf{k})=(0,\mathbf{0})$.} To this purpose, we compare
numerically the critical temperature of chiral symmetry restoration
in three different approximation: $T_c$ from (\ref{S21}) in the IR
limit, $T_{c}$ from (\ref{S28}) in the static limit and finally
$T_{c}$ from (\ref{SS32}) in the strong limit.
\subsection{Numerical analysis of $T_{c}$}
\par\noindent
In Sect. IV.B - IV.D, we have determined the ring improved critical
temperature of QED in the LLL using the ring potential from IR,
static and strong limits. They are given in (\ref{S21}), (\ref{S28})
and (\ref{SS32}), respectively. Note that, according to our
arguments in Appendix C, these results are indeed exact (full in
quantum corrections). They include all quantum correction through
the coupling $\tilde{G}^{\{\mbox{\tiny{IR, st., strong}}\}}$ [see
(\ref{D13}) for a mathematically rigorous definition of
$\tilde{G}^{\aleph}, \aleph={\{\mbox{{IR, st., strong}}\}}$]. In
this section, we will study the effect of the ring diagram in
decreasing the critical temperature that arises from the lowest
order of $\alpha$-correction (ladder approximation), \textit{i.e.}
$T_{c}^{(1)}(G^{(1)};T)$ from (\ref{SS10}).\footnote{Ring
contributions lead to nonperturbative correction to the critical
temperature $T_{c}^{(1)}$.} To do this let us consider
$T_{c}(\tilde{G}^{\{\mbox{\tiny{IR, st., strong}}\}};T)$, the
factors behind the ring contribution on the r.h.s. of (\ref{S21}),
(\ref{S28}) and (\ref{SS32}). Now consider
$T_{c}(\tilde{G}^{\{\mbox{\tiny{IR, st., strong}}\}};T)$, only in
the lowest order of $\alpha$ correction and denote the resulting
expression by $T_{c}(\tilde{G}^{(1)/\{\mbox{\tiny{IR, st.,
strong}}\}};T)$. Next, use the following approximation, which is
only reliable in the high temperature limit $m\beta\to 0$ and for
$eB\gg m^{2}$,\footnote{Note that according to our descriptions in
Appendix C.4, for different approaches to the ring potential
$\aleph=\{\mbox{IR},\mbox{static},\mbox{strong}\}$, the quantity
$\Delta
T_{c}^{\aleph}=T_{c}^{\aleph}(\tilde{G}^{(1)/\aleph};T)-T_{c}^{(1)}(G^{(1)};T)$
vanishes only in the limit of high temperature ($m\beta\to 0$) and
strong magnetic field ($eB\gg m^{2}$). Thus the above approximation
(\ref{SS0}) is only reliable in these limits. To check this for the
IR limit, for instance, let us consider $G^{(1)}$ and
$\tilde{G}^{(1)/\mbox{\tiny{IR}}}$ from (\ref{D6}) and (\ref{D16}),
respectively. In the lowest order of $\alpha$ correction
$G^{(1)}-\tilde{G}^{(1)/\mbox{\tiny{IR}}}=-\frac{1}{m_{dyn.}\Omega}\frac{\partial}{\partial\langle\bar{\psi}\psi\rangle}\left(\alpha
{\cal{R}}_{N=1}^{(n=0)}\right)$. This is a term that arises from the
two-loop diagram (a) in Fig. 3 ($N=1$) and includes only zero
Matsubara frequencies ($n=0$). As it turns out, in the limit of
$eB\gg m^{2}$, this term leads to a term proportional to
$(m\beta)^{-1}$. Adding this contribution to the gap equation
arising from one-loop effective potential (\ref{SS3}), this
contribution can indeed be neglected in the high temperature limit
$m\beta\to 0$ comparing to logarithmic divergent term including in
$K_{0}(n m \beta)$ in $m\beta\to 0$ limit [see our descriptions in
the paragraph following (\ref{SS3})].}
\begin{eqnarray}\label{SS0}
T_{c}(\tilde{G}^{\mbox{\tiny{(1)/IR}}};T)\approx
T_{c}(\tilde{G}^{\mbox{\tiny{(1)/st.}}};T)\approx
T_{c}(\tilde{G}^{\mbox{\tiny{(1)/strong}}};T)\approx
T_{c}^{(1)}(G^{(1)};T).
\end{eqnarray}
Using this approximation, we will be now able to compare the effect
of ring potential in different limits, the IR, the static and the
strong limits in decreasing $T_{c}^{(1)}$. To do this, let us first
define the ring improved critical temperature by
\begin{eqnarray}\label{SS0a}
{\cal{T}}_{c}\equiv
T_{c}^{(1)}\exp\left(-{\frac{m_{0}\kappa}{{\cal{T}}_{c}}}\right).
\end{eqnarray}
The argument in the exponent is the nonperturbative contribution
from ring potential. Comparing (\ref{SS0a}) with the critical
temperature arising from ring potential in IR, static and strong
limit from (\ref{S21}), (\ref{S28}) and (\ref{SS32}), respectively,
and using the approximation (\ref{SS0}),  we get
\begin{eqnarray}\label{S36}
\kappa^{\mbox{\tiny{IR}}}=\frac{35\zeta(3)}{24\pi^{2}}\ln(1+z_{0}^{2}),
\end{eqnarray}
for the IR limit,
\begin{eqnarray}\label{S37}
\kappa^{\mbox{\tiny{static}}}=0,
\end{eqnarray}
for the static limit, and
\begin{eqnarray}\label{S38}
\kappa^{\mbox{\tiny{strong}}}=\frac{7\zeta(3)\alpha}{4\pi^{2}}z_{0},
\end{eqnarray}
for the strong limit. Then, solving (\ref{SS0a}) as a function of
${{\cal{T}}_{c}}$, the ring improved critical temperature
${\cal{T}}_{c}$ in all three cases can be determined as a function
of the one-loop critical temperature in the ladder approximation,
$T_{c}^{(1)}$, the temperature independent mass, $m_{0}$, and the
parameter $\kappa$. Doing this, we get
\begin{eqnarray}\label{S39}
{\cal{T}}_{c}=-\frac{m_{0}\kappa}{W\left(-\frac{m_{0}}{T_{c}^{(1)}}\kappa\right)},
\end{eqnarray}
where the Lambert-function $W(z)$, is a function that satisfies
\cite{lambert}
$$W(z)e^{W(z)}=z.$$
To have a quantitative first estimate on the effect of the ring
potential on decreasing the one-loop critical temperature
$T_{c}^{(1)}$, we define further
\begin{eqnarray}\label{S40}
u\equiv\frac{T_{c}^{(1)}}{{\cal{T}}_{c}}=\frac{W(-a \kappa)}{-a
\kappa}\qquad \mbox{with}\qquad a\equiv \frac{m_{0}}{T_{c}^{(1)}},
\end{eqnarray}
using (\ref{S39}) and the efficiency factor
\begin{eqnarray}\label{S41}
\eta\equiv 1-\frac{1}{u}.
\end{eqnarray}
In Table I the values of $u$ and $\eta$ for various choices of
$eB\in [10^{-8},1]$ GeV$^{2}$ and for fixed $a=2$ are listed.
\begin{table}[ht]
\begin{tabular}{||c|c||c|c||c|c||c|c||}
\hline eB in GeV$^2$&$B$ in Gau\ss&
$u^{\mbox{\tiny{IR}}}$&$\eta^{\mbox{\tiny{IR}}}$ in \%&
$u^{\mbox{\tiny{static}}}$&$\eta^{\mbox{\tiny{static}}}$ in \%&
$u^{\mbox{\tiny{strong}}}$&$\eta^{\mbox{\tiny{strong}}}$ in \%\\
\hline\hline \hline
$10^{-8}$&$1.7\times 10^{12}$&$1.00007$&$0.01\%$&$1.$&$0\%$&$1.00004$&$0.004\%$ \\
\hline
$10^{-7}$&$1.7\times 10^{13}$&$1.00066$&$0.07\%$&$1.$&$0\%$&$1.0001$&$0.01\%$ \\
\hline
$10^{-6}$&$1.7\times 10^{14}$&$1.01$&$0.65\%$&$1.$&$0\%$&$1.0004$&$0.04\%$ \\
\hline
$10^{-5}$&$1.7\times 10^{15}$&$1.07$&$6.26\%$&$1.$&$0\%$&$1.0013$&$0.13\%$ \\
\hline
$3.2\times 10^{-5}$&$5.3\times 10^{15}$&$1.22$&$18.19\%$&$1.$&$0\%$&$1.0024$&$0.24\%$\\
\hline
$5.0\times 10^{-5}$&$8.5\times 10^{15}$&$1.38$&$27.60\%$&$1.$&$0\%$&$1.0030$&$0.30\%$ \\
\hline
$7.9\times 10^{-5}$&$1.3\times 10^{16}$&$1.77$&$43.39\%$&$1.$&$0\%$&$1.0038$&$0.38\%$ \\
\hline
$8.9\times 10^{-5}$&$1.5\times 10^{16}$&$2.07$&$50.17\%$&$1.$&$0\%$&$1.0040$&$0.40\%$ \\
\hline
$9.8\times 10^{-5}$&$1.6\times 10^{16}$&$2.67$&$62.62\%$&$1.$&$0\%$&$1.0042$&$0.42\%$ \\
\hline
$10^{-4}$&$1.7\times 10^{16}$&$2.65-0.45 i$&--&$1.$&$0\%$&$1.0043$&$0.43\%$ \\
\hline
$10^{-1}$&$1.7\times 10^{19}$&$-0.14-0.7 i$&--&$1.$&$0\%$&$1.1699$&$14.53\%$ \\
\hline
$1$&$1.7\times 10^{20}$&$-0.16-0.5 i$&--&$1.$&$0\%$&$2.13-1.24 i$&-- \\
\hline
\end{tabular}
\caption{The values of $u$ and the efficiency factor $\eta$ for
different values of $eB$ and different limits (IR, static and strong
limit). Here $a$, the proportionality factor between $m_{0}$ and
$T_{c}^{(1)}$ is chosen to be $a=2$. The efficiency factor $\eta$
increases by increasing the strength of magnetic field. For a given
value of $eB$, the IR limit is more efficient in decreasing the
critical temperature from its value arising from one-loop effective
potential $T_{c}^{(1)}$. }
\end{table}

The results are also drawn in the graphs of Fig. 2. Here, $eB$ is in
GeV$^2$ ($1$ GeV=$10^9$ eV) which is equivalent to $B=1.691\times
10^{20}$ in Gau\ss.\footnote{In this paper, we have worked in Planck
units, where $\hbar=c=1$. In these units $eB$ has the dimension of
energy, \textit{i.e.} Joule ($J$) and will be denoted $eB$ as
$[eB]_{\mbox{\tiny{J}}}$. To get a relation between
$[eB]_{\mbox{\tiny{J}}}$ and $B$ in Gau\ss, we have to convert $eB$
into SI units, where we get $eB=\frac{[eB]_{\mbox{\tiny{J}}}}{\hbar
c^2}$. Having in mind that $\hbar=1.054\times 10^{-34}$Js,
$e=1.602\times 10^{-19}$C and $c=2.998\times 10^8$ m/s, we get
$B=6.589\times 10^{35}\left([eB]_{\mbox{\tiny{J}}}\right)^2$ Tesla=
$6.589\times 10^{39} \left([eB]_{\mbox{\tiny{J}}}\right)^2$ in
Gau\ss, providing $[eB]_{\mbox{\tiny{J}}}$ is in Joule. Choosing,
for instance, $eB=1$ GeV$^{2}$, which is equivalent to
$\left([eB]_{\mbox{\tiny{J}}}\right)^2=2.567\times 10^{-20}$
J$^{2}$, we get $B=1.691\times 10^{20}$ in Gau\ss. Here, we have
used $1$J=$6.241\times 10^9$ GeV.} The above range corresponds
therefore to $B\in [1.7\times 10^{12}, 1.7\times 10^{20}]$ Gau\ss,
which is phenomenologically relevant in the astrophysics of neutron
stars, where it is believed that the strength of the magnetic field
is of the order $10^{13}-10^{15}$ Gau\ss. It is also relevant in the
heavy ion experiments, for example in RHIC, where it is believed
that the magnetic field in the center of gold-gold collision is
$10^2-10^3$ MeV$^{2}$ corresponding to $B\sim 10^{16}-10^{17}$
Gau\ss\ (Here, the center of mass energy is $\sim 200$ GeV per
nucleon pair) \cite{mclerran}. Defining further $m_{0}$ as the zero
temperature mass, \textit{i.e.} $m_{0}\equiv m(0)$, the above
choices for $a$ are indeed justified by the fact that the dynamical
mass at zero temperature $m(0)$ is proportional to the critical
temperature $T_{c}^{(1)}$ with a proportionality factor
$a={\cal{O}}(1)$ \cite{shovkovy, ebert}. To determine
$z_{0}^{2}=\frac{2\alpha}{\pi}\frac{eB}{m_{0}^{2}}$, we have fixed
$\alpha=\frac{1}{137}$ and chosen $m_{0}=0.5$ MeV, the electron mass
at zero temperature.

\begin{center}
\begin{figure}[h]
\includegraphics[width=5.5cm, height=4.125cm]{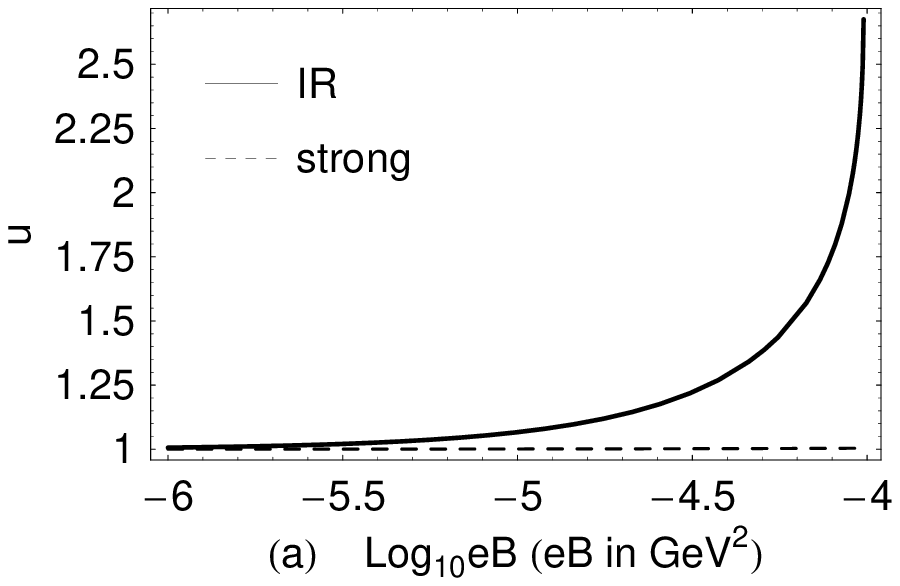}
\includegraphics[width=5.5cm, height=4.125cm]{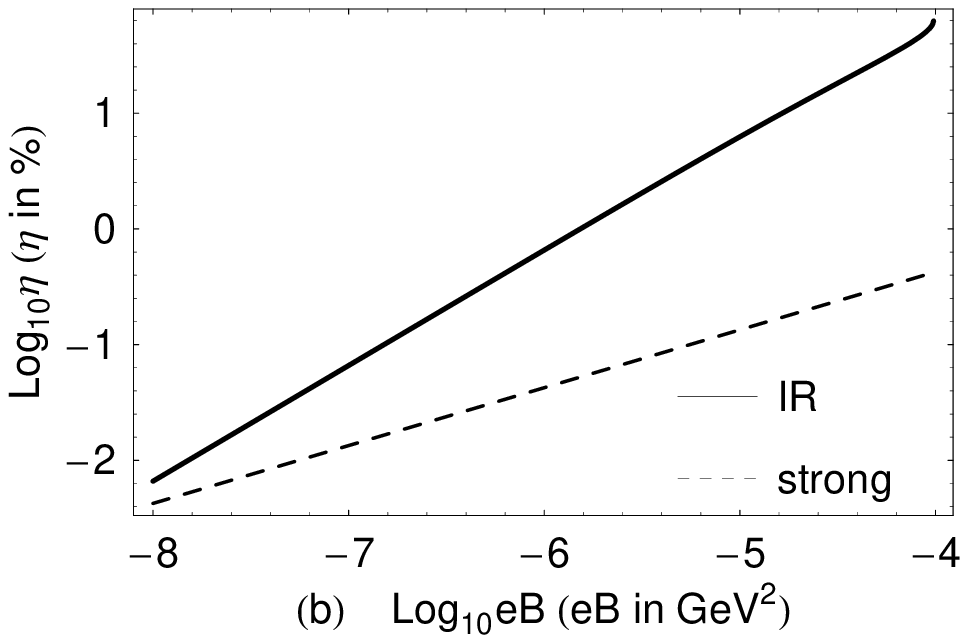}
\includegraphics[width=5.3cm, height=4.05cm]{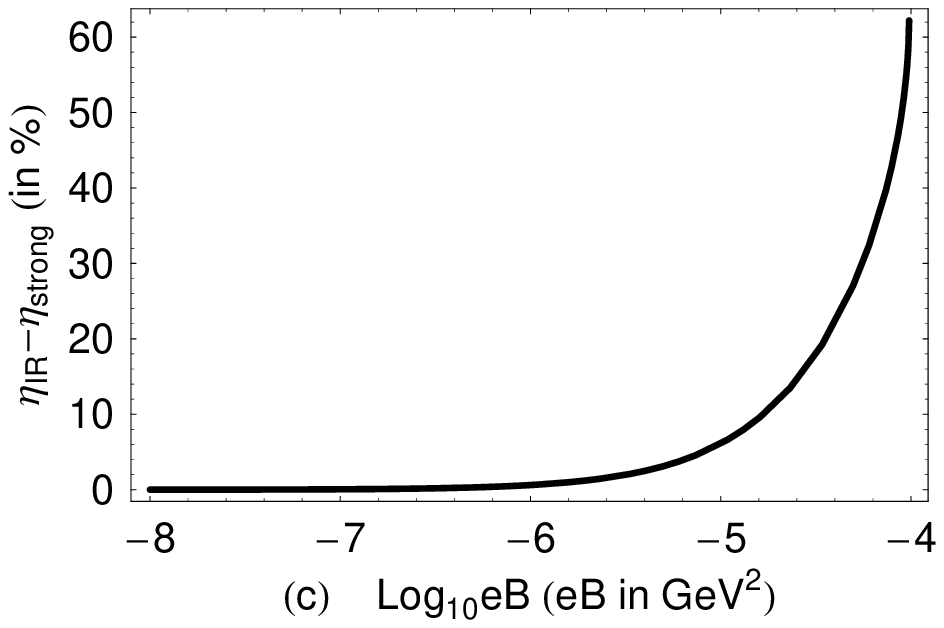}
\caption{(a) Fixing $a=2$, the ratio $u\equiv
T_{c}^{(1)}/{\cal{T}}_{c}$ for the IR and strong limit is considered
versus $\mbox{Log}_{10}eB$ ($eB$ in GeV$^{2}$). The effect of the
ring potential in the strong limit in decreasing $T_{c}^{(1)}$ is
minimal comparing to the effect of the ring potential in the IR
limit. (b) For $\eta$ the efficiency factor, $\mbox{log}_{10}\eta$
is considered versus $\mbox{log}_{10} eB$ for the improved IR and
the strong limit. The efficiency factor $\eta$ increases with
increasing eB. (c) The difference of the efficiency factor between
the improved IR and strong limits, $\Delta\eta\equiv
\eta^{\mbox{\tiny{IR}}}-\eta^{\mbox{\tiny{strong}}}$, is considered
versus $\mbox{Log}_{10}eB$. $\Delta\eta$ increases with increasing
$eB$. The maximum $\Delta\eta_{\star}$ is $\approx 62\%$ for
$B\approx 1.6\times 10^{16}$ Gau\ss. Note that $u$ for the static
limit is $u^{static}=1$. }
\end{figure}
\end{center}
As it can be seen in Table I, for every given values of $m_{0}$ and
$a$, there is always a certain value of $(eB)_{\star}$, for which
$u$ is imaginary and $\eta$ cannot be defined. This is due to the
fact that the Lambert $W$-function, $W(z)$ in (\ref{S40}), has a
branch cut discontinuity in the complex $z$ plane running from
$z=-\infty$ to $z=-1/e$. Here, $e$ is the Euler number. Using
(\ref{S40}), this threshold can be determined for the IR and the
strong limits as
\begin{eqnarray}\label{S42}
(eB)_{\star}^{\mbox{\tiny{IR limit}}}=\frac{\pi
m_{0}^{2}}{2\alpha}\left(-1+\exp\left(\frac{24\pi^{2}}{35 a
e\zeta(3)}\right)\right),\qquad (eB)_{\star}^{\mbox{\tiny{strong
limit}}}= \frac{8\pi^{5}m_{0}^{2}}{49
a^{2}e^{2}\alpha^{3}(\zeta(3))^{2}}.
\end{eqnarray}
For $a=2$ and $m_{0}=0.5$ MeV, we get therefore
\begin{eqnarray}\label{S43}
(eB)_{\star}^{\mbox{\tiny{IR limit}}}=9.77\times 10^{-5} \mbox{\
GeV}^{2}\qquad \mbox{or}\qquad B_{\star}^{\mbox{\tiny{IR
limit}}}=1.65\times 10^{16} \mbox{\ Gau\ss}.
\end{eqnarray}
The corresponding efficiency factor
$\eta^{\mbox{\tiny{IR}}}_{\star}\equiv
\eta^{\mbox{\tiny{IR}}}((eB)_{\star})=63.21\%$. This means a
variation from the corresponding efficiency factor in the strong
limit $\Delta\eta\equiv
\eta^{\mbox{\tiny{IR}}}-\eta^{\mbox{\tiny{strong}}}=62.79\%$. We
conclude therefore, that the IR limit, compared to the static and
the strong limit, leads to maximum efficiency factor $\eta$ for a
given value of $eB$ (see Fig. 2).
\section{Conclusion}
In the first part of this paper, using the vacuum polarization
tensor $\Pi_{\mu\nu}(k_0,\mathbf{k})$ in the IR limit $k_{0}\to 0$,
the general structure of the plasmon (ring) potential of QED is
determined in a constant magnetic field $B$. Then, taking the limit
of weak and strong magnetic field, the ring improved effective
potential including the one-loop and the ring potentials is
determined. In the weak magnetic field limit, the effective
potential consists of a $T^{4}\alpha^{5/2}$ term, in addition to the
expected $T^{4}\alpha^{3/2}$ contribution arising in the static
$(k_{0}\to 0,\mathbf{k}\to\mathbf{0})$ limit. The additional
corrections are potentially relevant for the study of electroweak
phase transition in the presence of weak magnetic field limit
\cite{ayala}. Note that similar terms of order $\alpha_{s}^{3/2}$
and $\alpha_{s}^{5/2}$ appear also in QCD effective potential at
finite temperature and without magnetic field. They are calculated
using the Hard Thermal Loop expansion \cite{braaten} (see also
\cite{alpha5-2} and the references therein). It would be interesting
to develop the same program for QED and QCD at finite temperature
and in the presence of weak/strong magnetic field.
\par
Next, QED ring potential is calculated in the strong magnetic field
limit. In this limit, QED dynamics is dominated by LLL and the
chiral symmetry of the theory is broken as a result of a dynamically
generated fermion mass. To study this well-known phenomenon of
magnetic catalysis for QED at finite temperature in the LLL, the
ring improved effective potential of the theory is determined in
strong magnetic field limit. In particular, the ring potential is
determined in the IR, $k_{0}\to 0$, as well as the static limit,
$(k_{0}\to 0, \mathbf{k}\to 0)$. In the IR limit, it includes a
novel term consisting of a dilogarithmic function
$(eB)\mbox{Li}_{2}\left(-\frac{2\alpha}{\pi}\frac{eB}{m^{2}}\right)$.
Similar term in the form $g_{s}^{4}\ln g_{s}$ appears also in QCD
ring potential at finite temperature and zero magnetic field
\cite{toimela}. As for the static limit in the presence of strong
magnetic field, there are indeed two different approaches leading to
different ring potentials in this limit [see III.C and III.D for
more details. Here, these  two limits are indicated by static and
strong limits]. Physically, the difference between these two results
lies in the dimensional reduction as one of the consequences of LLL
dynamics.
\par
In the second part of this paper, using the ring improved effective
potential in the IR, static and strong limits, the gap equation, the
dynamical mass and critical temperature $T_{c}$ of chiral symmetry
restoration of QED are determined. Note that the critical
temperature could only be determined by choosing a temperature
independent IR cutoff $m_{0}$ in the integrals leading to the ring
potential. Concerning the two different ring potentials in the
static and strong limits, we note that according to our arguments in
Sect. III.C and III.D, once we consider QED in the LLL at finite
temperature, we have to determine the full dynamical mass and
critical temperature using the ring potential in the static limit.
We have presented nevertheless the results arising from ring
potential in strong limit in Sect. IV.D and compared the results
from Sect. IV.B - IV.D in Sect. IV.E.
\par
To have an estimate on the efficiency of the IR limit in decreasing
the critical temperature from its value arising from the one-loop
effective potential, $T_{c}^{(1)}$, we have numerically determined
$u=T_{c}^{(1)}/{\cal{T}}_{c}$ for various magnetic fields and as a
function of $m_{0}$. Here, ${\cal{T}}_{c}$ is the ring improved
critical temperature defined in (\ref{SS0}). Further, to compare the
IR limit with the static and strong limit, we have defined an
efficiency factor $\eta=1-u^{-1}$ for the IR, static and strong
limits. As it turns out, for a given values of $eB$, the IR limit,
compared to the static and the strong limits, is more efficient in
decreasing the critical temperature $T_{c}^{(1)}$. The maximum
efficiency factor in the IR limit is $\eta^{\mbox{\tiny{IR}}}\approx
63\%$ for $B\approx 1.6\times 10^{16}$ Gau\ss.
\par
Apart from its importance in the framework of magnetic catalysis,
the above conclusion can be regarded as a promising result
concerning the problem of electroweak phase transition (EWPT) in the
electroweak SM in the presence of strong hypermagnetic field. There,
one is looking for a possibility to decrease the critical
temperature of EWPT in order to improve the baryogenesis condition
$\frac{\langle v\rangle }{T_{c}}>1-1.5$, where $\langle v\rangle$ is
the Higgs mass \cite{gavela-kajantie, ayala}. Note that the
existence of baryon number violation in the SM is realized by means
of its vacuum structure through sphaleron mediated processes. The
sphaleron transition between different topological distinct vacua is
associated to baryon number $n_{B}-n_{\bar{B}}$ violation and can
either induce of wash out a baryon asymmetry. In order to satisfy
the baryon asymmetry condition during the baryogenesis process the
rate of baryon violating transitions between different topological
vacua must be suppressed in the broken phase, when the universe
returns to thermal equilibrium. In other words, the sphaleron
transition must be slower than the expansion of the universe and
this in turn translates into the condition $\frac{\langle v\rangle
}{T_{c}}>1-1.5$ \cite{ayala}. Using the improved ring potential in
the IR limit in determining the critical temperature of EWPT in SM
may improve the results of \cite{bordag}-\cite{ayala-2} as one of
the possible solutions of baryon asymmetry problem within the
minimal SM \cite{sadooghi}.

\section{Acknowledgment}
We would like to thank the referee of this paper for instructive
questions and comments, that have led,in particular, to a
considerable improvement of the results in Sect. IV.
\begin{appendix}
\section{Ring improved effective potential of QED in LLL at $T=0$}
\par\noindent
In this appendix, using the composite effective action based on CJT
(Cornwall-Jackiw-Tomboulis) approach \cite{cornwall}, we will define
the ring improved effective potential of QED at zero temperature in
the LLL approximation. This fixes at the same time the notations
used in the following appendices, where the gap equation of QED at
zero and nonzero temperature are derived. The latter is used in
Sect. IV to determine the dynamical mass and critical temperature of
QED in the LLL approximation.
\par
Let us start with the composite effective action of QED (see
\cite{cornwall} and \cite{miransky-book} for more details)
\begin{eqnarray}\label{AB1}
\Gamma[{\cal{G}},{\cal{D}}_{\mu\nu}]=-i\mbox{Tr}\ln
{\cal{G}}^{-1}-i\mbox{Tr}\left(S^{-1}{\cal{G}}\right)+\Gamma_{2}\left({\cal{G}},{\cal{D}}_{\mu\nu}\right)+{\cal{F}}[{\cal{D}}_{\mu\nu}].
\end{eqnarray}
Here, $S$ is the free propagator of massless fermions. Further,
${\cal{G}}$ is the full fermion and ${\cal{D}}_{\mu\nu}$ is the full
photon propagator. As it is described in \cite{cornwall}, the
composite effective action (\ref{AB1}) includes the sum of two- and
higher-loop two-particle irreducible (2PI) diagrams. Note that
$\Gamma_{2}$ on the r.h.s. of (\ref{AB1}) includes the contribution
of diagrams which are two-particle irreducible with respect to
fermion lines only. In two loop order, the diagrams included in
$\Gamma_{2}$ are shown in Fig. 3.
\par\vspace{0.4cm}
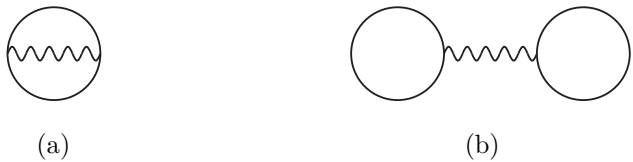
\begin{figure}[hbt]
\SetScale{0.35}
  \begin{picture}(20,20)(0,0)
    \SetWidth{2}
    \CArc(-300,0)(50,270,630)
    \Photon(-350,0)(-250,0){7.5}{5}
    \Text(-112,-40)[lb]{(a)}
    \CArc(70,0)(50,270,630)
    \Photon(120,0)(220,0){7.5}{5}
    \CArc(270,0)(50,270,630)
    \Text(50,-40)[lb]{(b)}
  \end{picture}
  \vspace{1.5cm}
\caption{Two-loop diagrams contributing to $\Gamma_{2}$ in
(\ref{AB1}). In the ladder LLL approximation, solid lines correspond
to free fermion propagator in the LLL with $m=m_{dyn.}$ from
(\ref{Z6})-(\ref{Z8}), and wavy lines correspond to free photon
propagator ${\cal{D}}_{\mu\nu}^{(0)}$ from (\ref{AB7}). Diagram (a)
is the same diagram appearing in (\ref{X6}) and can be regarded as
the diagram corresponding to $N=1$ in the ring potential (\ref{A9})
at zero and nonzero temperature. It is denoted by
$\alpha{\cal{R}}_{1}$ in (\ref{AB22}) and as
$\sum_{n}\alpha{\cal{R}}_{1}^{(n)}$ in (\ref{D4}).}
\end{figure}
Replacing the full fermion propagator ${\cal{G}}$ in (\ref{AB1}) by
the fermion propagator $\bar{S}_{\mbox{\tiny{LLL}}}$  of massive
fermions in the LLL approximation with mass $m=m_{dyn.}$, from
(\ref{Z6})-(\ref{Z9}), and, the full photon propagator,
${\cal{D}}_{\mu\nu}$ by the free photon propagator
$D_{\mu\nu}^{(0)}$ from (\ref{AB7}), the composite effective action
$\bar{\Gamma}_{\mbox{\tiny{LLL}}}\equiv
\Gamma[\bar{S}_{{\mbox{\tiny{LLL}}}},D_{\mu\nu}^{(0)}]$ in the
ladder LLL approximation reads\footnote{In the definition of
effective potential we have used $\Gamma=\Omega V$, where $\Gamma$
is the effective action, $V$ is the effective potential and $\Omega$
is the four-dimensional space-time volume.}
\begin{eqnarray}\label{AB18}
\bar{\Gamma}_{\mbox{\tiny{LLL}}}\simeq \Omega \tilde{V}(m_{dyn.},eB;
T=0)-i\mbox{Tr}\left(S_{{\mbox{\tiny{LLL}}}}^{-1}\bar{S}_{{\mbox{\tiny{LLL}}}}\right)
+\tilde{\Gamma}_{2}^{(\infty)}[\bar{S}_{{\mbox{\tiny{LLL}}}},
D_{\mu\nu}^{(0)}]+{\cal{F}}[D_{\mu\nu}^{(0)}].
\end{eqnarray}
Here, the free fermion propagator of massless fermions
$S_{\mbox{\tiny{LLL}}}$ is given in (\ref{Z6})-(\ref{Z8}) with
$m=0$. In (\ref{AB18}), the ring improved (one-loop) effective
potential $\tilde{V}(m_{dyn.},eB;T=0)$ in the ladder LLL
approximation is introduced. It is defined by the one-loop effective
potential ${V}^{(1)}(m_{dyn.},eB; T=0)$ and the ring potential
$V_{ring}(m_{dyn.},eB;T=0)$,
\begin{eqnarray}\label{AB19a}
\tilde{V}(m_{dyn.},eB; T=0)\equiv {V}^{(1)}(m_{dyn.},eB;
T=0)+V_{ring}(m_{dyn.},eB;T=0).
\end{eqnarray}
The one-loop effective potential in the LLL approximation is defined
by the first term in (\ref{AB1}) with ${\cal{G}}$ replaced by
$\bar{S}_{\mbox{\tiny{LLL}}}$,
\begin{eqnarray}\label{AB19}
V^{(1)}(m_{dyn.},eB; T=0)\equiv -i\Omega^{-1}\mbox{Tr}\ln
\bar{S}_{\mbox{\tiny{LLL}}}^{-1}.
\end{eqnarray}
In a constant magnetic field, (\ref{AB19}) is calculated in
\cite{sato} using the method of worldline formalism. In the LLL
approximation, it is given by
\begin{eqnarray}\label{AB17}
V^{(1)}(m,eB;
T=0)&=&-\frac{eB}{8\pi^{2}}\int_{\frac{1}{\Lambda^{2}}}^{\infty}\frac{ds}{s^{2}}\
e^{-sm^{2}}\nonumber\\
&=& -\frac{eB m^{2}}{8\pi^{2}}\
\Gamma\left(-1,\frac{m^{2}}{\Lambda^{2}}\right)\stackrel{\Lambda\to\infty}{\longrightarrow}
\frac{eB m^{2}}{8\pi^{2}}\left(\ln
\frac{m^{2}}{\Lambda^{2}}+{\cal{O}}(m^{0})\right),
\end{eqnarray}
[see also (\ref{A38}) for $T\neq 0$ case]. Here $m$ is a general
nonvanishing mass. On the last line of (\ref{AB17}),
$\Gamma(0,z)\stackrel{z\to 0}{\longrightarrow} -\ln z-\gamma$ is
used. Here, $\gamma$ is the Euler-Mascheroni constant.
\par
As for the ring potential, $V_{ring}$ from (\ref{AB19a}), it
receives contribution from the ring diagrams in (\ref{X6}) and Fig.
1. These diagrams are already included in $\Gamma_{2}$ from
(\ref{AB1}).\footnote{Using the fermion gap equation $\frac{\delta
{\Gamma}[\bar{S}_{{\mbox{\tiny{LLL}}}},{S}_{{\mbox{\tiny{LLL}}}},D^{(0)}]}{\delta\bar{S}_{{\mbox{\tiny{LLL}}}}}=0$,
it can be shown that
$\alpha^{N}{\cal{R}}_{N}[\bar{S}_{{\mbox{\tiny{LLL}}}},
D_{\mu\nu}^{(0)}]$ from (\ref{AB18}) leads, up to a constant factor,
to the same $\alpha^{N}$ correction in the resulting SD equation,
that arises from
$\frac{\delta\Gamma[{\cal{G}}_{{\mbox{\tiny{LLL}}}},{S}_{{\mbox{\tiny{LLL}}}},{\cal{D}}_{{\mbox{\tiny{LLL}}}}]}{\delta{\cal{G}}_{{\mbox{\tiny{LLL}}}}}=0$
where full photon propagator ${\cal{D}}_{\mu\nu}$ is used. In this
sense, ring diagrams are already included in a composite effective
action, where instead of free photon propagators, as in
(\ref{AB18}), full photon propagators are used. Other $\alpha^{N}$
corrections arising from quantum corrections of full fermion
propagator in the LLL, ${\cal{G}}_{\mbox{\tiny{LLL}}}$, are included
in $\bar{\Gamma}_{2}^{(\infty)}$ in (\ref{AB18}).} At zero
temperature, the ring potential of QED is given by
\begin{eqnarray}\label{AB22}
V_{ring}(m_{dyn.},eB;T=0)&\equiv&
\Omega^{-1}\sum\limits_{N=1}^{\infty}\alpha^{N}
{\cal{R}}_{N}[\bar{S}_{{\mbox{\tiny{LLL}}}},
D_{\mu\nu}^{(0)}]\nonumber\\
&=&-\frac{1}{2}\int\frac{d^{4}k}{(2\pi)^{4}}\sum\limits_{N=1}^{\infty}\frac{(-1)^{N}}{N}\bigg[D_{\mu\rho}^{(0)}(k)\Pi^{\rho\mu}(k)\bigg]^{N},
\nonumber\\
&=&-\frac{1}{2}\int\frac{d^{4}k}{(2\pi)^{4}}\
\ln\left(1+D_{\mu\rho}^{(0)}(k)\Pi^{\rho\mu}(k)\right).
\end{eqnarray}
[see also (\ref{A9}) for $T\neq 0$ case]. Here, ${\cal{R}}_{N}$
denotes the contribution of the $N$-th diagram in the ring series in
Fig. 1, where $N$ vacuum polarization tensor $\Pi_{\mu\nu}$ are
inserted in a photon loop. Note that in the LLL ladder (rainbow)
approximation, the vacuum polarization tensor in (\ref{AB22}) is
determined using \textit{free} propagator
$\bar{S}_{\mbox{\tiny{LLL}}}$ of \textit{massive} fermions in LLL
approximation with mass $m=m_{dyn.}$ from
(\ref{Z6})-(\ref{Z8}).\footnote{The ring potential $V_{ring}$ from
(\ref{AB22}) for zero external magnetic field and in the static
limit, \textit{i.e.} for $\Pi_{\rho\nu}\equiv\Pi_{\rho\nu}(k=0)$ in
(\ref{AB22}), is previously calculated by Akhiezer \textit{et al.}
\cite{akhiezer}. It leads apart from cutoff dependent terms to
nonperturbative $\alpha^{2}\log\alpha$ corrections to the effective
action.} As we have mentioned above, ring diagrams are already
included in $\Gamma_{2}$ from (\ref{AB1}). Thus
$\Gamma_{2}^{(\infty)}$ is (\ref{AB18}) is defined by
\begin{eqnarray}\label{AB21}
\tilde{\Gamma}_{2}^{(\infty)}[\bar{S}_{{\mbox{\tiny{LLL}}}},D_{\mu\nu}^{(0)}]\equiv
\Gamma_{2}[\bar{S}_{{\mbox{\tiny{LLL}}}},D_{\mu\nu}^{(0)}]-\Omega
V_{ring}(m_{dyn.},eB;T=0).
\end{eqnarray}
Here, the subscript $\infty$ in $\Gamma_{2}^{(\infty)}$ means that
infinitely many ring diagrams are subtracted from $\Gamma_{2}$ in
(\ref{AB1}).
\section{Gap equation of QED at zero
temperature}
\subsection{Gap equation from one-loop effective potential at zero
temperature}\par\noindent In the LLL, once the dynamical mass is
generated via the mechanism of magnetic catalysis, the chiral
symmetry of the originally massless QED breaks spontaneously. This
leads to the formation of a chiral condensate $\langle
\bar{\psi}\psi\rangle$. For a nonvanishing fermion mass, $m$, the
nonvanishing chiral condensate can be most easily calculated by (see
\cite{schwinger} and Eq. (19) in \cite{miransky-2} for more details)
\begin{eqnarray}\label{AB23}
\langle \bar{\psi}\psi\rangle\equiv -\lim\limits_{x\to
y}\mbox{tr}\left(\bar{S}_{\mbox{\tiny{LLL}}}(x,y;m)\right)\simeq
\frac{eB
m}{4\pi^{2}}\left(\ln\frac{m^{2}}{\Lambda^{2}}+{\cal{O}}\left(m^{0}\right)\right).
\end{eqnarray}
Here, $\Lambda$ is a large momentum cutoff. Comparing now the value
of the chiral condensate (\ref{AB23}) with the one-loop effective
potential (\ref{AB17}), we arrive at\footnote{Relation (\ref{AB25})
can generally be derived for $eB=0$ using the definition of one-loop
effective action \cite{ebert}. }
\begin{eqnarray}\label{AB25}
\langle\bar{\psi}\psi\rangle=\frac{\partial
V^{(1)}(m,eB;T=0)}{\partial m}.
\end{eqnarray}
The above chiral condensate can be used to determine the gap
equation of QED from composite effective action. To do this in the
LLL approximation, let us consider the composite effective action
$\bar{\Gamma}_{\mbox{\tiny{LLL}}}$ from (\ref{AB18}) and use the
fermion gap equation $\delta \bar{\Gamma}_{\mbox{\tiny{LLL}}}/\delta
\bar{S}{\mbox{\tiny{LLL}}} =0$, to arrive at
\begin{eqnarray}\label{AB30}
\frac{\partial \bar{\Gamma}_{\mbox{\tiny{LLL}}}}{\partial
\langle\bar{\psi}\psi\rangle}=\int d^{4}x\
\mbox{tr}\left(\frac{\delta \bar{\Gamma}{\mbox{\tiny{LLL}}}}{\delta
\bar{S}{\mbox{\tiny{LLL}}}(x)}\ \frac{\partial
\bar{S}{\mbox{\tiny{LLL}}}(x)}{\partial
\langle\bar{\psi}\psi\rangle}\right)=0.
\end{eqnarray}
Using now the gap equation $\frac{\partial
\bar{\Gamma}_{\mbox{\tiny{LLL}}}}{\partial
\langle\bar{\psi}\psi\rangle}=0$ and the definition of
$\bar{\Gamma}_{\mbox{\tiny{LLL}}}$  from (\ref{AB18}), we arrive at
the gap equation of QED arising from one-loop effective potential
(\ref{AB19}) [see also (\ref{S1})]
\begin{eqnarray}\label{AB26}
\frac{\partial V^{(1)}\left(m_{dyn.},eB;T=0\right)}{\partial
\langle\bar{\psi}\psi\rangle}\equiv G_{0}m_{dyn.}.
\end{eqnarray}
Here, the ``effective coupling'' of QED in the LLL dominant regime,
$G_{0}$, is defined by
\begin{eqnarray}\label{AB28}
G_{0}\equiv - \frac{1}{m_{dyn.}\Omega}\frac{\partial }{\partial
\langle\bar{\psi}\psi\rangle}\left(-i\mbox{Tr}\left(S_{{\mbox{\tiny{LLL}}}}^{-1}\bar{S}_{{\mbox{\tiny{LLL}}}}\right)+\alpha
{\cal{R}}_{1}[\bar{S}_{{\mbox{\tiny{LLL}}}},
D_{\mu\nu}^{(0)}]+\tilde{\Gamma}_{2}^{(1)}[\bar{S}_{{\mbox{\tiny{LLL}}}},
D_{\mu\nu}^{(0)}]+{\cal{F}}[D_{\mu\nu}^{(0)}] \right),
\end{eqnarray}
where the definition of $\bar{\Gamma}_{\mbox{\tiny{LLL}}}$ from
(\ref{AB18}) is used. In general $G_{0}$ is a complicated function
of $\alpha$ (see below). It can be determined order by order in
$\alpha$ by computing the r.h.s. of (\ref{AB28}). In the lowest
order of $\alpha$ correction, for instance, it receives contribution
from $\alpha{\cal{R}}_{1}$ [diagram (a) in Fig. 3], and diagram (b)
in Fig 3, which is included in $\tilde{\Gamma}_{2}^{(1)}$. To
determine $G_{0}$ in the lowest order of $\alpha$ correction, let us
first give the gap equation (\ref{AB26}) in a more appropriate form
by making use of the identity
\begin{eqnarray}\label{AB38}
\frac{\partial}{\partial
\langle\bar{\psi}\psi\rangle}=\left(\frac{\partial^{2}V^{(1)}}{\partial
m_{dyn.}^{2}}\right)^{-1}\frac{\partial}{\partial m_{dyn.}},
\end{eqnarray}
that can be derived from $\frac{\partial V^{(1)}}{\partial
m}=\langle\bar{\psi}\psi\rangle$. As for the gap equation
(\ref{AB26}), it is given by
\begin{eqnarray}\label{AB31}
\frac{\partial V^{(1)}\left(eB,m_{dyn.};T=0\right)}{\partial
m_{dyn.}}=G_{0}m_{dyn.}\frac{\partial^{2}V^{(1)}\left(eB,m_{dyn.};T=0\right)}{\partial
m^{2}}.
\end{eqnarray}
The above gap equation (\ref{AB31}) can now be used to ``fix''
$G_{0}$ order by order as a function in $\alpha$. To show this, let
us consider the one-loop effective potential from (\ref{AB17}) with
$m=m_{dyn.}$ and plug it into (\ref{AB31}). We arrive at
\begin{eqnarray}\label{AB32}
\left(1-G_{0}\right)\Gamma\left(0,\frac{m^{2}_{dyn.}}{\Lambda^{2}}\right)+2G_{0}\Gamma\left(1,\frac{m^{2}_{dyn.}}{\Lambda^{2}}\right)=0.
\end{eqnarray}
Setting $\Lambda= \Lambda_{B}\equiv \sqrt{eB}$ and assuming that
$m_{dyn.}\ll \Lambda_{B}$, the $\Gamma$-functions on the r.h.s. of
(\ref{AB32}) can be expanded as
\begin{eqnarray}\label{AB33}
\Gamma\left(0,z\right)\stackrel{z\to 0}{\longrightarrow}-\gamma-\ln
z,\qquad\mbox{and}\qquad\Gamma(1,z)\stackrel{z\to
0}{\longrightarrow} 1.
\end{eqnarray}
Plugging these relations into (\ref{AB32}), we arrive at the
following gap equation
\begin{eqnarray}\label{AB34}
\ln\frac{m_{dyn.}}{\Lambda_{B}}=-\frac{\gamma}{2}+\frac{G_{0}}{1-G_{0}},
\end{eqnarray}
that leads to
\begin{eqnarray}\label{AB35}
m_{dyn.}(G_{0};T=0)={\cal{C}}\Lambda_{B}\exp\left(\frac{G_{0}}{1-G_{0}}\right),
\qquad\mbox{with}\qquad {\cal{C}}=e^{-\gamma/2}.
\end{eqnarray}
This is indeed a general structure of the dynamical mass as a
function of $G_{0}$. The latter includes all higher loop
contributions through its definition from  (\ref{AB28}). To
determine $G_{0}$ in the lowest order of $\alpha$ correction, we use
the result of the dynamical mass $m_{dyn.}^{(1)}$ in ladder
approximation from (\ref{Z16}) and compare (\ref{AB35}) with it.
Thus, the ``effective coupling of QED in the LLL'', $G_{0}$, can be
``fixed'' in this lowest order of $\alpha$ correction as
\begin{eqnarray}\label{AB36}
G_{0}^{(1)}\equiv \frac{1}{1-\sqrt{\frac{\alpha}{\pi}}}\approx
1+\sqrt{\frac{\alpha}{\pi}}.
\end{eqnarray}
Here, similar to $m_{dyn.}^{(1)}$ the superscript (1) denotes that
$G_{0}^{(1)}$ receives contribution from two-loop diagrams of order
$\alpha$ that are shown in Fig. 3. These diagrams contribute to the
composite effective action as it can be checked from (\ref{AB18}).
It would be interesting to perform a bottom-up calculation of
$G_{0}^{(1)}$ from its definition (\ref{AB28}) in the one-loop
level. Calculating the expression in the parentheses in (\ref{AB28})
up to order $\alpha$ and replacing $m_{dyn.}(G_{0};T=0)$ from
(\ref{AB35}) leads to a nontrivial equation for $G_{0}$, whose
solution leads to an expression for $G_{0}$ that can be compared
with $G_{0}^{(1)}$ from (\ref{AB36}) in the lowest order of $\alpha$
correction.\footnote{This calculation will be performed elsewhere
\cite{sadooghi-2}.}

\subsection{Gap equation from the ring improved effective potential at zero temperature}
\par\noindent
To determine explicitly the contribution of ring diagrams to the
dynamical mass and critical temperature, we will use in Sect. IV.B
an alternative gap equation. It arises from the ring improved
effective potential $\tilde{V}(eB,m_{dyn.};T\neq 0)$ at finite
temperature. At zero temperature, the ring improved effective
potential is given in (\ref{AB19a}). The corresponding ring improved
gap equation reads then
\begin{eqnarray}\label{AB45}
\frac{\partial \tilde{V}(m_{dyn.},eB;T=0)}{\partial
\langle\bar{\psi}\psi\rangle}=\tilde{G_{0}}m_{dyn.},
\end{eqnarray}
where
\begin{eqnarray}\label{AB46}
\tilde{G_{0}}=- \frac{1}{m_{dyn.}\Omega}\frac{\partial}{\partial
\langle\bar{\psi}\psi\rangle}\left(-i\mbox{Tr}\left(S_{{\mbox{\tiny{LLL}}}}^{-1}\bar{S}_{{\mbox{\tiny{LLL}}}}\right)+\tilde{\Gamma}_{2}^{(\infty)}[\bar{S}_{{\mbox{\tiny{LLL}}}},
D_{\mu\nu}^{(0)}]+{\cal{F}}[D_{\mu\nu}^{(0)}]\right).
\end{eqnarray}
Comparing to $G_{0}$ from (\ref{AB28}) we get
\begin{eqnarray}\label{AB47}
G_{0}=\tilde{G_{0}}-\frac{1}{m_{dyn.}}\frac{\partial
V_{ring}(m_{dyn.},eB;T=0)}{\partial \langle\bar{\psi}\psi\rangle},
\end{eqnarray}
where $V_{ring}$ is given in (\ref{AB22}).
\section{Gap equation of QED at finite temperature}
\subsection{Composite effective action in the LLL approximation at
finite temperature}
\par\noindent
Let us consider the ladder LLL composite effective action
(\ref{AB18}) at zero temperature. Its generalization to finite
temperature is indicated as $\bar{\Gamma}_{T}\equiv
\Gamma_{\mbox{\tiny{LLL}}}[\bar{S}_{\mbox{\tiny{LLL}}},D_{\mu\nu}^{(0)};T]$,
which is defined as
\begin{eqnarray}\label{D1}
\bar{\Gamma}_{\mbox{\tiny{LLL}}}^{T}\equiv
\Omega^{-1}\tilde{V}\left(m_{dyn.},eB;T\right)-i\mbox{Tr}\left(
S^{-1}_{\mbox{\tiny{LLL}}}\bar{S}_{\mbox{\tiny{LLL}}}\right)_{T}+\tilde{\Gamma}_{2}^{(\infty)}[\bar{S}_{\mbox{\tiny{LLL}}},D_{\mu\nu}^{(0)};T]+
{\cal{F}}[D_{\mu\nu}^{(0)};T].
\end{eqnarray}
Here, $m_{dyn.}\equiv m_{dyn.}(T)$  is the temperature dependent
dynamical mass. The ring improved effective potential,
$\tilde{V}(m_{dyn.},eB;T)$, is defined as
\begin{eqnarray}\label{D2}
\tilde{V}(m_{dyn.},eB;T)=V^{(1)}(m_{dyn.},eB;T)+V_{ring}\left(m_{dyn.},eB;T\right).
\end{eqnarray}
Here, the one-loop effective potential at finite temperature is
defined as in (\ref{AB19}) by
\begin{eqnarray}\label{D3}
V^{(1)}\left(m_{dyn.},eB;T\right)\equiv -i\Omega^{-1}\mbox{Tr}\ln
\bar{S}_{LLL}^{T},
\end{eqnarray}
and the ring potential is given as in (\ref{AB22}) by
\begin{eqnarray}\label{D4}
V_{ring}\left(m_{dyn.},eB;T\right)\equiv
\sum\limits_{n=-\infty}^{\infty}\alpha^{N}{\cal{R}}_{N}^{(n)}[\bar{S}{\mbox{\tiny{LLL}}},D_{\mu\nu}^{(0)};T].
\end{eqnarray}
In (\ref{D3}) the bare fermion propagator in the LLL is the
generalization of (\ref{Z6})-(\ref{Z8}) to finite temperature. For a
general nonzero mass, the free fermion propagator at finite
temperature including the contribution of all Landau levels is given
by (\ref{Z18}). The ring potential in (\ref{D4}) is defined in
(\ref{A9}), where the contribution of all diagrams in Fig. 1 and the
diagram in (\ref{X6}) at finite temperature\footnote{This is the
$N=1$ term in (\ref{A9}).} is taken into account. As it can be seen
in (\ref{A9}) to determine the ring potential, we have to add over
all $n\in ]-\infty,\infty[$ that label the Matsubara frequencies
$\omega_{n}$. The summation over $n$ in (\ref{D4}) denotes the same
summation over $n$ of Matsubara frequencies and
${\cal{R}}_{N}^{(n)}$ is the corresponding contribution of the
$N$-th ring diagram to the $n$-th Matsubara frequency.
\subsection{Gap equation from one-loop effective potential at finite temperature}
\par\noindent
In Sect. IV.A, the gap equation arising from one-loop effective
potential is used to determine the dynamical mass (\ref{SS5}) and
the critical temperature (\ref{SS8}) at finite temperature. The gap
equation from one-loop effective potential at finite $T$ is given by
\begin{eqnarray}\label{D5}
\frac{\partial V^{(1)}\left(m_{dyn.},eB;T\right)}{\partial
\langle\bar{\psi}\psi\rangle}\equiv G_{}m_{dyn.}.
\end{eqnarray}
It is indeed a generalization of (\ref{AB26})-(\ref{AB28}) to finite
temperature. Using the ladder LLL composite effective action
(\ref{D1}), the effective coupling of QED in the LLL dominant regime
at finite temperature, $G_{}$, can be defined as
\begin{eqnarray}\label{D6}
G_{}\equiv - \frac{1}{m_{dyn.}}\frac{\partial }{\partial
\langle\bar{\psi}\psi\rangle}\left( -i\mbox{Tr}\left(
S^{-1}_{\mbox{\tiny{LLL}}}\bar{S}_{\mbox{\tiny{LLL}}}\right)_{T}+\alpha{\cal{R}}_{1}[\bar{S}_{\mbox{\tiny{LLL}}},D_{\mu\nu}^{(0)};T]+\tilde{\Gamma}_{2}^{(1)}[\bar{S}{\mbox{\tiny{LLL}}},D_{\mu\nu}^{(0)};T]+
{\cal{F}}[D_{\mu\nu}^{(0)};T]\right).\nonumber\\
\end{eqnarray}
In Sect. IV.A, we will use this gap equation to determine the
dynamical mass $m_{dyn.}^{(1)}(G_{};T)$ and the critical temperature
$T_{c}^{(1)}(G_{})$ in the lowest order of $\alpha$ correction in
ladder (rainbow) LLL approximation [see (\ref{SS5}) for the
dynamical mass and (\ref{SS10}) for the critical temperature].
\subsection{Gap equation from ring improved effective potential at finite temperature}
\par\noindent
The gap equation arising from the ring improved effective potential
at finite temperature can be given most easily as a generalization
of (\ref{AB45})-(\ref{AB46}) to finite temperature. It is given by
\begin{eqnarray}\label{D7}
\frac{\partial \tilde{V}(m_{dyn.},eB;T)}{\partial
\langle\bar{\psi}\psi\rangle}=\tilde{G}_{}m_{dyn.},
\end{eqnarray}
where $\tilde{G}_{}$ is defined by
\begin{eqnarray}\label{D8}
\tilde{G}_{}=- \frac{1}{m_{dyn.}\Omega}\frac{\partial}{\partial
\langle\bar{\psi}\psi\rangle}\left(-i\mbox{Tr}\left(S_{{\mbox{\tiny{LLL}}}}^{-1}\bar{S}_{{\mbox{\tiny{LLL}}}}\right)_{T}
+\tilde{\Gamma}_{2}^{(\infty)}[\bar{S}_{{\mbox{\tiny{LLL}}}},
D_{\mu\nu}^{(0)};T]+{\cal{F}}[D_{\mu\nu}^{(0)};T]\right).
\end{eqnarray}
Comparing to $G_{}$ from (\ref{D6}) we get
\begin{eqnarray}\label{D9}
G_{}=\tilde{G}_{}-\frac{1}{m_{dyn.}}\frac{\partial
V_{ring}(m_{dyn.},eB;T)}{\partial \langle\bar{\psi}\psi\rangle}.
\end{eqnarray}
\subsection{Gap equations used in sections IV.B- IV.D}\par\noindent
The general form of the gap equations (\ref{D7})-(\ref{D8}) are not
exactly the gap equations that are used in Sect. IV.B - IV.D to
determine the full dynamical mass and critical temperature of QED in
the ladder LLL approximation. There, we have used the ring potential
in three different approaches
$\aleph=\{\mbox{IR},\mbox{static},\mbox{strong}\}$. To give the
general definition of the gap equation corresponding to all these
three approaches, we will indicate the ring potential in the
$\aleph$-approach by $V_{ring}^{\aleph}(m_{dyn.},eB;T)$ and use the
separation
\begin{eqnarray}\label{D10}
V_{ring}(m_{dyn.},eB;T)\equiv
V_{ring}^{\aleph}(m_{dyn.},eB;T)+\bar{V}_{ring}^{\aleph}(m_{dyn.},eB;T).
\end{eqnarray}
In Sect. IV.B, IV.C and IV.D, the superscript $\aleph=\mbox{IR}$,
$\aleph=\mbox{static}$ and $\aleph=\mbox{strong}$, respectively. We
define further the ring improved effective potential corresponding
to $\aleph$-approach by one-loop effective potential (\ref{D2}) and
the ring potential in $\aleph$-approach
\begin{eqnarray}\label{D11}
\tilde{V}^{\aleph}(m_{dyn.},eB;T)=V^{(1)}\left(m_{dyn.},eB;T\right)+V_{ring}^{\aleph}(m_{dyn.},eB;T).
\end{eqnarray}
The gap equation in the corresponding $\aleph$-approach is therefore
given by
\begin{eqnarray}\label{D12}
\frac{\partial \tilde{V}^{\aleph}(m_{dyn.},eB;T)}{\partial
\langle\bar{\psi}\psi\rangle}=\tilde{G}^{\aleph}\ m_{dyn.},
\end{eqnarray}
where $m_{dyn.}=m_{dyn.}^{\aleph}$ and the corresponding coupling,
$\tilde{G}_{}^{\aleph}$ is given by
\begin{eqnarray}\label{D13}
\tilde{G}^{\aleph}&=&-\frac{1}{m_{dyn.}\Omega}\frac{\partial}{\partial
\langle\bar{\psi}\psi\rangle}\left\{-i\mbox{Tr}\left(S_{{\mbox{\tiny{LLL}}}}^{-1}\bar{S}_{{\mbox{\tiny{LLL}}}}\right)_{T}\right.\nonumber\\
&&\left.+\Omega^{-1}\bar{V}_{ring}^{\aleph}(m_{dyn.},eB;T)+\tilde{\Gamma}_{2}^{(\infty)}[\bar{S}_{{\mbox{\tiny{LLL}}}},
D_{\mu\nu}^{(0)};T]+{\cal{F}}[D_{\mu\nu}^{(0)};T]\right\}.
\end{eqnarray}
We have therefore
\begin{eqnarray}\label{DA}
\tilde{G}_{}=\tilde{G}^{\aleph}-\frac{1}{m_{dyn.}}\frac{\partial
V_{ring}^{\aleph}(m_{dyn.},eB;T)}{\partial
\langle\bar{\psi}\psi\rangle}.
\end{eqnarray}
Let us finally give an example to clarify the above notation. In
Sect. IV.B for instance, we have used the ring potential in the IR
limit. In particular, this is determined by $n=0$ in (\ref{D4}).
Using the definitions
\begin{eqnarray}\label{D14}
V_{ring}^{\mbox{\tiny{IR}}}(m_{dyn.},eB;T)&\equiv&
\Omega^{-1}\sum\limits_{N=1}^{\infty}\alpha^{N}{\cal{R}}^{(n=0)}_{N}[\bar{S}_{{\mbox{\tiny{LLL}}}},
D_{\mu\nu}^{(0)};T],\nonumber\\
\bar{V}_{ring}^{\mbox{\tiny{IR}}}(m_{dyn.},eB;T)&\equiv&\Omega^{-1}
\sum\limits_{\stackrel{n=-\infty}{n\neq
0}}^{\infty}\sum\limits_{N=1}\alpha^{N}{\cal{R}}^{(n)}_{N}[\bar{S}_{{\mbox{\tiny{LLL}}}},
D_{\mu\nu}^{(0)};T],
\end{eqnarray}
to separate zero and nonzero Matsubara frequencies, the gap equation
in the IR approach is given by
\begin{eqnarray}\label{D15}
\frac{\partial \tilde{V}^{\mbox{\tiny{IR}}}(m_{dyn.},eB;T)}{\partial
\langle\bar{\psi}\psi\rangle}=\tilde{G}^{\mbox{\tiny{IR}}}\
m_{dyn.},
\end{eqnarray}
where the corresponding coupling is given by
\begin{eqnarray}\label{D16}
\tilde{G}_{}^{\mbox{\tiny{IR}}}&=&-
\frac{1}{m_{dyn.}\Omega}\frac{\partial}{\partial
\langle\bar{\psi}\psi\rangle}\left\{-i\mbox{Tr}\left(S_{{\mbox{\tiny{LLL}}}}^{-1}\bar{S}_{{\mbox{\tiny{LLL}}}}\right)_{T}\right.\nonumber\\
&&\left.+\Omega^{-1}\bar{V}_{ring}^{\mbox{\tiny{IR}}}(m_{dyn.},eB;T)+\tilde{\Gamma}_{2}^{(\infty)}[\bar{S}_{{\mbox{\tiny{LLL}}}},
D_{\mu\nu}^{(0)};T]+{\cal{F}}[D_{\mu\nu}^{(0)};T]\right\}.
\end{eqnarray}
\end{appendix}

\end{document}